\documentclass[10pt]{article}

\usepackage{amsmath}
\usepackage{array}
\usepackage{appendix}
\usepackage{graphicx}
\usepackage{amsfonts}
\usepackage{amssymb}
\usepackage{mathrsfs}
\usepackage{yfonts}
\usepackage{euscript}
\usepackage{upgreek}
\usepackage{slantsc}
\usepackage{calligra}
\usepackage[T1]{fontenc}
\usepackage{epsf}
\usepackage{latexsym}

\usepackage{tipa}

\def\be{\begin{equation}}
\def\ee{\end{equation}}
\def\beq{\begin{equation}}
\def\eeq{\end{equation}}
\def\bea{\begin{eqnarray}}
\def\eea{\end{eqnarray}}

\def\ni{\noindent}
\def\foo{\footnote}

\def\!{\hspace{-1.6667em}}

\def\mA{\mbox{A}}   
  
\def\mC{\mbox{C}}   
\def\mD{\mbox{D}}
\def\mE{\mbox{E}}
\def\mF{\mbox{F}}
\def\mG{\mbox{G}}
 
\def\mI{\mbox{I}}

\def\mL{\mbox{L}}
\def\mM{\mbox{M}}
\def\mN{\mbox{N}} 
\def\mO{\mbox{O}}
\def\mP{\mbox{P}}
\def\mQ{\mbox{Q}}
\def\mR{\mbox{R}}
\def\mS{\mbox{S}}
 
\def\mU{\mbox{U}}
\def\mV{\mbox{V}}

\def\md{\mbox{d}} 
\def\me{\mbox{e}}
\def\mf{\mbox{f}}
\def\mg{\mbox{g}}
\def\mh{\mbox{h}}

\def\ml{\mbox{l}}   
   
\def\mn{\mbox{n}}   
\def\mo{\mbox{o}}
\def\mp{\mbox{p}}

\def\ms{\mbox{s}}

\def\muu{\mbox{u}}

\def\sFrU{\mbox{\boldmath\scriptsize$\mathfrak{U}$}}
\def\sFrV{\mbox{\boldmath\scriptsize$\mathfrak{V}$}}
\def\sFrW{\mbox{\boldmath\scriptsize$\mathfrak{W}$}}

\def\brho{\mbox{\boldmath$\rho$}}          
\def\bupSigma{\mbox{\boldmath$\Sigma$}}                 

\def\fC{\mbox{\sffamily C}}
\def\fD{\mbox{\sffamily D}}

\def\r{\underline{r}}

\def\q{\underline{q}}

\def\bh{\underline{\underline{\mbox{h}}}  }            
\def\sbh{\underline{\underline{\mbox{\scriptsize h}}}  }     

\def\barp{\bar{\tt p}}

\def\bp{\mbox{\bf p}}

\def\bM{\mbox{\bf M}}

\def\bM{\mbox{{\bf M}}}
\def\bM{\mbox{{\bf M}}}

\def\bh{\mbox{{\bf h}}}

\def\br{\mbox{{\bf r}}}

\def\sumnlm{\sum\mbox{}_{\mbox{}_{\mbox{\scriptsize $\sn, \sll, \sm$}}}}

\def\scD{\mbox{\scriptsize ${\cal D}$}}          

\def\scG{\mbox{\scriptsize ${\cal G}$}}          
\def\scH{\mbox{\scriptsize ${\cal H}$}}          

\def\scL{\mbox{\scriptsize ${\cal L}$}}          
\def\scM{\mbox{\scriptsize ${\cal M}$}}          

\def\scP{\mbox{\scriptsize ${\cal P}$}}

\def\bigr{\mbox{\boldmath$\mathfrak{R}$}}

\def\Fr{\mbox{\Large $\mathfrak{r}$}}

\def\FP{\mbox{\Large $\mathfrak{p}$}}
\def\FrQ{\mbox{\Large $\mathfrak{q}$}}
\def\FrE{\mbox{$\mathfrak{E}$}}                   
\def\FrT{\mathfrak{T}}                            
\def\FrN{\mathfrak{N}}                            
\def\FrY{\mathfrak{Y}} 
\def\FrA{\mbox{\boldmath$\mathfrak{A}$}} 
\def\FrB{\mbox{\boldmath$\mathfrak{B}$}}          
\def\FrU{\mbox{\boldmath$\mathfrak{U}$}}
\def\FrP{\mbox{\Large\boldmath$\mathfrak{p}$}}    
\def\FrH{\mbox{\boldmath$\mathfrak{H}$}}          
\def\FrV{\mbox{\Large\boldmath$\mathfrak{v}$}}          

\def\FrX{\mathfrak{X}}

\def\FrT{\mbox{\boldmath$\mathfrak{T}$}}                        

\def\sFrG{\mbox{\boldmath\small$\mathfrak{g}$}}
\def\sFrQ{\mbox{\boldmath\small$\mathfrak{g}$}}

\def\FrF{\mbox{$\mathfrak{F}$}}                                 
 
\def\FrM{\mbox{\Large $\mathfrak{m}$}}                         
\def\FrMgen{\mbox{\boldmath$\mathfrak{M}$}}                     

\def\FrR{\mbox{\Large $\mathfrak{r}$}}

\def\FrS{\mbox{\Large $\mathfrak{s}$}}

\def\FrG{\mbox{\Large $\mathfrak{g}$}}                            
 
\def\bFrB{\mbox{\boldmath$\mathfrak{B}$}}         
\def\bFrE{\mbox{\boldmath$\mathfrak{E}$}}         

\def\bFrF{\mbox{\boldmath$\mathfrak{F}$}}         

\def\lFrg{\mbox{\Large \textfrak{g}}}             
                                                  												  
\def\bFrl{\mbox{\Large $\mathfrak{l}$}}               
												  
\def\bFrM{\mbox{\boldmath$\mathfrak{M}$}}         

\def\FrO{\mbox{\Large $\mathfrak{o}$}}
   
\def\FA{\mbox{\Large $\mathfrak{a}$}}

\def\sa{\mbox{\scriptsize a}}
\def\sb{\mbox{\scriptsize b}}
\def\scc{\mbox{\scriptsize c}}

\def\se{\mbox{\scriptsize e}}

\def\sg{\mbox{\scriptsize g}} 
\def\sh{\mbox{\scriptsize h}}

\def\sll{\mbox{\scriptsize l}}  
\def\sm{\mbox{\scriptsize m}}
\def\sn{\mbox{\scriptsize n}} 
\def\so{\mbox{\scriptsize o}} 
\def\sp{\mbox{\scriptsize p}}
\def\sq{\mbox{\scriptsize q}}
\def\sr{\mbox{\scriptsize r}}

\def\su{\mbox{\scriptsize u}}
\def\sv{\mbox{\scriptsize v}}

\def\sy{\mbox{\scriptsize y}} 

\def\sA{\mbox{\scriptsize A}} 

\def\sC{\mbox{\scriptsize C}}

\def\sF{\mbox{\scriptsize F}}

\def\sP{\mbox{\scriptsize P}}

\def\sS{\mbox{\scriptsize S}}

\def\sfA{\mbox{\sffamily{\scriptsize A}}}      
\def\sfB{\mbox{\sffamily{\scriptsize B}}}      
\def\sfC{\mbox{\sffamily{\scriptsize C}}}      
\def\sfD{\mbox{\sffamily{\scriptsize D}}}      
\def\sfV{\mbox{\sffamily{\scriptsize V}}}      
\def\sfW{\mbox{\sffamily{\scriptsize W}}}      

\def\sbh{\mbox{{\bf \scriptsize h}}}

\def\sbM{\mbox{{\bf \scriptsize M}}}

\def\tn{\mbox{\tiny n}}

\def\tfC{\mbox{\sffamily{\tiny C}}}

\def\bFrP{\mbox{\Large $\mathfrak{p}$}}         
\def\K{Kucha\v{r} }

\def\pa{\partial}
\def\d{\textrm{d}}

\def\5Star{\mbox{\Large$\star$}}              
\def\cr{\mbox{\scriptsize{\bf $\mbox{ } \times \mbox{ }$}}}

\def\sumi2{\sum\mbox{}_{\mbox{}_{\mbox{\scriptsize $i$=1}}}^2}
\def\sumi3{\sum\mbox{}_{\mbox{}_{\mbox{\scriptsize $i$=1}}}^3}

\def\sumj3{\sum\mbox{}_{\mbox{}_{\mbox{\scriptsize $j$=1}}}^3}
\def\sumk3{\sum\mbox{}_{\mbox{}_{\mbox{\scriptsize $k$=1}}}^3}

\begin{document}

\begin{titlepage}

\begin{center}

{\Large{\bf CONFIGURATION SPACES IN FUNDAMENTAL PHYSICS}}

\vspace{.1in}

{\bf Edward Anderson} 

\vspace{.1in}

{\em DAMTP, Centre for Mathematical Sciences, Wilberforce Road, Cambridge CB3 OWA.} \normalsize

\end{center}

\begin{abstract}

I consider configuration spaces for $N$-body problems, gauge theories and for GR in both geometrodynamical and Ashtekar variables forms, 
including minisuperspace and inhomogeneous perturbations thereabout in the former case.
These examples include many interesting spaces of shapes (with and without whichever of local or global notions of scale). 
In considering reduced configuration spaces, stratified manifolds arise. 
Three strategies to deal with these are `excise', `unfold' and `accept'. 
I show that spaces of triangles arising from various interpretations of 3-body problems already serve as model arena for all three.
I furthermore argue in favour of the `accept' strategy on relational grounds.  
This approach requires sheaf methods (which go beyond fibre bundles and general bundles, which I contrast with sheaves and presheaves in some appendices).
Sheaf methods are also required for the stratifold construct that pairs some well-behaved stratified manifolds with sheaves.
I apply arguing against `excise' and `unfold' to GR's superspace and thin sandwich, and to the removal of collinear configurations in mechanics.
Non-redundant configurations are also useful in providing more accurate names for various spaces and theories.

\end{abstract}

\end{titlepage}

\section{Introduction} 

Given a physical system, configuration space $\FrQ$ \cite{Lanczos} is the space of all its possible configurations $Q^{\sfA}$. 
The corresponding morphisms -- the coordinate transformations of $\FrQ$ -- form the {\it point transformations}. 
In some settings, these are the {\it scleronomic} (time-independent) morphisms of the form $q^{\sfA} = f^{\sfA}(Q^{\sfB})$, the space of which I denote by Point. 
In other settings, these are the {\it rheonomic} morphisms --time-dependent in the sense of $q^{\sfA} = f^{\sfA}(Q^{\sfB}, t)$ but $t$ itself not transforming -- 
                                                                                                                            the space of which I denote by Point$_t$.

\subsection{Configurations and configuration spaces in Mechanics}

\ni Example 1) Consider first `constellations' of $N$ labelled (possibly superposed) material point particles in $\mathbb{R}^d$ with coordinates $\underline{q}_I$: 
Fig \ref{Relative-Coordinates}.a).
These are taken together to form an $Nd$-dimensional configuration, the space of possible values of which form the configuration space  $\FrQ(N, d) = \mathbb{R}^{Nd}$.\footnote{I use 
underline, and also sometimes lower-case Latin indices, to denote spatial vectors.  
I use upper-case Latin indices $I, J, K$ for particle labels ranging from 1 to $N$.
I use lower-case Latin indices $A, B, C$ for bases of relative separation labels, ranging from 1 to $n := N - 1$.}
%
The corresponding mass matrix -- alias kinetic metric -- is 

\ni\beq 
M_{iIjJ} := m_I\delta_{IJ}\delta_{ij} \mbox{ } .
\eeq
\ni Example 2) Taking the centre of mass to be meaningless, there arise relative inter-particle separation vectors 

\ni $\underline{r}_{IJ} := \underline{q}_{J} - \underline{q}_{J}$.
{\it Lagrange coordinates} (Fig \ref{Relative-Coordinates}.b) are then some basis set made from the, which form an $nd$-dimensional configuration space $\FrR(N, d) = \mathbb{R}^{nd}$.
However, the kinetic metric is no longer diagonal in these coordinates. 
This can be rectified by passing to {\it Jacobi coordinates} \cite{Marchal, LR97} $\underline{\rho}_A$ [Fig \ref{Relative-Coordinates}.c)].  
These are a basis of relative inter-particle {\sl cluster} separation vectors.  
%
{            \begin{figure}[ht]
\centering
\includegraphics[width=0.8\textwidth]{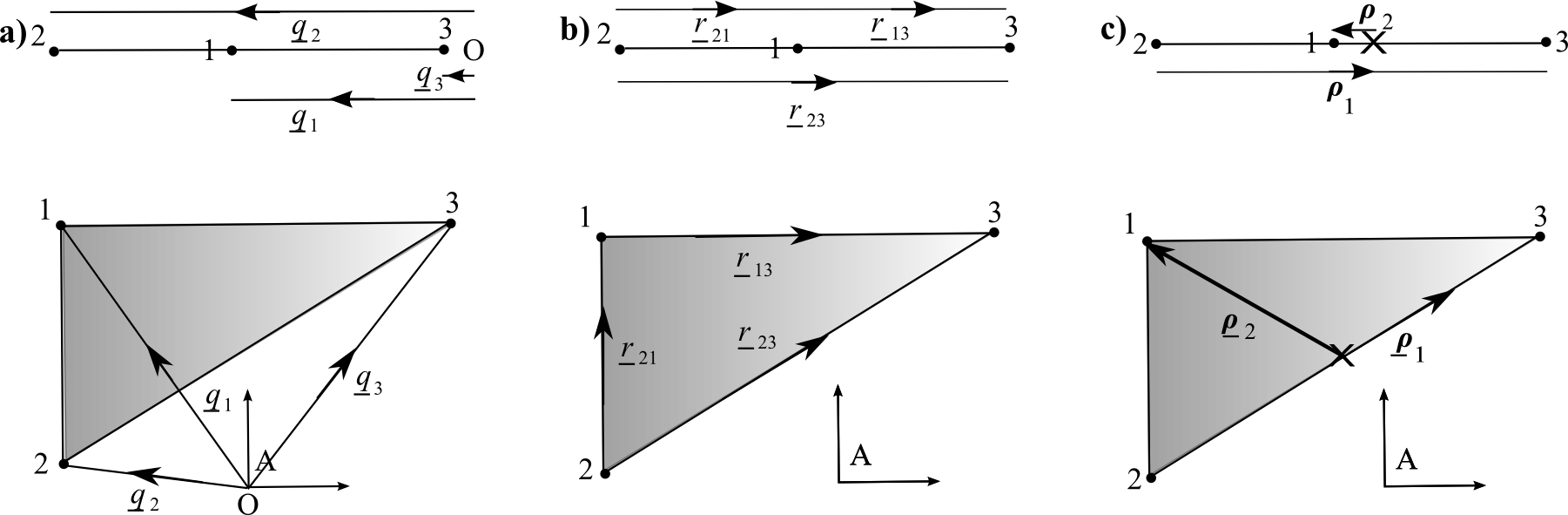}
\caption[Text der im Bilderverzeichnis auftaucht]{        \footnotesize{Coordinate systems for 3 particles in each of 1- and 2-$d$; 
Sec \ref{RPM-Q-Geom} justifies concentrating on these two particular cases, though the notions presented here for now indeed trivially extends to arbitrary $N$ and $d$.  
I consider 3 particles because {\it relational nontriviality} requires for one degrees of freedom to be expressed in terms of another, by which 2 particles are not enough.  
\ni a) Absolute particle position coordinates ($\q_1$, $\q_2$, $\q_3$) in 1- and 2-$d$.  
These are defined with respect to, where they exist, fixed axes A and a fixed origin O.
\ni b) Relative inter-particle (Lagrange) coordinates $\br = \{\r^{IJ}, I > J\}$.
Their relation to the $\q^I$ are obvious: $\r^{IJ} = \q^J - \q^{I}$.  
In the case of 3 particles, any 2 of these form a basis. 
No fixed origin enters their definition, but they are in no way freed from A.
\ni c) Relative particle inter-cluster mass-weighted Jacobi coordinates $\brho$, which are more convenient but still involve A. 
$\times$ denotes the centre of mass of particles 2 and 3.} }
\label{Relative-Coordinates}\end{figure}            }

\ni Example 3) Use of other coordinatization for the configurations $Q^{\sfA}$ that form $\FrQ$.  
For the examples given so far, these are curvilinear coordinates, e.g. spherical polar coordinates for each $\mathbb{R}^3$ factor in Example 1.

\ni Example 4) More generally, configuration spaces can be curved manifolds, with configurations then being coordinates thereupon.  
A well-known example is the rigid rotor, 
for which natural configurational coordinates are spherical angles $\theta$ and $\phi$ that trace out the surface of the configuration space sphere $\mathbb{S}^2$.
This example also illustrates that configuration coordinates are not necessarily globally defined -- $\phi$ is not defined at the poles ($\theta = 0, \pi$); 
indeed it is well known that {\sl no other} coordinates on spheres are globally defined: at least two different coordinate charts are required to cover the sphere.  

\ni Example 5) Constrained system \cite{Lanczos, Goldstein, Dirac, HTBook}.\footnote{One can see from the forms of the given constraints that these are defined on {\it phase space}: 
the space of all possible values of the coordinates {\sl and} the momenta.
On the other hand, at least for the range of examples considered in this Article, reduced configuration spaces make good sense upon having taken into account (some subset of) constraints.
Configuration space itself is already useful in a number of contexts (Sec \ref{Q-Motiv}) and rather under-studied as compared to phase space.}
%
One can in fact view `taking out the centre of mass' as imposing a zero total momentum constraint $\underline{\scP} := \sum_I\underline{p}_{I} = 0$ that eschews absolute translations, 
and the rigid rotor as a particle confined to move on the surface of a sphere.

\ni Example 6) One can furthermore eschew absolute rotations (Fig \ref{Relative-Coordinates}'s A), 
by far most simply done in the case of zero total angular momentum                 $\underline{\scL} := \sum_I\underline{q}_{I} \cr \underline{p}_{I} = 0$.
These leads to relational configuration coordinates \cite{LR97, FileR} as outlined in Sec 2.

\ni Example 7) One can furthermore eschew absolute scale \cite{Kendall, FORD, Cones, FileR}, 
corresponding either to a) separating out scale or b) to rendering scale meaningless by imposing zero dilational momentum $\scD := \sum_I\underline{q}_{I} \cdot \underline{p}_{I} = 0$.  
Pure-shape configuration coordinates \cite{Kendall, FileR} ensue, as also outlined in Sec 2.  
From the perspective of Examples 7) and 8), $\FrQ(N, d)$ is rather physically redundant as a configuration space.

\ni Example 8) It is clear from considering jointed rods that eschewing absolute rotation does not get rid of all forms of rotation or of angular momentum, 
for such models retain meaningful rotation of some rods relative to others, and of corresponding relative angular momenta.
This is also clear within Example 6) and 7)'s models, by considering e.g. the base and median parts of a relational triangle of particles, 
which are free to rotate relative to each other and to possess a corresponding relative angular momentum.  

\mbox{ } 

\ni Some settings in which some of the given examples are useful are as follows. 

\mbox{ } 

\ni A) The $N$-body problem, studied in particular in the case of Celestial Mechanics \cite{Marchal, Moeckel}, 
e.g. the Earth--Moon--Sun system or the solar system, but also with larger numbers of bodies in modelling globular clusters, and in `medium particle number' Newtonian cosmology \cite{BGS}.
Most of these are usually modelled using point particles.

\ni B) Molecular Physics \cite{LR97} usually starts from a classical point-particle model, which is subsequently quantized. 
One situation often modelled assumes that the nuclei form a fixed scaled shape, 
whilst the lighter electrons fluctuate on a faster time-scale (Born--Oppenheimer and adiabatic approximations \cite{Merzbacher}).

\ni C) Relational Particle Mechanics (RPM) \cite{BB82, B03, FileR, QuadI}: whole-universe point-models, which interpretation leads to a number of distinctions from A) and B) \cite{FileR}
due to there no longer being any notion of surrounding large.
RPM comes in scaled and pure shape versions, as per Examples 7) and 8) respectively.  
In the scaled case, only relative angles and relative separations are meaningful, whereas in the pure-shape case only relative angles and {\sl ratios of} relative separations are.
In each case also only relative times are meaningful. 
These are model arenas for a number of closed universe, quantum cosmological and background independent features.

\ni D) The rigid rotor is but one example of a problem involving rigid bodies.
On the other hand, jointed rod problems are a type of non-rigid body problem, which can be used e.g. to model `the falling cat' \cite{Mont93}, 
and have applications to robotics and manufacturing \cite{AG}.

\subsection{Configuration spaces in field theory and GR}

The values of a field over space at a given instant of time are another example of configuration.

\mbox{ }
 
\ni Example 9) Consider Electromagnetism, and Yang--Mills Theory; configurations here are a 1-form $\mA_i$ and a set of internally-indexed 1-forms $\mA_{iP}$ respectively.\footnote{In 
this Article, I use the convention of straight font for field quantities and slanty font for finite quantities.
I also use $\uppi^i$ and $\uppi^{iP}$ for the conjugate momenta to $\mA_i$ and $\mA_{iP}$. 
In the Yang--Mills case, I denote the coupling constant by $g$, structure constants by $C_{PQR}$ and covariant derivative in the fibre bundle sense by $\fD_i$, 
whose action on a 1-form is as per (\ref{YMGauss}).}
%
While gauge theories involving extra scalars and/or fermions can be associated with these, Electromagnetism and Yang--Mills theories are already `vacuum' gauge theories in their own right, 
with $U(1)$ and a more general Lie group gauge invariances respectively.
The above configurations are then gauge fields in the Dirac alias data sense of gauge, as opposed to spacetime or the whole dynamical path senses, which involve 
more extended (spacetime) indices and domain of definition.
See Sec \ref{Q-Fields} for the corresponding configuration spaces.
In gauge theories, the gauge field is not only a 1-form but can furthermore be interpreted as a fibre bundle connection.\footnote{This article assumes that the reader 
is familiar with fibre bundle mathematics and its application to Theoretical Physics (see e.g. \cite{IshamBook, Nakahara} if not).
An outline is now provided in Appendix \ref{Bundles} due to popular demand for a brief outline comparison of these with general bundles, presheaves and sheaves: Appendix \ref{PreSheaves}.}  
%
These gauge theories are also examples of constrained theories, well-covered as such in \cite{HTBook}.  
E.g. in vacuo the Gauss constraint of Electromagnetism is

\ni\be
{\scG} := \pa_i\pi^{i} = 0 \mbox{ } ,
\label{gauss}
\ee
and the Yang--Mills--Gauss constraint is 

\ni\be
{\scG}_P := \fD_i\uppi^i_P = \pa_i\uppi^i_P - g C_{RPQ}\mA^Q_i\pi^{iR} = 0 \mbox{ } ,
\label{YMGauss}
\ee
One could then additionally consider working with more physical (in this case gauge-invariant) configuration variables after reducing out the corresponding version of Gauss constraint.
This amounts to quotienting out the corresponding gauge group; see Sec \ref{Q-Red-Fields} for resultant configuration spaces (orbit and loop spaces).

\mbox{ } 

\ni Example 10) Major motivation for configuration space study comes from GR \cite{Battelle, DeWitt67, Fischer70, Magic, York74, Fischer86, Giu95, Giu95b, FM96, Giu09}.
Whereas GR's spacetimes are pairs ($\FrM$, $g_{\mu\nu}$) of topological spacetime manifolds and indefinite spacetime metrics respectively, 
GR's configurations themselves are pairs ($\bupSigma$, $\mh_{ij}$) for $\bupSigma$ the spatial topological manifold and $\mh_{ij}$ a positive-definite spatial metric thereupon.
Topological manifold $\bupSigma$, is standardly taken to be fixed, and, in the current article to additionally be compact without boundary on Machian grounds 
(and often furthermore $\mathbb{S}^3$ in concrete examples).  
From a configuration space primality perspective \cite{Battelle, RWRAM13}, the evolving $\mh_{ij}$ themselves {\sl form} spacetime; 
from a spacetime primality perspective, they are a piece of $\mg_{\mu\nu}$ under the ADM decomposition \cite{ADM}.  
The configuration space formed from the $\mh_{ij}$ on a given $\bupSigma$ is then named Riem($\bupSigma$) (after Riemann); see Sec \ref{GR-Config} for details.   
This is another example of highly redundant configuration space, since GR is also a constrained theory \cite{ADM}, with\footnote{$\mh_{ij}$ has determinant $\mh$, 
inverse $\mh^{ij}$, covariant derivative $\mD_i$, Ricci scalar $\mR$ and conjugate momentum $\mp^{ij}$.  
$\Lambda$ is the cosmological constant.}

\ni \beq
\mbox{\it GR momentum constraint } \mbox{ } \scM_{i} := - 2\mD_{j}{\mp^{j}}_{i} = 0  \mbox{ } , 
\label{Momm}
\eeq

\ni \beq
\mbox{\it GR Hamiltonian constraint } \mbox{ }  \scH := \mN_{abcd}\mp^{ab}\mp^{cd} - \sqrt{\mh}\{  \mR - 2\Lambda\} = 0
\label{Hamm}
\eeq
for 

\ni\beq
\mbox{\it DeWitt supermetric } \mbox{ } \mN_{abcd} := \mbox{$\frac{1}{\sqrt{\sh}}$} \big\{ \mh^{ac}\mh^{bd} - \mbox{$\frac{1}{2}$}\mh^{ab}\mh^{cd} \big\} 
\label{N-DeWitt}
\eeq
which is the inverse of the

\ni\beq
\mbox{\it GR configuration space metric } \mbox{ } \mM_{abcd} := \sqrt{\mh}\{\mh^{ac}\mh^{bd} - \mh^{ab}\mh^{cd}\} \mbox{ } . 
\label{M-GR}
\eeq
Now the GR momentum constraint can be straightforwardly interpreted in terms of Diff($\bupSigma$) freedom.
The information contained in $\mh_{ij}$ can then be considered as split into 3 degrees of freedom per space point (dofpsp) of unphysical Diff($\bupSigma$) information 
and a core of 3 dofpsp of partly-physical information: the {\it 3-geometry}. 
This is still but partly-physical due to the GR Hamiltonian constraint not yet having been taken into account; 
moreover, it is far less clear how to take this into account \cite{K92I93APoT123, ABook}.

\ni \beq
\mbox{Superspace($\bupSigma$) := Riem($\bupSigma$)/Diff($\bupSigma$)} \mbox{ }  
\eeq
is then the space of 3-geometries, which Wheeler greatly encouraged the study of \cite{Battelle};\footnote{Moreover, 
I argue that this `call for arms' is really for as wide a range of configuration spaces as necessary, as per this Article and \cite{I91I03, ASoS}.}
see Sec \ref{GR-Config-2} for an outline of what has been determined about this configuration space since.
As part of that, Wheeler asked the very natural follow-up question of what is ``{\it 2/3 of superspace}"?

\mbox{ } 

\ni Two geometrically natural possibilities for this `2/3 of superspace' subsequently considered by York are as follows. 

\ni 1) {\it Conformal superspace} \cite{York74, FM96} CS($\bupSigma$) is the space of all conformal 3-geometries on a fixed $\bupSigma$; 
it corresponds to the {\it maximal slice condition} $\mp = 0$ being imposed;\footnote{Conf($\bupSigma$) are the conformal transformations on $\bupSigma$.
Finally, the {\it semidirect product group} $\sFrG = \FrN \rtimes \FrH$ is given by 
$(n_1, h_1)\circ(n_2, h_2) = (n_1 \varphi_{h_1}(n_2), h_1\circ h_2)$ for $\FrN \lhd \sFrG$, $\FrH$ a subgroup of $\sFrG$ 
                                                                     and $\varphi:\FrH \rightarrow \mbox{Aut}(\FrN)$ a group homomorphism.\label{Conf-Foo}}
 
\ni\beq
\mbox{CS($\bupSigma$) = Superspace($\bupSigma$)/Conf($\bupSigma$) = Riem($\bupSigma$)/Conf($\bupSigma$)$\rtimes$ Diff($\bupSigma$)} \mbox{ } .
\eeq

\ni 2) \{CS + V\}($\bupSigma$) \cite{York72, ABFKO} adjoins to this a solitary global degree of freedom -- the spatial volume of the universe; 
it corresponds to the {\it constant mean curvature slicing condition} $\mp/\sqrt{\mh} = const$ being imposed.  

\ni It is also then natural to consider the simpler if more redundant configuration space {\it conformal Riem} 

\ni\beq
\mbox{CRiem($\bupSigma$) := Riem($\bupSigma$)/Conf($\bupSigma$)} \mbox{ } :
\eeq 
the space of conformal equivalence classes of Riemannian metrics \cite{DeWitt67}, which can be represented e.g. by the {\it unit-determinant metrics} 

\ni \beq 
\muu_{ij} := \mh_{ij}/\mh^{1/3} \mbox{  } .
\label{u}
\eeq
For further mathematical detail, see Sec \ref{CRiem-Geom} concerning CRiem($\bupSigma$), and Sec \ref{GR-Config-2} as regards CS($\bupSigma$) and \{CS + V\}($\bupSigma$).
N.B. also that 1) and 2) bear close ties to by far the most successful methods for approaching the initial-value problem in GR \cite{York72, IVP}.  
Finally, bear in mind, that the `2/3 of superspace' picked out by each of 1) and 2)  might not be directly related to the `2/3 of superspace' picked out by $\scH$ itself. 
Let us call the latter True($\bupSigma$) whilst acknowledging that for now this is but a formal naming rather than a space of known and understood geometry.  

\mbox{ } 

\ni Example 11) Homogeneous GR configurations form configuration spaces widely known as minisuperspaces.
Relational nontriviality [Fig 1] then dictates that isotropic such solutions contain at leas one matter degree of freedom; 
with quantization in mind this is to be not phenomenological matter but fundamental matter \cite{BI75HH83} -- most straightforwardly a minimally-coupled scalar field.   
Solutions with anisotropic degrees of freedom \cite{Mis68, Magic}, however, can be considered in vacuo.

\mbox{ } 

\ni Example 12) Perturbative inhomogeneous GR configurations are required for structure formation -- relevant to observational cosmology -- 
and to tap into numerous diffeomorphism-nontrivial matters \cite{SIC1}.
A particular such model consists of inhomogeneous perturbations about the spatially-$\mathbb{S}^3$ isotropic minisuperspace with single minimally-coupled scalar field matter; 
see Sec \ref{SIC-Q-Geom} for an unreduced treatment and Sec \ref{SIC-Red-Geom} for a reduced one.
This particular model becomes the {\it Halliwell--Hawking model} \cite{HallHaw} at the semiclassical level.  

\mbox{ } 

\ni Example 13) GR in Ashtekar variables.
These are related to the above geometrodynamical variables by a canonical transformation \cite{Ashtekar} 
(and various extensions: this is now a complexified GR with degenerate 3-metrics allowed).  
The new configurational variable is now a $SU(2)$ Yang--Mills 1-form ${\mathbb{A}_{i}}^{\mbox{\scriptsize\tt PQ}}$, 
again reinterpretable as a connection, and often then furthermore recast as a holonomy or loop (Sec \ref{Loop-G}).\foo{The typewriter face capital indices here denote spinorial $SU(2)$ indices. 
Its conjugate momentum ${\mathbb{E}^{i}}_{\mbox{\scriptsize\tt PQ}}$ is now a {\it 3-bein}: related to the 3-metric by $\mh_{ij} = - \mbox{tr}(\mathbb{E}_{i}\mathbb{E}_{j})$. 
This is now indeed a conjugate momentum, despite its relation to the previous configurational variable $\mh_{ab}$ because a canonical transformation has been applied. 
tr denotes the trace over these.
$|[\mbox{ }, \mbox{ } ]|$ denotes the classical Yang--Mills-type commutator.
Note that due to the specific form of $\mathbb{A}$ and $\mathbb{E}$, $h_{ij}$ is in fact complexified, i.e. pointwise in $GL(3, \mathbb{C})$ rather than in $GL(3, \mathbb{R})$.
Real Ashtekar variables have a more complicated form of $\scH$, but loop variables still apply to these.}  
%
This formulation's constraints are then

\ni \be
\scG_{\mbox{\scriptsize\tt PQ}} := \mathbb{D}_{i}  {\mathbb{E}^{i}}_{\mbox{\scriptsize\tt PQ}} 
                                := \pa_{i}{\mathbb{E}^{i}}_{\mbox{\scriptsize\tt PQ}}  +  |[\mathbb{E}_{i}, \mathbb{E}^{i}]|_{\mbox{\scriptsize\tt PQ}} = 0 \mbox{ } , 
\label{ashgauss} 
\ee

\ni \be
\scM_i := \mbox{tr}(\mathbb{E}^{j} \mathbb{F}_{ij}) = 0 \mbox{ } ,
\label{ashmom}  
\ee

\ni \be
\scH  := \mbox{tr}(\mathbb{E}^{i}\mathbb{E}^{j}\mathbb{F}_{ij}) = 0 \mbox{ } .
\label{ashham} 
\ee
(\ref{ashgauss}) is an $SU(2)$ Yang--Mills--Gauss constraint.  
(\ref{ashmom}) and (\ref{ashham}) are the polynomial forms now taken by the GR momentum and Hamiltonian constraints respectively. 
One can see that (\ref{ashmom}) is indeed associated with momentum since it is the condition for a vanishing Yang--Mills--Poynting vector.  
Again, this formulation's version of the Hamiltonian constraint (\ref{ashham}) lacks such a clear-cut interpretation; 
it is simpler due to being polynomial in this approach's canonical variables.

\ni Here loops and knots \cite{KauffmanBook} are increasingly reduced configurations, taking into account (\ref{ashgauss}) and also (\ref{ashmom}) respectively.
See Sec 11 for an outline of the corresponding configuration spaces.

\subsection{Stratified manifolds arise, and three attitudes to them}

I next further motivate the paper on a common theme observed in studying the above examples that pertain to Fundamental Physics (or useful model arenas thereof).
Namely, that stratified manifolds appear (first widely studied in \cite{Fischer70}: superspace is a stratified manifold).
Note that this also happens in study of reduced phase spaces; thus much of what is said here carries over to symplectic stratified manifolds.
These are a type of quotient, see Appendix B for quotient spaces and more specifically for stratified manifolds themselves.  
Manifolds are Hausdorff, second-countable and locally Euclidean. 
In general stratified manifolds are none of these, though this paper mostly focuses on stratified manifolds that happen to be Hausdorff and second-countable.  

\mbox{ } 

\ni Stratified manifolds having appeared, three strategies for dealing with them are as follows.  

\mbox{ }  

\ni Strategy A) {\it Excise}. This consists of discarding all bar the principal stratum.
While this simplifies the remaining mathematics to handle, it is a crude approximation and an unphysical consideration.
E.g. this strategy is often used in the context of removing the collinearities from the 3-$d$ $N$-body problem.

\ni Strategy A) {\it Excise}. This strategy consists of discarding all bar the principal stratum; 
while this simplifies the remaining mathematics to handle, it is a crude approximation and an unphysical consideration.
E.g. this strategy is often used in the context of removing the collinearities from the 3-$d$ $N$-body problem.  

\ni Strategy B) {\it Unfold}. I.e. unfold non-principal strata so that these end up possessing the same dimension as the principal stratum.
This was considered e.g. by 
%
%
Fischer \cite{Fischer86}.  
However, issues here are whether such an unfolding is physically meaningful and mathematically unique. 

\mbox{ }  

\ni The mathematical advantages of strategies A and B are to remain within the conventional mathematics of Manifold Geometry and Fibre Bundle Theory.

\mbox{ }  

\ni Strategy C) {\it Accept}.  This points to harder mathematics being required. 
Prima facie, it is `accept' that is accord with Leibniz's Identity of Indiscernibles.  
If one takes this path, then fibre bundle theory does not suffice due to heterogeneity amongst what had been homogeneous fibres. 
One needs at least general bundles (Appendix \ref{Bundles}), and perhaps furthermore sheaves (Appendix \ref{Sheaves}.  
This article's main relational program favours C); further quantum-level reasons to favour C) are outlined in \cite{ABook}.

\mbox{ } 

\ni First note that that the 3-body problem already serves as an arena for this; this justifies Sec 2 being quite extensive.
[Model arenas do well to i) exhibit the desired feature and ii) elsewise be as simple as possible].
Furthermore, analysing various works in terms of the above classification, and favouring C) adds useful interpretation to a number of GR results.
E.g. Fischer's work \cite{Fischer86} is an unfold strategy.

Also, the Bartnik--Fodor Thin Sandwich Theorem \cite{TSC2} involves two locality conditions: potential factor zeros and staying away from metrics with Killing vectors.
Due to the latter, this theorem can be viewed as an excision result, and hence as relationally undesirable. 
The Thin Sandwich, moreover, has further significance as one of the Problem of Time facet: \cite{K92I93APoT123}.

\mbox{ } 

\ni Finally, returning to the breakdown of the scope of Fibre Bundle Methods, I note that more general sheaf methods can be applied (Appendices \ref{Stratifolds} and \ref{PreSheaves}).
These are rather new methods in the range of theories considered in this Article.

\subsection{Further motivation for configuration space}\label{Q-Motiv}

Stratified manifolds are likely to become widely significant in Theoretical Physics along the below lines.

\mbox{ }

\ni Motivation 0) Understanding configurations and configuration spaces, 
especially kinematically or dynamically non-redundant ones, is useful in providing more accurate names for various spaces and theories.

\ni Motivation I) Addressing very natural questions along the lines of `which shapes are more alike than others?' or `how can one quantify that one space is more inhomogeneous than another?'
These can be approached by notions of distance between shapes in shape space, or, more generally, by notions of distance on configuration spaces. 
For an outline, see Sec \ref{Dist} for positive-definite $\FrQ$) and \ref{Gdyn-Dist} for indefinite $\FrQ$. 

\ni Motivation II) Addressing questions along the lines of `how probable are particular (ranges of) shapes?'.
These can be approached by viewing configuration space as a sampling space, or Kolmogorovian probability space, upon which to build theories of Probability and Statistics. 
E.g. Kendall's (pure) Shape Statistics \cite{Kendall} is a particular instance of geometrical statistics, 
based on particular geometries that so happen to be \cite{FORD, Records} the configuration spaces for Barbour's pure-shape RPM \cite{B03}.  
Also note that such studies are not just applicable to Theoretical Physics situations; 
to date many such considerations have been in the fields of Biology and Archaeology \cite{Kendall, Small}.

\ni Motivation III) Addressing questions such as `how much information is contained in shapes?', via considering notions of information on configuration spaces \cite{ABook}. 

\ni Motivation IV) Configuration spaces are useful in timeless approaches for use in whichever of closed-universe or quantum gravitational situations.
This is the case firstly in pure solipsism \cite{NSIPW83B94IIEOTPage12, Records}, in which the $Q^{\sfA}$ are all.  
It is also the case in approaches for which there is no time for the universe as a whole at the primary level \cite{B94I, FileR, Rovelli}. 
These nevertheless allow for a notion of time to emerge from change in configuration at the secondary level (so $\d Q^{\sfA}$ makes sense alongside the $Q^{\sfA}$ themselves).   
Note that in both these cases, it is Point rather than Point$_t$ which are appropriate morphisms.
In fact, Motivation I)--III) are how to equip a solipsist worldview with concrete mathematics, 
though such timeless calculations can also be part of doing Physics within the more extensive worldviews of b) and VI).

\ni Motivation V) Reduced configuration variables appear in the configuration space restriction of some notions of observables or beables \cite{APoB}. 

\ni Motivation VI) Histories \cite{GMHHartleILH03H09}  
-- used in a further range of problem of time strategies and foundational approaches to Quantum Cosmology -- can be viewed as strings of configurations.

\ni Motivation VII) Dynamics can be re-envisaged as a path on configuration space \cite{Lanczos, Magic, B94I, FileR} (this was already used historically by Jacobi and extended by Synge).
Then by knowing the geometrical meaning of the configuration to configuration space correspondence, one can read off from such a path the sequence of shapes a given evolution goes through.

\ni Motivation VIII) QM in fact unfolds on configuration space.\footnote{{\it Polarizations}, more generally, are choices of a suitable half-set of phase space variables \cite{Woodhouse}.
Moreover, arguments for $\sFrQ$ primality \cite{Battelle, I84} may extend to the possibility of configuration space being a privileged polarization.}
%
With configuration space acquiring a geometrical character, then such as {\it geometric quantization} is to be used; 
indeed this approach can be heavily centred upon configuration space mathematics \cite{I84}.
In reduced quantization approaches, quotient configuration spaces that are stratified manifolds feature directly prior to quantization.
On the other hand in Dirac quantization approaches, quotienting out linear constraints is {\sl postponed} until after these have been promoted to quantum equations.
Choices of operator-ordering have also been tied to priorly understanding the underlying configuration space geometry \cite{DeWitt57, Magic, ABook}.

\ni Motivation IX) {\it Generalized} configuration spaces \cite{I91I03, ASoS, ABook} enter into consideration upon letting a wider range of structures than usual be dynamical 
and consequently quantum-mechanically fluctuate.
These have a wider still range of mathematical structures than just geometries (e.g. the space of topological spaces on a fixed set form a lattice).
In this way, an even wider range of notions of distance and information, probability and statistics theories, timeless formulations, dynamics and quantum theories arise.

\section{Configuration space geometry: Mechanics}\label{Q-Geom}

\subsection{Mechanics and RPM configuration spaces}\label{RPM-Q-Geom}

In the simplest (and relationally redundant) approach to Mechanics, one's incipient notion of space (NoS) is {\it absolute space} $\FA(d)$ of dimension $d$.   
This is usually taken to be $\mathbb{R}^{d}$ equipped with standard Euclidean inner product alias metric. 
The corresponding configuration space $\FrQ(N, d)$ is then just $\mathbb{R}^{N d}$, i.e. itself a Euclidean space but now of dimension $N d$.  
In this Article, I consider just the case of equal masses (see \cite{FileR} for discussion of other cases).

As regards less physically redundant presentations, various possibilities for physically-irrelevant groups of transformations $\FrG$ are as follows.
Translations $\mbox{Tr}(d)$ forming the noncompact Abelian Lie group $(\mathbb{R}^{d}, +)$ of dimension $d$,   
rotations $\mbox{Rot}(d) = SO(d)$ i.e. the compact non-Abelian special orthogonal group $SO(d)$ of $d \times d$ matrices of dimension $d$\{$d$ -- 1\}/2,  
and dilations $\mbox{Dil}$ forming the $d$-independent noncompact commutative group $(\mathbb{R}^+, \cdot)$ of dimension 1. 
Particular further combinations of these then include the Euclidean  group Eucl($d$) := Tr($d$) $\rtimes$   Rot($d$) 
                                                  and the similarity group Sim($d$)  := Tr($d$) $\rtimes$ \{Rot($d$) $\times$ Dil($d$)\}.  
Strictly speaking, Eucl($d$) and Sim($d$) are the `proper' versions of these groups due to not being taken to include the discontinuous reflections.
Then dim(Eucl($d$)) = $d\{d + 1\}/2$ and dim(Sim($d$)) = $d\{d + 1\}/2 + 1$. 
See \cite{AMech} for a much wider range of geometrically significant groups, corresponding notions of shape, corresponding shape spaces (quotient configuration spaces), 
and relational theories of mechanics thereupon.

The subsequent quotient spaces $\FrQ/\FrG$ are as follows.\footnote{See
Appendix \ref{Quotients} for a general outline of quotient spaces.} 
%
Absolute position is rendered meaningless by passing to {\it relative space} $\Fr(N, d) = \FrQ(N, d)/Tr(d) = \mathbb{R}^{n d}$ for $n := N - 1$.
Moreover, this quotienting is devoid of mathematical structure or any extra analogy with GR, so starting from $\Fr(N, d)$ makes for a clearer presentation in Fig \ref{Q-RPM-GR}.
The diagonal form for the kinetic matrix for this in relative Jacobi coordinates is $\mu_{ij AB} := \mu_A\delta_{ij}\delta_{AB}$, 
for $\mu_A$ are the corresponding Jacobi inter-particle cluster reduced masses $\mu_A$.
E.g. for 3-body case, these take the form  

\ni\beq
\mu_1 = \frac{m_2m_3}{m_2 + m_3} \mbox{ } \mbox{ and } \mbox{ } \mu_2 = \frac{m_1\{m_2 + m_3\}}{m_1 + m_2 + m_3} \mbox{ } . 
\eeq
The $\underline{\rho}_A$ I use are furthermore mass-weighted, so $\sqrt{\mu_A}$ has already been absorbed into each, 
and so the final kinetic metric is just an identity array with components $\delta_{ij}\delta_{AB}$.

If absolute axes are also to have no meaning, the configuration space one is left with is 

\ni\beq
\mbox{\it relational space }  \mbox{ } \bigr(N, d) :=  \FrR(N, d)/\mbox{Rot($d$)} \mbox{ } [ \mbox{ }  = \FrQ(N, d)/\mbox{Eucl($d$)} \mbox{ } ] \mbox{ } .   
\eeq
I denote configuration space dimension by $k$.\footnote{In general this refers to na\"{i}ve or largest dimension, 
since the outcome of quotienting in general has strata with a variety of dimensions.}
%
In the above case, $k= nd$ -- $d\{d - 1\}/2$ = d\{2$n$ + 1 -- $d$\}/2, i.e. $N - 1$ in 2-$d$, $2N - 3$ in 2-$d$ and $3N - 6$ in 3-$d$.
If, instead, absolute scale is also to have no meaning, then the configuration space is  {\it preshape space} \cite{Kendall} $\FP(N, d) := \Fr(N, d)/\mbox{Dil}$, with $k = nd - 1$.  
If both absolute orientation and absolute scale are to have no meaning, then the configuration space is Kendall's \cite{Kendall} 

\ni\beq
\mbox{\it shape space } \mbox{ } \FrS(N, d) := \FrQ(N, d)/\mbox{Sim}(d) \mbox{ } . 
\eeq
Now $k = Nd - \{d\{d + 1\}/2 + 1\} = d\{2n + 1 - d\}/2 - 1$, i.e. $N - 2$ in 1-$d$, $2N - 4$ in 2-$d$ and $3N - 7$ in 3-$d$.
Note also that $\FP(N, 1) = \FrS(N, 1)$, since there are no rotations in 1-$d$.  
The above quotient spaces are taken to be not just sets but also normed spaces, metric spaces, topological spaces, and, where possible, Riemannian geometries.      
Their analogy with GR's configuration spaces is laid out in Fig \ref{Q-RPM-GR}, along with a summary of the specific geometrical forms these take in various simpler cases. 

{            \begin{figure}[ht]
\centering
\includegraphics[width=0.8\textwidth]{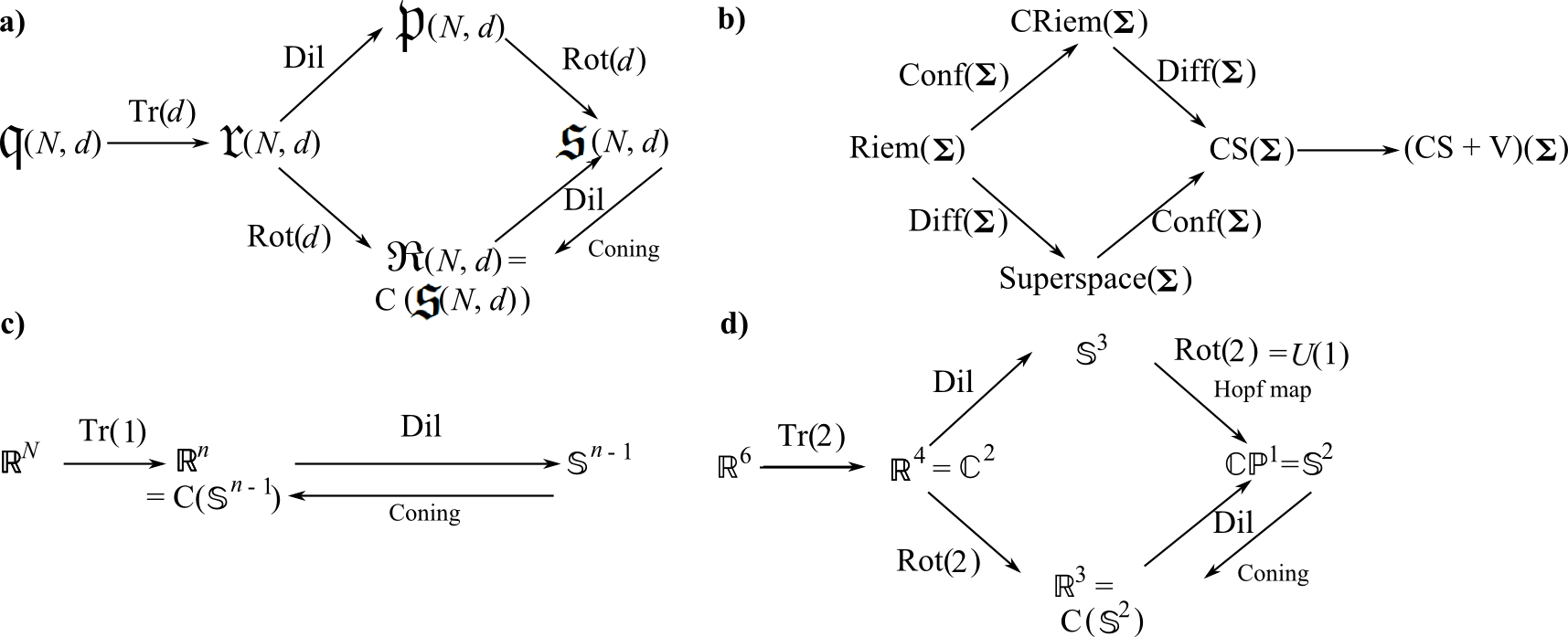}
\caption[Text der im Bilderverzeichnis auftaucht]{        \footnotesize{a) This Sec's specific sequence of configuration spaces, as a useful model arena for GR's in b).  
c) to e) cover, respectively, the furtherly simplified cases of 1-$d$, 2-$d$, and 3 particles in 2-$d$, whose forms are derived in the next Subsection.} }
\label{Q-RPM-GR} \end{figure}          }

\subsection{Picking out the triangleland example}\label{Tri-Sel}
%

I begin with the pure-shape RPM's, since these are geometrically simpler than scaled RPM's, and furthermore occur as subproblems within the latter. 
Fig \ref{Allow}.a)-b) tabulates configuration space dimension $k$ in , so as to display inconsistency, triviality, and relational triviality by shading.
I follow this up identifying tractable topological manifolds and metric geometries in Fig \ref{Allow}.c)-d) \cite{FORD}.
%
{            \begin{figure}[ht]
\centering
\includegraphics[width=0.7\textwidth]{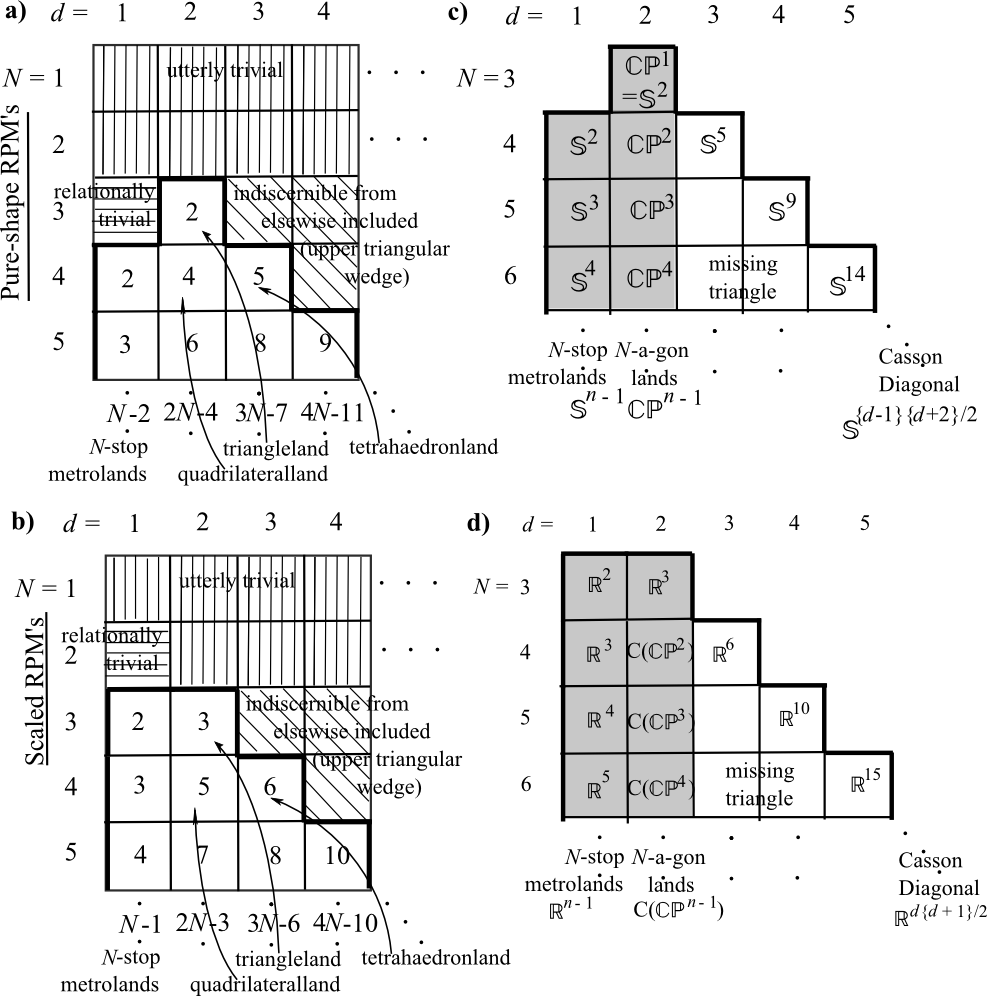}
\caption[Text der im Bilderverzeichnis auftaucht]{        \footnotesize{a) and b) are pure-shape and scaled RPM's configuration space dimensions $k$ respectively.
c) and d) are the corresponding topological manifolds (\cite{FileR} summarizes further topological results about RPM configuration spaces).  
Whilst this gives 3 tractable series -- see \cite{Kendall}, including for the `Casson diagonal' -- only the shade two column groups admit tractable metrics as well.  
I term 1-$d$ RPM universe models {\it N-stop metrolands} since their configurations look like an underground train line. 
I term 2-$d$ RPM universe models {\it N-a-gonlands} since each point in them is a planar $N$-sided polygon. 
The mathematically highly special $N$ = 3 case of this is {\it triangleland}, and the first mathematically-generic $N = 4$ case is {\it quadrilateralland} \cite{QuadI}.   
See \cite{FileR} for the basic Algebraic Topology of basic RPM configuration spaces.} }
\label{Allow}\end{figure}            }

\ni Note that $N$-stop metrolands already possesses notions of localization, clumping, inhomogeneity, structure and hence structure formation.
Contrast with how in GR these only appear in much more complicated Midisuperspace models.
$N$-a-gonlands have only distance-ratio structure but also relative-angle structure, as well as the further Midisuperspace-like feature of nontrivial linear constraints.
Contrast with how in GR, linear constraints on the one hand, and localization, clumping, inhomogeneity and structure on the other, are interlinked.
This is because in GR both follow from spatial derivatives being nontrivial.
On the other hand, these two sets of notions are separable in RPM's, with 1-$d$ RPM's then serving to study the former in isolation from the latter.

The $N$-stop metroland shape space possesses the {\it hyperspherical metric}

\ni \beq
\d s^2 = \sum\mbox{}_{\mbox{}_{\mbox{\scriptsize $p$ = 2}}}^{n - 1} \prod\mbox{}_{\mbox{}_{\mbox{\scriptsize $m$ = 1}}}^{p - 1} \mbox{sin}^2\theta_m \d\theta_p^2 \mbox{ } .
\label{HS}
\eeq
These $\theta_{\barp}$'s are related to ratios of the $\rho_A$ in the usual manner in which hyperspherical coordinates are related to Cartesian ones \cite{FileR}.

On the other hand, the $N$-a-gonland shape space has the {\it Fubini--Study metric} \cite{Kendall}

\ni \beq
\d s^2 = 
\left.
\{\{1 + ||\mbox{\boldmath$Z$}||_{\sC}^2\} ||\d\mbox{\boldmath$Z$}||_{\sC}^2 -  |(\mbox{\boldmath$Z$} \cdot \d \mbox{\boldmath$Z$})_{\sC}|^2\}
\right/ 
{\{1 + ||\mbox{\boldmath$Z$}||_{\sC}^2\}^2} 
\label{FS}
\eeq
for $\sC$ here denoting complex inner product and norm, with \mbox{\boldmath$Z$} running over $n - 1$ copies of the complex plane.

\ni $Z_{\bar{p}} = {\cal R}_{\bar{p}} \mbox{exp}(i\Phi_{\bar{p}})$ -- a multiple $\mathbb{C}$ plane polar coordinates version of ratios of the $\underline{\rho}_i$, 
with $\Phi_{\bar{p}}$ relative angles between $\underline{\rho}_A$ and ${\cal R}_{\bar{p}}$ ratios of magnitudes $\rho_A$ \cite{FileR}.

N.B. both of the above metrics are written in coordinates standard to each of these geometries (hyperspherical angles and inhomogeneous coordinates \cite{Nakahara} respectively). 
Moreover, in the present RPM setting these coordinates can be traced back to the spatial coordinates describing the particles themselves: see \cite{FileR}.  
N.B. also that, as mechanical theories, RPM's have positive-definite kinetic arc elements, which are significantly different from GR's indefinite one. 
[This feature is then inherited by this Article's other principal model arenas: minisuperspace and inhomogeneous perturbations thereabout.]

Next, a generalized notion of {\it cone} over some topological manifold $\FrMgen$ is denoted by C($\FrMgen$) and takes the form 

\ni \beq
\mbox{C(\FrMgen) = \FrMgen $\times$ [0, $\infty$)/\mbox{ }$\widetilde{\mbox{ }}$} \mbox{ } . 
\eeq
$\widetilde{\mbox{ }}$ here means that all points of the 
form $\{\mp \in \FrX, 0 \in [0, \infty)\}$ are `squashed' or identified to a single point termed the {\it cone point}, 0. 
Then at the metric level, given a manifold $\FrMgen$ with a metric with line element $\d s$, the corresponding cone has a natural metric of form 

\ni\beq
\d s^2_{\scc\so\sn\se} := \d \rho^2 + \rho^2 \d s^2 \mbox{ } . 
\eeq
\ni \mbox{ } \mbox{ } Then relational space is just the cone over shape space \cite{Cones, FileR}, which cone structure makes clear the {\it scale--shape split} formulation of scaled RPM. 
Furthermore, $\mC(\FrS(N, 1))$ are just $\mathbb{R}^{n}$.

For triangleland, the additional coincidence $\mathbb{CP}^1 = \mathbb{S}^2$ `doubles' the amount of geometry and linear methods available 
(and the spherical ones are both simpler and better-known than complex-projective ones).  
Here, 

\ni\beq
\d s^2 = \d \Theta^2 + \mbox{sin}^2\Theta \, \d\Phi^2 \mbox{ } ;
\label{Tri-Sphericals}
\eeq
see Fig \ref{Relational-Coordinates} for the meanings of these coordinates.
%
{           \begin{figure}[ht]
\centering
\includegraphics[width=0.5\textwidth]{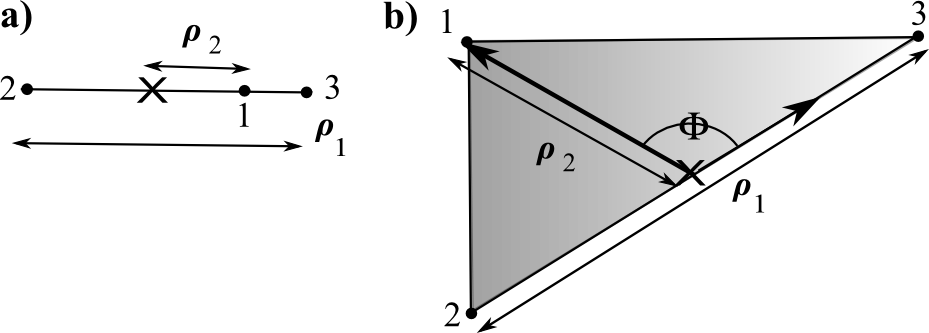}
\caption[Text der im Bilderverzeichnis auftaucht]{  \footnotesize{a) For 3 particles in 1-$d$, just use the magnitudes of the two Jacobi coordinates.
b) For 3 particles in 2-$d$, use the magnitudes of the two Jacobi coordinates and define $\Phi$ as the `Swiss army knife' angle 
$\mbox{arccos}\big( \brho_1 \cdot \brho_3 / \rho_1 \rho_3 \big)$.
This is a relative angle, so, unlike the $\brho$, these three coordinates {\sl do not} make reference to absolute axes A.  
Pure-shape coordinates are then the relative angle $\Phi$ and some function of the ratio $\rho_2/\rho_1$.  
In particular, $\Theta := 2\,\mbox{arctan}(\rho_2/\rho_1)$ is then the azimuth to $\Phi$'s polar angle.} }
\label{Relational-Coordinates} \end{figure}         } 

The scaled case is just the cone over the pure-shape case's configuration space, allowing for that case to be covered also. 

\ni\beq
\d s^2 = \d \rho ^2 + \rho^2\{\d \Theta^2 + \mbox{sin}^2\Theta_2\d\Phi^2\}/4 = \big\{\d I^2 + I^2\{\d \Theta^2 + \mbox{sin}^2\Theta\d\Phi_2^2\}\big\}/4I \mbox{ } ;
\label{Tri-Scale}
\eeq
Here, the {\it configuration space radius} $\rho := \sqrt{\rho^2_1 + \rho^2_2}$ (alias {\it hyperradius} in the Molecular Physics literature); 
this is also the square root of the moment of inertia, $I$.
$\mC(\FrS(3, 2))$ is also $\mathbb{R}^3$, albeit not with the flat metric.  
It is, however, conformally flat \cite{FileR} -- just apply conformal factor $4I$ to the second form of (\ref{Tri-Scale}).

Another useful observation is the known forms of the corresponding isometry groups, 
Isom($\FrS(N, 1)$) = Isom($\mathbb{S}^{n - 1}$) =  $SO(n)$, 
Isom($\FrS(N, 2)$) = Isom($\mathbb{CP}^{n - 1}$) = $SU(n)/\mathbb{Z}_n$ -- among which triangleland is further distinguished by 
Isom($\FrS(3, 2)$) = Isom($\mathbb{CP}^{1}$) = $SU(2)/\mathbb{Z}_2$ = $SO(3)$ = Isom($\mathbb{S}^{2}$) -- and 
Isom($\FrR(N, 1)$) = Isom($\mathbb{R}^{n}$) =  Eucl$(n)$.

\ni Finally, Atomic and Molecular Physics provide a number of useful parallels for the spherical cases \cite{FileR}. 
On the other hand, $N$-a-gonlands can draw from \cite{QuadI} Geometrical Methods, Shape Statistics, and the standard Representation Theory of $SU(N)$.

\subsection{3-particle configuration spaces in more detail}\label{CPST}

Consider first the topological-level configurations.
Here the only distinct 3-particle shapes are the double collision D and the generic other configuration.  
If the particles are labelled, the D can furthermore be distinguished by which particle remains out of the collision.
One can also choose whether mirror images are identified provided that the dimension is low enough that the configurations cannot be rotated into each other.
These considerations give four topological 3-stop metrolands (Fig \ref{Top-Config}) and four topological trianglelands (Fig \ref{Tutti-Tri-2c}).  

{            \begin{figure}[ht]
\centering
\includegraphics[width=1.0\textwidth]{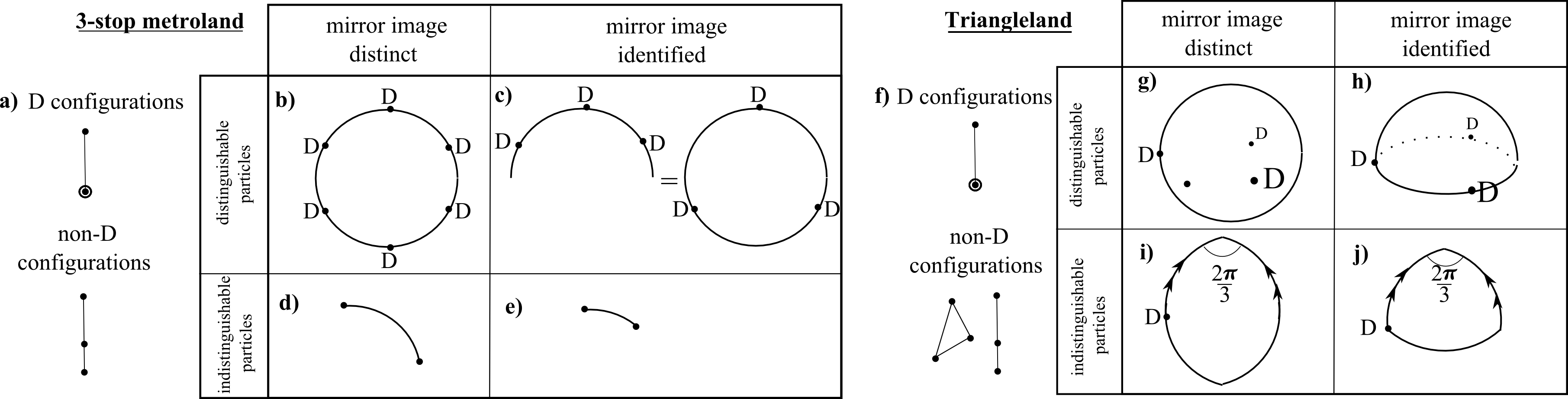}
\caption[Text der im Bilderverzeichnis auftaucht]{        \footnotesize{Topological-level configurations and configuration spaces for 3 particles.
The double arrows indicate topological identification.
The kinetic geometry itself does not present any reasons to excise the D points.
However if a potential singular at the D's is employed, e.g. in the familiar case of Newtonian Gravitation, then mathematical study would often proceed by excising the D's. 
If g)'s D's are excised, one obtains what Montgomery termed the `pair of pants' \cite{ArchRatMontgomery2}.} }
\label{Top-Config} \end{figure}          }

\ni One can additionally choose whether to model the particles as indistinguishable.
Finally, for small enough physical dimension, one additionally has the choice of whether to identify mirror image configurations.  
These two features account for the quadrupling of configuration space types in Fig \ref{Top-Config}.   

\mbox{ } 

\ni Consider next the metric level configurations.  
3 particles on a line now have continua of distinguishable non-D configurations .
These include a further distinguished notion of {\it merger} M: a configuration in which the third particle coincides with the centre of mass of the other two: Fig \ref{Top-Config}.a).
In configuration space, these sit in the mid-points of the arcs between adjacent D's, so e.g. the most extensive 3-stop metroland forms a `clock-face'.

\mbox{ }

\ni On the other hand, triangles additionally have  
i) a notion of collinear configurations C, either side of which the triangle is ordered clockwise or anticlockwise: Fig \ref{Top-Config}.b).
ii) A notion of isosceles configurations I, either side of which the triangle is right or left leaning: Fig \ref{Top-Config}.c). 
iii) A notion of regular configurations ($I_1 = I_2$: equality of base and median partial moments of inertia, or equality of base and median themselves in mass-weighted coordinates). 
Either side of these, the triangle is sharp or flat: Fig \ref{Top-Config}.d).
%
{            \begin{figure}[ht]
\centering
\includegraphics[width=0.9\textwidth]{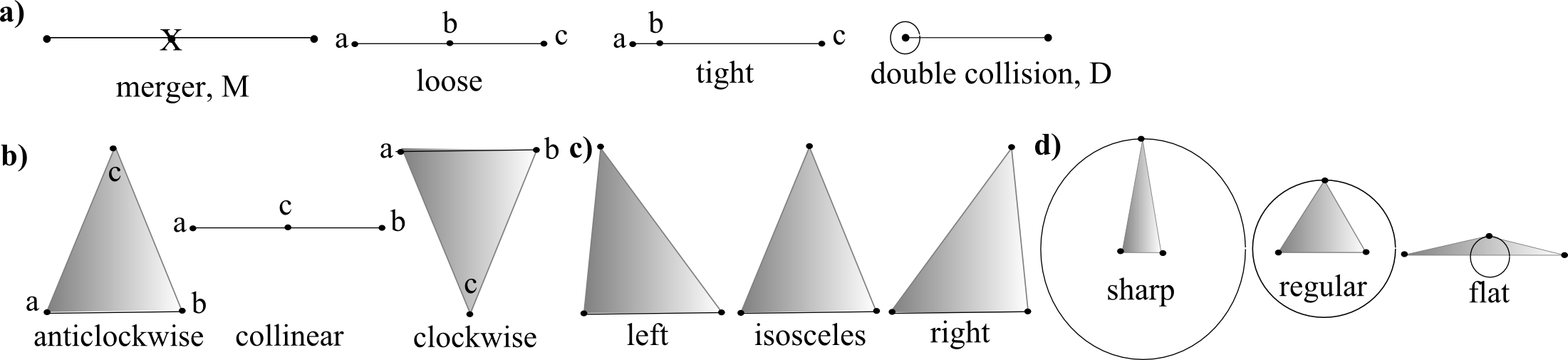}
\caption[Text der im Bilderverzeichnis auftaucht]{        \footnotesize{Metric level types of configuration for 3 particles in 1- and 2-$d$. 
`Tight' is used as in `tight binary'.} }
\label{Tutti-Tri-1} \end{figure}          }

Then in triangleland, C is the equator great circle, dividing the shape sphere into hemispheres of clockwise and anticlockwise ordered triangles. 
There are then 3 notions of each of isosceles I and regular R, corresponding to the 3 ways of labelling `base' and `apex'.
These are all meridian great circles, alternating between I and R and evenly spaced out to form the pattern of the `zodiac or 12-segmented orange'.  
Each I divides the shape sphere into hemispheres of left and right leaning triangles, and each R into hemispheres of sharp and flat triangles.
All 6 of these great circles intersect at the poles, which are equilateral triangles E (I denote the orientation reversed equilateral triangle pole by $\overline{\mE}$). 
These make for a very natural and significant choice of poles for triangleland: Fig \ref{Tutti-Tri-2c}.a).
Next, C $\bigcap$ I is D at one end and M at the other. 
Finally, C $\bigcap$ R has no further special features, so I denote these by S for `spurious'.  
Note that C and E are labelling-independent notions, unlike I, R, D, M or S.

The overall pattern \cite{+Tri, FileR} is then that of an orange cut in half perpendicular to its segments (or a zodiac additionally split into northern and southern skies).
Due to its regularity, this is an example of a {\it tessellation}: a partition of a space into a number of equal shaped regions (`tiles').
Faces, edges and vertices therein being physically significant in the present context, one is really dealing with a {\sl labelled} tessellation.
The 4-stop metroland sphere carries an even more elaborate tessellation based on the cube--octahaedron group \cite{AF, FileR}.
Quadrilateralland -- much more typical of an $N$-a-gonland -- is given a comparable configuration space analysis in \cite{QuadI}.
Such tesselations are useful as `interpretational back-cloth' for dynamical trajectories, probability distributions and quantum wavefunctions.   
This method was originally applied in the Shape Statistics setting by Kendall \cite{Kendall89} (his spherical blackboard, c.f. Fig \ref{Tutti-Tri-2c}.f).
%
{            \begin{figure}[ht]
\centering
\includegraphics[width=1.0\textwidth]{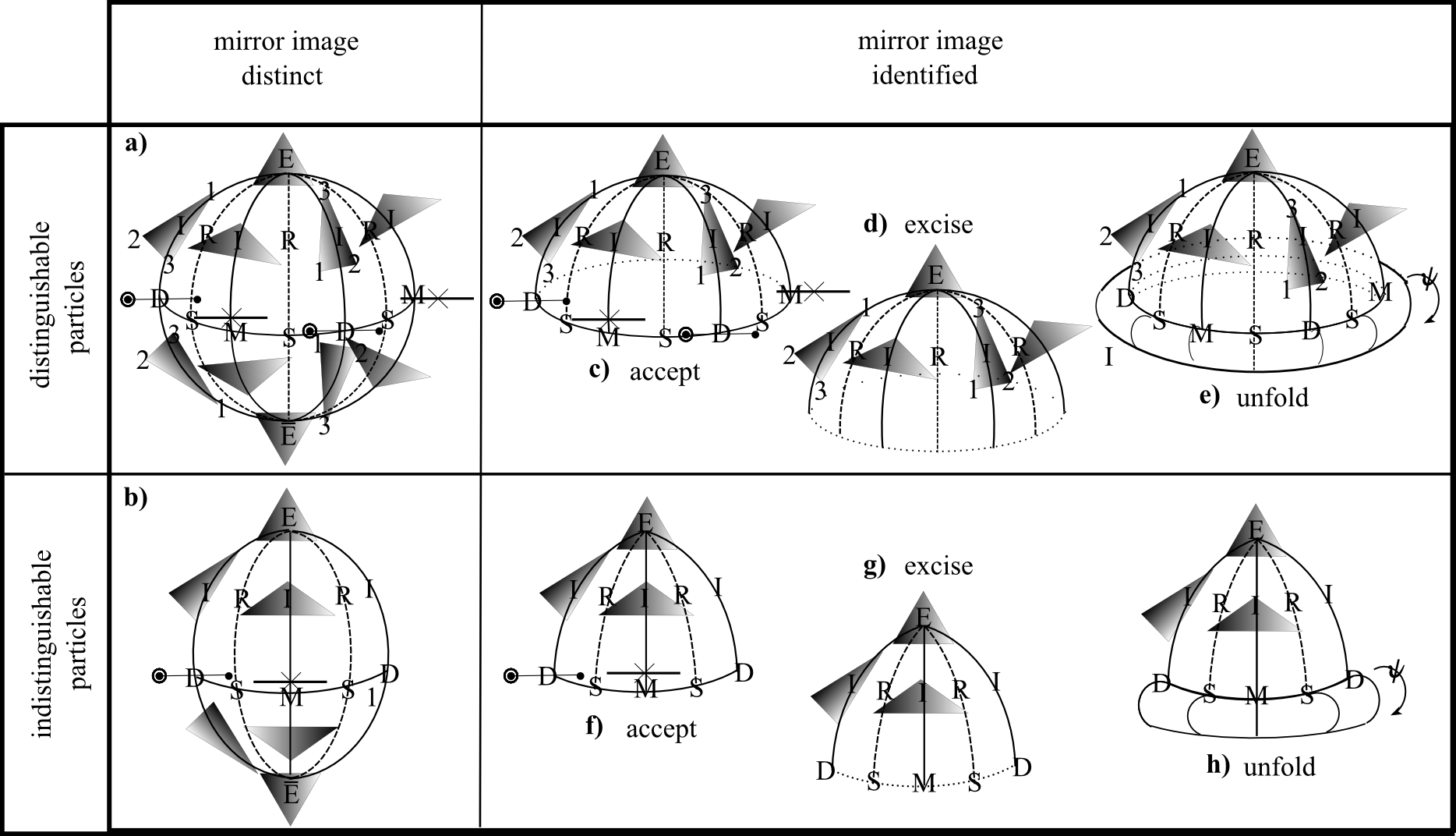}
\caption[Text der im Bilderverzeichnis auftaucht]{        \footnotesize{Triangleland configuration spaces at the metric level.  
a) The sphere. 
b) The $\mathbb{S}^2/\mathbb{Z}_3$ orbifold. 
c) The hemisphere with edge: an example of stratified manifold. 
f) The hemisphere with edge quotiented by $\mathbb{Z}_3$ is a stratified orbifold.
The coordinate range involved here is also Kendall's spherical blackboard \cite{Kendall89}. 
Note that orbifolds and especially stratified manifolds play a significant role in Appendix \ref{Quotients} and Sec \ref{GR-Config-2}, 
alongside Sec 1.3's deliberation of whether to excise, unfold or accept strata, which is sketched out here in b), d), e), f), g) and h).} }
\label{Tutti-Tri-2c} \end{figure}          }

\ni In 2-$d$, mirror image identification is optional: a) and b) are both viable options.
In 3-$d$, however, rotation out of the plane sends one mirror image to the other, so a) ceases to be a valid option.  
As regards stratification, 1-$d$ has no capacity for isotropy groups of different dimension, 
whereas shape spaces for 2-$d$ shapes avoid stratification issues due to only involving $SO(2) = U(1)$, which acts the same on C and non-C configurations.  
However, in 3-$d$ the C have only an $SO(2)$ subgroup of the $SO(3)$ acting upon them, so there is stratification
In 3-$d$ also, the inertia tensor has zero eigenvalues for the C, causing mathematical complications (these prevent inversion of kinetic metric and lead to curvature singularities). 
That is one mathematical reason for excision (Fig \ref{Tutti-Tri-2c}.d), along with a physical reason to not want to: the C configurations that are quite clearly physically acceptable.

A second option is to accept the stratification (Fig \ref{Tutti-Tri-2c}.c).

A third possibility is to unfold the equator, by introducing an extra angular coordinate that parametrizes the hitherto unused rotation about the collinearity axis. 
At the level of configuration space, this has the effect of blowing up the equator into a torus: the `hemisphere with thick edge' of Fig \ref{Tutti-Tri-2c}.e).
However, within the point-particle model setting, 
the value of this extra angular coordinate is not physically meaningful, providing physical and philosophical reasons not to take this path.

Moreover, a gap now becomes apparent in the assumption made so far that unfolding is bereft of physical content, due to the following possibility.  

\mbox{ }

\ni Strategy D) unfolding purely by enhanced physical modelling.  

\mbox{ }

\ni This is clear at this stage through contemplating cases in which the point particles are but modelling approximations for more general bodies of finite extent. 
Then their centres of mass being aligned does not alter the isotropy group in question.   
However, enhanced physical modelling would not be expected to get round how quotienting in general does not preserve local Euclideanness (or Hausdorffness or second countability).
I.e. there is no guarantee that increasing modelling accuracy will be reflected by a successful unfolding of the reduced configuration space stratified manifold into a manifold. 
See Sec \ref{starr} for an outline of GR counterparts of this.

\mbox{ } 

\ni Moreover, a gap becomes apparent in the assumption made so far that unfolding is bereft of physical content, due to the possibility of the following fourth strategy.  

\mbox{ } 

\ni As regards the corresponding relationalspaces, 3-stop metroland's is trivially $\mathbb{R}^2$, indeed N-stop metroland's is $\mathbb{R}^{n}$. 
In each case it is entirely clear how to represent an $n$-sphere within $\mathbb{R}^n$.  
The $\underline{n}_{A}$ play the role of Cartesian directions.
What plays this role for $\mathbb{S}^2$ within $\mathbb{R}^4$?  
Here there are four components of $\underline{n}_A$; how does one relate these to an $\mathbb{R}^3$? 
It turns out that $\mathbb{R}^4 \rightarrow \mathbb{S}^3 \rightarrow \mathbb{S}^2 \rightarrow \mathbb{R}^3$ occurs, where the second step is the Hopf map. 
Thus the Hopf--Dragt quantities \cite{DragtIwai87, LR97} arise (Dragt is the name used in the Molecular Physics literature): 

\ni\beq
dra_x = \mbox{sin}\,\Theta\,\mbox{cos}\,\Phi = 2 n_1 n_2 \ ,\mbox{cos}\,\Phi = 2 \{\underline{n}_1 \cr \underline{n}_2\}_3 \mbox{ } ,
\label{ dragt1}
\eeq

\ni\beq
dra_y = \mbox{sin}\,\Theta\,\mbox{sin}\,\Phi = 2 n_1 n_2\,\mbox{sin}\,\Phi = 2 \underline{n}_1 \cdot \underline{n}_2 \mbox{ } ,
\label{dragt2}
\eeq

\ni\beq
dra_z = \mbox{cos}\,\Theta = n_2\mbox{}^2 - n_1\mbox{}^2 \mbox{ } .   
\label{dragt3}
\eeq
[The 3 component in the first of these indicates the component in the fictitious third dimension of this cross-product, and $n_{A} := \rho_{A}/\rho$.]
These appear as `ubiquitous quantities' \cite{08I} in studying the relational triangle, and are indeed \K observables for that problem \cite{APoB}.  
They can be interpreted as follows.

$dra_x$ is a measure of `anisoscelesness' $aniso$: departure from the underlying clustering's notion of isoscelesness, c.f. anisotropy in Sec \ref{MSS-Q-Geom}. 
It is specifically a measure of anisoscelesness in that Aniso per unit base length in mass-weighted space is the $\ml_1 - \ml_2$ indicated in Fig \ref{Tutti-Tri-3}.a).                   
I.e., it is the amount by which the perpendicular to the base fails to bisect it (which it would do were the triangle isosceles).

dra$_y$ is a measure of noncollinearity.  
Moreover this is actually clustering-independent, known in Molecular Physics as a `democracy invariant' \cite{LR97}.
It is furthermore equal to $4 \mbox{ } \times$ $area$ (the area of the triangle per unit $I$ in mass-weighted space), which is lucid enough to use as notation for this quantity. 
In comparison, in the equal-mass case 
\ni\beq
\mbox{$physical \mbox{ } area$} = \mbox{$\frac{I\sqrt{3}}{m}$} area \mbox{ }  \label{area} \mbox{ } . 
\eeq
\mbox{ } \mbox{ } Finally, $dra_z$ is an ellipticity, ellip: 
the difference of the two `normalized' partial moments of inertia involved in the clustering in question, i.e. that of the base and that of the median.
In contrast to $aniso$, this is clearly a function of pure ratio of relative separations rather than of relative angle.  
%
{            \begin{figure}[ht]
\centering
\includegraphics[width=0.8\textwidth]{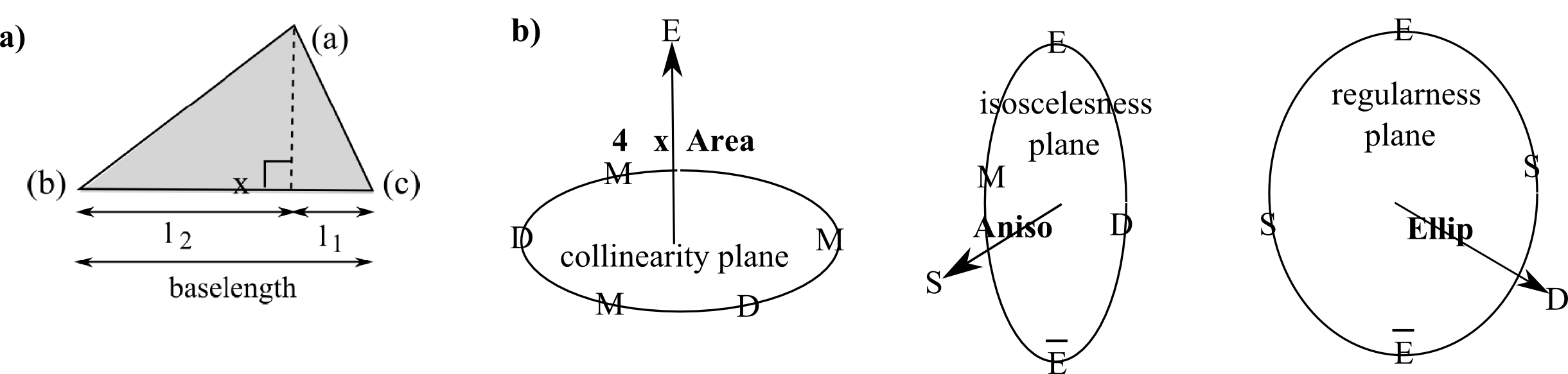}
\caption[Text der im Bilderverzeichnis auftaucht]{        \footnotesize{a) Setting up the definition of anisoscelesness quantifier, 
and b)-d) interpreting the three Hopf--Dragt axes in terms of the physical significance of the planes that they are perpendicular to.} }
\label{Tutti-Tri-3} \end{figure}          }

\ni Maximal collisions are singular for both 2- and 3-$d$ RPM's.
E.g. for scaled triangleland, the Ricci scalar is $R = 6/I$.
%

\ni Finally, pure-shape triangleland has the maximal three Killing vectors, the `axial' $\pa/\pa\Phi$ now corresponding to invariance under change of relative angle.
Scaled triangeland has six conformal Killing vectors: 3 $\pa/\pa \, dra^i$ and 3 $dra^j\pa/\pa \, dra^i - dra^i\pa/\pa \, dra^j$.

\section{Notions of distance on configuration spaces}\label{Dist}

One can build various such from $\FrQ$'s kinetic metric's norm and inner product $\mbox{\boldmath$M$}$ (at least if this is positive definite) \cite{Kendall, BB82, DeWitt70, FileR}   

\ni\be
\mbox{(Kendall Dist)} = (\mbox{\boldmath$Q$}, \mbox{\boldmath$Q$})\mbox{}_{\mbox{\scriptsize\boldmath$M$}} \mbox{ } ,    
\label{ProtoKenComp}
\ee

\ni\be
\mbox{(Barbour Dist)} = ||\d \mbox{\boldmath$Q$}||_{\mbox{\scriptsize\boldmath$M$}}\mbox{}^2 \mbox{ } , 
\label{ProtoBarComp}
\ee

\ni\be
\mbox{(DeWitt Dist)} = (\d \mbox{\boldmath$Q$}, \d \mbox{\boldmath$Q$}^{\prime})_{\mbox{\scriptsize\boldmath$M$}} \mbox{ } .  
\label{ProtoDeWittComp}
\ee

One general if formal and indirect method to incorporate $\FrG$-invariance involves acting with $\FrG$ on a given object, 
and then using a `$\FrG$-all' operation which involves the whole of $\FrG$.
Perhaps the most well-known example of this two-step procedure is group averaging.
Then

\ni\be
\mbox{(Kendall $\FrG$-Dist)} = (\mbox{\boldmath$Q$} \cdot \stackrel{\longrightarrow}{\FrG_{g}} \mbox{\boldmath$Q$}^{\prime})_{\mbox{\scriptsize\boldmath$M$}} \mbox{ } ,  
\label{Kend}
\ee 

\ni\be
\mbox{(Barbour $\FrG$-Dist) } =  ||\d_{\sg}\mbox{\boldmath$Q$}||_{\mbox{\scriptsize\boldmath$M$}}\mbox{}^2 \mbox{ } \mbox{ and }  
\label{Barb}
\ee

\ni\be
\mbox{(DeWitt $\FrG$-Dist)} = 
(\stackrel{\longrightarrow}{\FrG}_{\d g} \mbox{\boldmath$Q$}, \stackrel{\longrightarrow}{\FrG}_{\d g} \mbox{\boldmath$Q$}^{\prime})_{\mbox{\scriptsize\boldmath$M$}} 
\mbox{ } .		
\label{DeWi}
\ee
(Here $\d_{\sg}\mbox{\boldmath$Q$} := \d \mbox{\boldmath$Q$} - \stackrel{\longrightarrow}{\FrG}_{\d g}\mbox{\boldmath$Q$}$ is the {\it best-matched derivative} \cite{BB82}, 
amounting to comparing two shapes by keeping one fixed and shuffling the other by $\FrG$'s group action on the configuration space $\FrQ$.) 
One can then apply a suitable $\FrG$-all move to each of these, such as sum, integrate, average, inf, sup or extremum.
Note that (\ref{Kend}) differs from the other two in using a {\sl finite} group action to the other two cases' infinitesimal ones.
On the other hand, (\ref{Kend}) and (\ref{DeWi}) compare two distinct inputs whereas (\ref{Barb}) works around a single input.  
Comparers have a further issue: if $\mbox{\boldmath$M$} = \mbox{\boldmath$M$}(\mbox{\boldmath$Q$})$, 
does one use $\mbox{\boldmath$Q$}$ or $\mbox{\boldmath$Q$}^{\prime}$ in evaluating $\mbox{\boldmath$M$}$ itself? 
This situation did not arise in Kendall's context, but it did in DeWitt's; 
he resolved it (Sec \ref{Gdyn-Dist}) in the symmetric manner, i.e. making used of $Q_1$ and $Q_2$ to equal extents.

In some cases, one might instead be able to work directly with, or reduce down to, $\FrQ/\FrG$ objects, in which case there is no need for the above indirect construct. 
One would then make use of the relational or reduced configuration space geometry $\widetilde{\mbox{\boldmath$M$}}$ itself.

\mbox{ } 

\ni Note also the dichotomy between direct comparisons of two configurations as per above, and performing intrinsic computations from each configuration piecemeal, e.g.

\ni\beq
\iota: \FrQ \longrightarrow \mathbb{R}^{p} \mbox{ } ,                                                                                        
\eeq
and then comparing these computations.\footnote{This is motivated e.g. by the preceding comparers failing to give distances when $\mbox{\scriptsize\boldmath$M$}$ is indefinite
-- losing the non-negativity and separation properties of bona fide distance -- which we know will occur for GR.}  
%
In the latter case, one can consider using norms in the space of computations that is mapped into (the $p$-dimensional Euclidean metric in the above example).  
Note however that the outcome of doing this may well depend on the precise quantity under computation. 
Also $\iota$ will in general has a nontrivial kernel, by which the candidate $\iota$-Dist would miss out on the separation property of bona fide distances.
If this separation fails, one can usually (see e.g. \cite{Gromov}) quotient so as to pass to a notion of distance. 
[Though sometimes this leaves one with a single object so that the candidate notion of distance has collapsed to a trivial one.]
Also it is sometimes limited or inappropriate to use such a distance if it is the originally intended space $\FrX$ and not the quotient that has deeper significance attached to it.

A range of candidate $\iota$'s for the GR case are provided in Sec \ref{Gdyn-Dist}.
$\iota$'s can again be directly or indirectly $\FrG$-invariant; indeed that is one way to select amongst the vast number of possibilities for $\iota$'s. 
Other selection criteria include extendability to unions of configuration spaces, physical naturality, and recurrence of the structure used in other physical computations 
E.g. a notion of distance that is, or at least shares structural features in common with, such as a classical action, an entropy or a notion of information, a quantum path integral 
or a statistical mechanical partition function.

\section{Field Theory: unreduced configuration space geometry}\label{Q-Fields}

Scalar field theory's configuration space is a space of scalar field values $\phi(\underline{x})$, which space I denote by Sca 
[implicitly Sca($\mathbb{R}^3$) in the most standard flat space case]. 

\mbox{ } 

\ni Electromagnetism's configuration space is a space of 1-forms $\mA_i(\underline{x})$,                which space I      denote by $\Lambda_1$ [implicitly $\Lambda_1(\mathbb{R}^3)$]. 
Yang--Mills theory's configuration space is a larger space of 1-forms $\mA_i^P(\underline{x})$,         which space I also denote by $\Lambda_1$ 
[thus this notation is also dropping reference to the corresponding gauge group].

\mbox{ } 

\ni In modelling the above, one can start off working with ${\cal L}^2$: the square-integrable functions.
One can furthermore pass to e.g. the Fr\'{e}chet spaces of Appendix A, 
which are useful in subsequent curved-space and GR-coupled versions (now with $\bupSigma$ in place of $\mathbb{R}^3$).
The scalar field version is then useful in cosmological modelling, and easily appended to the GR configuration space, at least in the minimally-coupled case and at the unreduced level.

\section{GR: unreduced configuration space geometry}\label{GR-Config}

\subsection{Topology of Riem($\bupSigma$)}\label{Riem-Top}

The space of Riemannian geometries Riem($\bupSigma$) can be modelled as an open positive convex cone\footnote{This is a Linear Algebra characterization of a space $\FrS$  
\cite{Fischer70, RS}, that is not itself linear but obeys $\FrS + \FrS \subset \FrS$ and $m \FrS \subset \FrS$ for $m \in \mathbb{R}_+$. 
See \cite{Fischer70} for more on this and for consideration of why Fr\'{e}chet spaces are appropriate.
Do not confuse this use of `cone' with Sec \ref{Q-Geom}'s topological and geometrical uses.} 
in the Fr\'{e}chet space (see Appendix A) Fre$_{\sS\sy\sm(0, 2)}$(${\cal C}^{\infty}$) for Sym(0, 2) the symmetric rank-2 tensors. 

\mbox{ }

\ni Riem($\bupSigma$) can furthermore be equipped with a metric space notion of metric \cite{Fischer70}, Dist; 
this can additionally be chosen such that it is preserved under Diff($\bupSigma$). 
Thus Riem($\bupSigma$) is a metrizable topological space.
Consequently it obeys all the separation axioms (including Hausdorffness), and is also paracompact.
Riem($\bupSigma$) is additionally second countable \cite{Giu09}, 
and additionally has an infinite-dimensional analogue of the locally Euclidean property, by which a single type of chart suffices for it.  
Thus Riem($\bupSigma$) is a manifold that is infinite-dimensional in the sense of Fr\'{e}chet(${\cal C}^{\infty}$).

\subsection{Riem($\bupSigma$) at the level of geometrical metric structure}\label{Riem-Geom}

In studying of GR, Riem($\bupSigma$) is usually taken to carry the infinite-$d$ indefinite Riemannian metric provided by GR's kinetic term, 
i.e. the inverse DeWitt supermetric $\mM^{abcd}$ of (\ref{M-GR}).  
More generally, one might consider other members of the family of {\it ultralocal supermetrics} \cite{Giu95b, KieferBook}

\ni\beq
\mM_{\beta}^{abcd} := \sqrt{\mh}\{\mh^{ac}\mh^{bd} - \beta\mh^{ab}\mh^{cd}\} \mbox{ } .
\label{M-beta}
\eeq
These split into three cases: $\beta < 1/3$ are positive-definite, $\beta = 1/3$ is degenerate, and $\beta > 1/3$ are indefinite (heuristically \{--+++++\}$^{\infty}$).
Due to ultralocality, it makes sense to study these pointwise; the more problematic degenerate case is usually dropped from the study.
Pointwise, then, they arise from positive-definite symmetric matrices (the $\mh_{ab}$ at that point), which is diffeomorphic to the homogeneous space \cite{Giu95b} 
$GL(3, \mathbb{R})/ SO(3) \, \cong \, \mathbb{R}^6$.
On the other hand, the $\muu_{ab}$ of (\ref{u}) are pointwise on $SL(3, \mathbb{R})/ SO(3)$; $\mh_{ab}$ can be decomposed into this and a scale part taking values in $\mathbb{R}_+$.
[The split corresponds to $GL(3, \mathbb{R})/ SO(3) \, \cong SL(3, \mathbb{R})/ SO(3) \times \mathbb{R}_+$.]
The pointwise structures then uplift to Riem($\bupSigma$) due to ultralocality.
The scale-free part gives rise to 8 Killing vectors and the scale part to a homothety \cite{Giu95b}.  
The local Riemannian geometry of this was covered by DeWitt \cite{DeWitt67}, including the form of the geodesics.
This exhibits various difficulties: curvature singularities and geodesic incompleteness.

\subsection{Conformal variants}\label{CRiem-Geom}

Expanding on footnote \ref{Conf-Foo}, the {\it conformal transformations} Conf($\bupSigma$) are smooth positive functions on $\bupSigma$ form .
These form an infinite-dimensional Lie group; moreover this is Abelian under pointwise multiplication.  

\mbox{ }

\ni Then {\it conformal Riem} CRiem($\bupSigma$) := Riem($\bupSigma$)/Conf($\bupSigma$) = `$\{SL_3(\mathbb{R})\}^{\infty}$' heuristically in parallel to the preceding Subsection.  
This is simpler and better-behaved  \cite{DeWitt67} than Riem($\bupSigma$) at the level of metric geometry.
This is firstly in the sense that the natural supermetric thereupon is  

\ni\beq
\mU^{abcd} := \muu^{ac}\muu^{bd} \mbox{ } ,
\label{U}
\eeq
which is {\it positive-definite} and thus furthermore the basis of a bona fide notion of distance.  
Note that the part of the GR configuration space metric that causes it to be indefinite is the (local) scale part \cite{DeWitt67}.
Secondly, geodesics are better-behaved upon CRiem($\bupSigma$) as compared to Riem($\bupSigma$).  

\mbox{ } 

\ni Conformal Riem has also been termed `pointwise conformal superspace' \cite{FM96}
However, this name is confusing in various ways.
Firstly, the name can only be understood if conformal superspace itself has already been introduced.
Yet CRiem($\bupSigma$) is a simpler space, and a strong case can be made for simpler entities to be introduced on their own terms rather than by reference to more complicated ones.
Secondly, Sec \ref{Riem-Geom} already made a {\sl distinct} use of `pointwise', to mean `looking at a field at just one point', which is a very clear use. 
The current use, on the other hand, would appear to be along the following lines.
Take a space that involves quotienting out Conf($\bupSigma$) and Diff($\bupSigma$) [`conformal superspace'] but now do not quotient out Diff($\bupSigma$) after all [`pointwise'].
However, this can be contracted to just `take a space that involves quotienting out Conf($\bupSigma$)', 
i.e. making no mention, rather than two cancellatory implicit mentions, of the concept that is unnecessary for the definition: Diff($\bupSigma$).  
On these grounds, I use instead the name `conformal Riem'.  
I denote this by `CRiem', making use of `C' for `conformal' parallelling the habitual use in `CS' for `conformal superspace', 
noting that `C' standing for `conformal' can {\sl just as well} be introduced prior to any mention of superspace or the associated Diff($\bupSigma$).

On the other hand, while passing to equivalence classes is mathematically convenient, the equivalence classes themselves can be considered to be more primary.
If CRiem($\bupSigma$) were viewed in this way, 
it would then make more sense for it and Riem($\bupSigma$) to be renamed so that now Riem($\bupSigma$)'s new name derives from CRiem($\bupSigma$)'s by a `locally-scaled' addendum.
One might go as far as viewing CS($\bupSigma$) as primary (for all that this is unlikely to be motivated by the final form of the `true degrees of freedom' of the gravitational field. 
Such primality amounts to assuming not geometrodynamics but {\it conformogeometrodynamics}.
In this case, a good primary name would be shape space, Shape, for the conformal 3-geometry notion of shape. 
Whence Superspace would be known as `locally-scaled Shape', CRiem as `diffeomorphism-redundant Shape' and Riem as `locally-scaled diffeomorphism-redundant Shape.

Moreover, the conventional approach to conformogeometrodynamics is that, in solving the Lichnerowicz--York equation \cite{York73}, 
one finally passes from $\{\mC\mS + \mV\}(\bupSigma)$ to True($\bupSigma$) by the solution fixing a particular form of the local scalefactor to be the physically realized one.  
Whereas traditionally, Conformogeometrodynamics is viewed as a convenient decoupling leading to substantial mathematical and numerical tractability, 
from the relational perspective, one can take 

\ni $\FrG$ = Conf($\bupSigma$) $\rtimes$ Diff($\bupSigma$). 

\mbox{ } 

\ni Next, \{CRiem + V\}($\bupSigma$) 's metric is `$\{-+++++\}^{3\infty}$', which is actually hyperbolic rather than pointwise hyperbolic. 
The -- direction here corresponds to a global scale variable, such as indeed the global spatial volume, or the cosmological scalefactor $a$ when applicable.

Both for GR and RPM's, many of the configuration spaces have physically-significant singular points. 
In particular, $a = 0$ is the Big Bang and $\mI = 0$ is the maximal collision, which are furthermore analogous through each involving scale variables.

\mbox{ } 

\ni Finally, quotienting out conversely the overall (constant) dilations Dil alone from Riem($\bupSigma$) gives a VPRiem($\bupSigma$) configuration space (volume-preserving Riem).

\subsection{GR alongside minimally-coupled matter}\label{RIEM}

This case is taken to include fundamental-field second-order minimally-coupled bosonic matter, 
covering e.g. minimally-coupled scalars, Electromagnetism, Yang--Mills Theory and scalar gauge theories.
Then the redundant configuration space metric splits according to the direct sum \cite{Teitelboim}

\ni\beq
\bM = \bM^{\sg\sr\sa\sv} \oplus \bM^{\sm\scc\sm} \mbox{ } .
\label{grav-oplus-mcm}
\eeq

\mbox{ } 

\ni In the case of a minimally-coupled scalar field, I denote this configuration space by RIEM($\bupSigma$).\footnote{In this Article, 
I make wider use of such a capping convention for versions including a minimally-coupled scalar field.}
%
The (undensitized) metric on this takes the blockwise form 
$\mbox{\boldmath $\cal M$}(\mh) := \mbox{\Huge{(}} \stackrel{\mbox{\normalsize \, \, 1 \, \, \, 0}} {\mbox{\normalsize 0  \,\, $M(\mh)$}} \mbox{\Huge{)}}$ and $\bM(\mh)$ 
the GR configuration space metric itself.

\mbox{ } 

\ni It is usually additionally assumed that $\bM$ is independent of the matter fields.  
This is well-known, and held to secure freedom to `add in' scalar fields in cosmological modelling.

\mbox{ } 

\ni Similar considerations apply throughout to extending CRiem($\bupSigma$), \{CRiem  + V\}($\bupSigma$) and VPRiem($\bupSigma$).

\section{Minisuperspace: homogeneous GR}\label{MSS-Q-Geom}

The vacuum case of minisuperspace, Mini($\bupSigma$) \cite{Magic}, is the space of homogeneous positive-definite 3-metrics on $\bupSigma$. 
Each corresponds to a notion of space in which every point is the same.  
Here full GR's ${\mM}^{ijkl}(\mh_{mn}(x^i))$ collapses to an ordinary $6 \times 6$ matrix, ${M}_{\sfA\sfB}(h_{ab})$: 
an overall -- rather than independently per space point -- curved (--+++++) `minisupermetric'.  
A particular simpler subcase are the {\it diagonal minisuperspaces}, with $3 \times 3$ matrix $M_{\sfA\sfB}$ \cite{Magic}.

Minisuperspaces are classified by the isometry groups Isom($\bupSigma, \bh$) of their spatially homogeneous surfaces. 
This leads to two cases according to whether Isom($\bupSigma, \bh$) acts simply transitively. 
If this is not the case, it turns out that \cite{MacCallum}
 there is a single case: $SO(3) \times \mathbb{R}$ acting upon the cylindrical 3-space $\mathbb{S}^2 \times \mathbb{R}$; this gives the {\it Kantowski--Sachs model}.
The other case gives the family of {\it Bianchi models}.
The $I$ being Lie groups, they are in turn characterized by the form of their structure constants.  
They are subdivided according to 

\ni\beq
{C^{i}}_{ij} = 0 \mbox{ for type A and $\neq 0$ for type B } . 
\eeq
A finer classification of ${C^{k}}_{ij}$ yields a subdivision into nine kinds of Bianchi models, labelled by I to IX \cite{MacCallum}.  
The general case's spatial metric can be represented as

\ni\beq
\d s^2 = h_{ij}\d \sigma^i \d \sigma^j
\eeq 
for 1-forms $\d \sigma^k$ obeying $\d^2 \sigma^k = {C^k}_{ij} \d \sigma^i \wedge \d \sigma^j$. 

\mbox{ } 

\ni Example 1) The spatially closed $\mathbb{S}^3$ isotropic case has spatial metric $\d s^2 = a(t)^2\d s_{\mathbb{S}^3}^2$.

\ni Example 2) Diagonal Bianchi IX models have spatial metrics

\ni\beq 
\d s^2 = \mbox{exp(2\{$-\Omega + \beta_+ + 3\sqrt{3}\beta_-\}$)} \d \Omega^2 
       + \mbox{exp(2\{$-\Omega + \beta_+ - 3\sqrt{3}\beta_-$\})} \d \beta_+^2  +  \mbox{exp(2\{$-\Omega - 2\beta_+$\})}  \d \beta_-^2  
\eeq																		   
on $\mathbb{S}^3$.
Diagonal Bianchi IX models are potentially of great importance through being conjectured to be the generic GR behaviour near cosmological singularities \cite{BKL}.

These also have a nontrivial potential term inherited from the GR Ricci scalar potential term, 

\ni\beq
V = \mbox{exp}(-4\Omega)\{V(\beta) - 1\} \mbox{ } , \mbox{ } \mbox{ for}
\eeq

\ni \beq
V(\beta) = \mbox{$\frac{\mbox{\scriptsize exp}(-8\beta_+)}{3}$} - \mbox{$\frac{4 \, \mbox{\scriptsize exp}(-2\beta_+)}{3}$}\mbox{cosh}\,(2\sqrt{3}\,\beta_-) + 1 
         + \mbox{$\frac{2\,\mbox{\scriptsize exp}(4\beta_+)}{3}$}\{\mbox{cosh}(4\sqrt{3}\beta_-) - 1\} \mbox{ } ,
\eeq
which is an open-ended equilateral triangular cross-section well (Fig \ref{Anisotropy}.c). 

{            \begin{figure}[ht]
\centering
\includegraphics[width=0.9\textwidth]{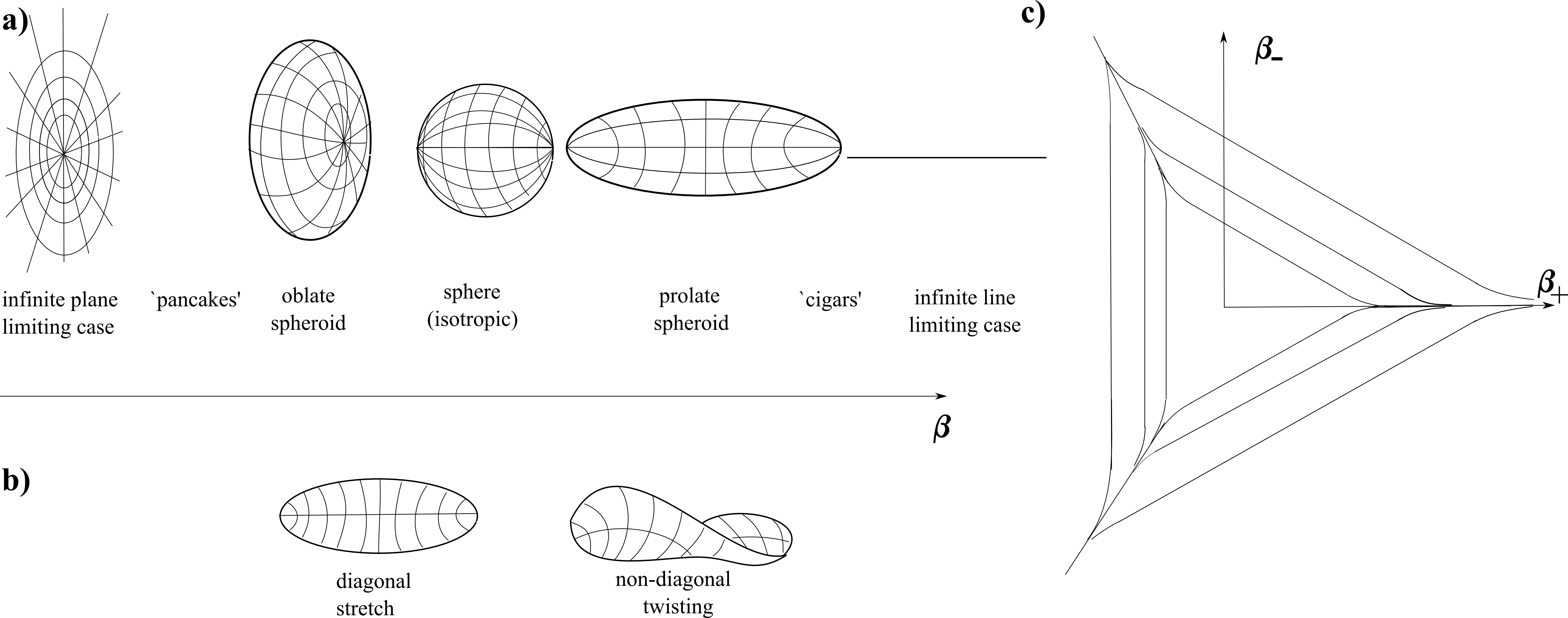}
\caption[Text der im Bilderverzeichnis auftaucht]{        \footnotesize{a) Visualization of diagonal anisotropy in 2-$d$.  
b) Non-diagonal anisotropy allows for the hypersurface to be `twisted' as well.
In 3-$d$ there are then two independent anisotropic stretches and three twistings. 
c) The diagonal Bianchi IX model's potential well.} }
\label{Anisotropy} \end{figure}          }

\ni The above uses the {\it Misner variable} 
\beq
\Omega := \mbox{ln} \, a \mbox{ } , 
\label{Misner} 
\eeq 
and also Misner's \cite{Mis68} parametrization of anisotropy -- a type of GR shape variable -- by writing the tracefree spatial metric $u_{ab}$ as 

\ni\beq
\mbox{exp}(2\beta)_{ab} \mbox{ } \mbox{ for } \mbox{ } \beta_{ab} \mbox{ } \mbox{ a tracefree symmetric matrix } .
\eeq
In the case of diagonal $u_{ab}$ diagonal, 

\ni\beq 
\beta_{ab} = \mbox{diag}(\beta_1, \{\sqrt{3}\beta_2 - \beta_1\}/2, -\{\sqrt{3}\beta_2 + \beta_1\}/2) \mbox{ } . 
\eeq 
These are related to $\beta_{\pm}$ by $\beta_1 = \beta_+ + \sqrt{3}\beta_-, \beta_2 = \beta_+ - \sqrt{3}\beta_-$.
Fig \ref{Anisotropy} provides a simple conceptual outline of the meaning of anisotropy for a 2-$d$ hypersurface.   

\mbox{ }

\ni Then in the homogeneous case the configuration spaces Riem,   CRiem + V,   Superspace and CS + V coincide as {\it Minisuperspace}, Mini 
and                          CRiem, VPRiem, Sec \ref{CS-Geom}'s VPSuperspace,             and CS     coincide as {\it Anisotropyspace}, Ani.
Ani is yet another example of pure shape space.

\mbox{ }

\ni Then for diagonal Bianchi class A, Mini = $\mathbb{M}^3$ with configuration space metric 

\ni\beq
\d s^2 = - \d \Omega^2 + \d\beta_+^2 + \d\beta_-^2 \mbox{ } .
\label{Bianchi-A}
\eeq
Ani = $\mathbb{R}^2$ with shape metric 

\ni\beq
\d s^2 = \d\beta_+^2 + \d\beta_-^2 \mbox{ } .
\eeq  

\ni In the general non-diagonal minisuperspace, full configurations can be represent by elements of $GL_3$($\mathbb{R}, \bupSigma$) and 
                                                and pure shapes (anisotropies)       by elements of $SL_3$($\mathbb{R}, \bupSigma$).  

\mbox{ } 

\ni Example 3) Upon inclusion of a single minimally-coupled scalar field, I use the corresponding capitalized notation MINI and ANI.
The undensitized configuration space metric on MINI is 

\ni\beq
\d s^2 = - \d\Omega^2 + \d\phi^2 
\label{M-MINI}
\eeq
and the undensitized potential is

\ni\beq 
V : = -\mbox{exp}(-2\Omega) + V(\phi) + 2\Lambda \mbox{ } . 
\label{MSS-Action}
\eeq
for $V(\phi)$ an unrestricted function.

\section{Perturbations about minisuperspace: unreduced formulation}\label{SIC-Q-Geom}

\ni As regards incipient (redundant) configuration variables, the 3-metric and scalar field are expanded as \cite{HallHaw}

\ni \beq
\mh_{ij} = \mbox{\scriptsize exp(2$\Omega(t)$)}\{S_{ij}(t) + \upepsilon_{ij}(\underline{x}, t)\} \mbox{ } , \mbox{ } \mbox{ } 
\upphi = \sigma^{-1}
\left\{
\phi(t) + \sumnlm f_{\sn\sll\sm} \, \mQ^{\sn}_{\sll\sm}(\underline{x}) 
\right\} 
\mbox{ } . 
\label{Sepsi}
\eeq
Here, $S_{ij}$ is the standard hyperspherical $\mathbb{S}^3$ metric. 
$\upepsilon_{ij}$ are inhomogeneous perturbations of the form 

\ni $$
\upepsilon_{ij}=\sumnlm
\big\{ 
\sqrt{\mbox{$\frac{2}{3}$}} a_{\sn\sll\sm}S_{ij}\mQ^{\sn}\mbox{}_{\sll\sm} + \sqrt{6} \, b_{\sn\sll\sm}\{\mP_{ij}\}^{\sn}\mbox{}_{\sll\sm}         
$$

\ni \beq
+  \sqrt{2}\{c^{\so}_{\sn\sll\sm}\{\mS^{\so}_{ij}\}^{\sn}\mbox{}_{\sll\sm}    +         c^{\se}_{\sn\sll\sm}\{\mS^{\se}_{ij}\}^{\sn}\mbox{}_{\sll\sm}\} 
+        2 \{d^{\so}_{\sn\sll\sm}\{\mG^{\so}_{ij}\}^{\sn}\mbox{}_{\sll\sm}    +         d^{\se}_{\sn\sll\sm}\{\mG^{\se}_{ij}\}^{\sn}\mbox{}_{\sll\sm}\}  
\big\} .
\label{abcd}
\eeq
The superscripts `$\mo$' and `$\me$' for stand for `odd' and `even' respectively.  
I subsequently use n indices as a shorthand for nlm.  
Let $x_{\sn}$ be a collective label for the 6 gravitational modes per n $a_{\sn}$, $b_{\sn}$, $c_{\sn}$ and $d_{\sn}$, 
and $y_{\sn}$ for these alongside the $f_{\sn}$.
The $y_{\sn}$ are all functions of just the coordinate time (which is also label time for GR) $t$.
The     $\mQ_{\sn}(\underline{x})$                                                   are the                      $\mathbb{S}^3$             scalar (S) harmonics, 
        $\mS_{i\,\sn}^{\so}(\underline{x})$  and $\mS^{\se}_{i,\sn}(\underline{x})$  are the transverse           $\mathbb{S}^3$             vector (V) harmonics, 
and the $\mG_{ij\,\sn}^{\so}(\underline{x})$ and $\mG^{\se}_{ij,\sn}(\underline{x})$ are the transverse traceless $\mathbb{S}^3$ symmetric 2-tensor (T) harmonics.
The $\mS_{ij\,\sn}(\underline{x})$ are then given by $\mD_j\mS_{i\,\sn} + \mD_i\mS_{j\,\sn}$ (for each of the suppressed $\mo$ and $\me$ superscripts). 
The $\mP_{ij\,\sn}(\underline{x})$ are traceless objects given by $\mP_{ij\,\sn} := \mD_j\mD_i\mQ_{\sn}/\{\mn^2 - 1\} + S_{ij}\mQ_{\sn}/3$. 
$\sigma := \sqrt{2/3\pi}/m_{\sP\sll}$ is a normalization factor.  

\mbox{ } 

\ni Then in the vacuum case, the redundant configuration space is Riem$_{0,1,2}$($\mathbb{S}^3$); the 0, 1 and 2 subscripts refer to the orders in perturbation theory that feature in it.  
This is the $1 + 6 \times \{\mbox{countable } \infty\}$ space of scale variable alongside the $x_{\sn}$.
In the minimally-coupled scalar field case, the redundant configuration space is RIEM$_{0,1,2}$($\mathbb{S}^3$). 
This is the $2 + 7 \times \{\mbox{countable } \infty\}$ space of scale variable, homogeneous scalar field mode and the $y_{\sn}$.
The first form in Fig \ref{SIC-Metric} displays the latter for one mode to second order overall in $y_{\sn}$, $\d y_{\sn}$.
By the direct sum split of Sec \ref{RIEM}, RIEM$_{0,1,2} = \mbox{Riem}_{0,1,2}(\mathbb{S}^3) \oplus \mbox{Sca}_{0,1,2}(\mathbb{S}^3)$, for Sca standing for scalar field configuration space.
Thus the former configuration space can readily be read off the figure as a sub-block.
%
{\begin{figure}[ht]
\centering
\includegraphics[width=0.9\textwidth]{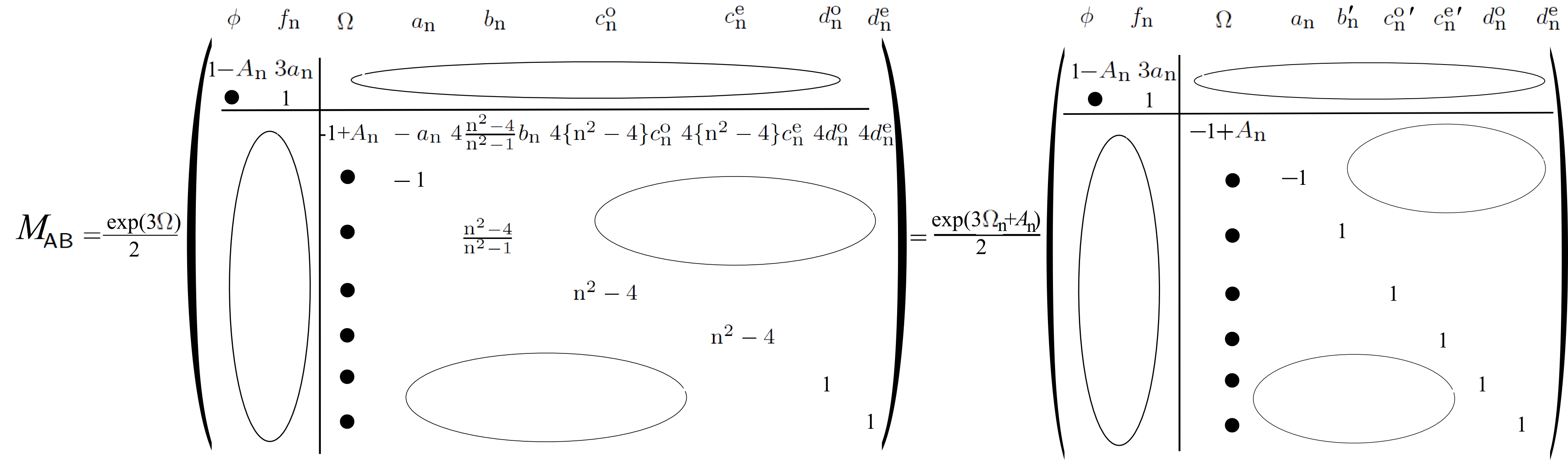}
\caption[Text der im Bilderverzeichnis auftaucht]{\footnotesize{a) Slightly inhomogeneous cosmology's configuration space metric \cite{SIC1}.  
The heavy dot denotes `same as the transposed element' since metrics are symmetric.  
N.B. this is the blockwise corrections' configuration space metric rather than the full one.}} 
\label{SIC-Metric}\end{figure}} 

\ni Blockwise-simplifying coordinates can additionally be found. 
Considering for instance the modewise case, 

\ni\beq
\Omega_{\sn} = \Omega - A_{\sn}/3    \mbox{ }  ,
\label{Straightener}
\eeq
removes the gravitational sector's off-diagonal terms, for \cite{AHH2}

\ni \beq
A_{\sn} := - \mbox{$\frac{3}{2}$}
\left\{
a_{\sn}^2 - 4
\left\{ 
\mbox{$\frac{\sn^2 - 4}{\sn^2 - 1}$}b_{\sn}^2 + \{\mn^2 - 4\}c_{\sn}^2 + d_{\sn}^2
\right\}
\right\} \mbox{ } , 
\label{A-n}
\eeq
By this and trivial rescalings $b_{\sn}^{\prime} := \sqrt{\mbox{$\frac{\sn^2 - 4}{\sn^2 - 1}$}}b_{\tn}$ and $c_{\sn}^{\prime} := \sqrt{\mn^2 - 4} \, c_{\tn}$, 
one arrives at the second form of the metric in Fig \ref{SIC-Metric}.

The above configuration space geometry for slightly inhomogeneous cosmology is curved; nor is it conformally flat. 
For $x_{\sn}$ small, its Ricci scalar $R$ has no singularities away from the Big Bang.    
In the minimally coupled scalar field case, $\pa/\pa\phi$ and $\pa/\pa f_{\sn}$ are Killing vectors for slightly inhomogeneous cosmology's configuration space geometry. 
This corresponds to the `adding on' status of scalar fields at this level.
Additionally, $\pa/\pa\Omega$ is a conformal Killing vector.

\mbox{ } 

\ni Finally consider these model arenas' scale-free spaces of inhomogeneities.
I use $w_{\sn}$ as a collective label for the 5 positive-definite gravitational modes $b_{\sn}$, $c_{\sn}$, $d_{\sn}$, 
and   $z_{\sn}$ for the 6 positive-definite modes $w_{\sn}$ and $f_{\sn}$.
Then CRiem$_{2}$($\mathbb{S}^3$) is the space of the $w_{\sn}$, which is straightforwardly just $\mathbb{R}^5$.
Thus it has the obvious 15 Killing vectors built from the Cartesian rescalings of the $w_{\sn}$ coordinates: 
5 $\pa/\pa w_{\sn}^{\sfW}$ and 10 $w^{\sfW}_{\sn}\pa/\pa w_{\sn}^{\sfW^{\prime}} - w^{\sfW^{\prime}}_{\sn}\pa/\pa w_{\sn}^{\sfW}$.  
On the other hand, CRIEM$_{0,1,2}$($\mathbb{S}^3$) -- the space of homogeneous scalar field modes $\phi$ alongside the $z_{\sn}$ -- is neither flat nor conformally flat.

\section{Reduced configuration spaces for Field Theory}\label{Q-Red-Fields}

\subsection{Gauge Theory's orbit spaces}\label{GOS}

For gauge theories, as well as the configuration space of connections Conn itself, there is a gauge group $\FrG$ acting upon Conn.
It is a Lie group, and in the more usual cases such as Electromagnetism and Yang--Mills Theory, it acts internally.
In this way, $\Lambda_1/\FrG$ arise as reduced configuration spaces (more concretely, gauge orbit spaces). 

\mbox{ } 

\ni Fibre Bundle Theory supports this to some extent, as follows.  

\mbox{ } 

\ni 1) Modelling using principal fibre bundles: with the above $\FrG$ entering as both structure group and fibres.

\ni 2) A wider range of associated fibre bundles with $\FrG$ as structure group and distinct fibres, by which coupling of gauge fields to a number of further (gauged) fields can be modelled.

\mbox{ } 

\ni On the other hand, the space of orbits itself is in general heterogeneous, and thus not itself amenable to a fibre bundle description. 
Moreover, due to the group action in question being smooth and proper, orbit spaces are separable -- and thus in particular Hausdorff -- 
as well as metrizable, second-countable and paracompact.  
See e.g. \cite{KR86, RSV02, Schmidt03} for more on the topology and geometry of gauge orbit space, 
and the corresponding symplectic spaces, including from a stratified manifolds perspective.
Some particular theorems of note that apply are as follows.  

\mbox{ } 

\ni {\it Gauge Theory's Slice Theorem} \cite{KR86}.  
The action of the gauge group on space of $\mA_i$ admits a slice at every point; this applies in a principal fibre bundle setting. 

\ni {\it Gauge Theory's Stratification Theorem} \cite{KR86}.  
The decomposition of $\Lambda_1/\FrG$ by orbit type is a regular stratification. 

\mbox{ } 

\ni Note that $L^2$ (`square integrable') mathematics suffices for the above workings, 
though one can uplift to more general function spaces \cite{RSV02} including so as to attain compatibility with the GR case (see Sec \ref{GOS}).

\subsection{Loops and loop spaces for gauge theory}\label{Loop-G}

Another approach is to make use of {\it Wilson loop variables}; these contain an equivalent amount of information to the $\mA_i$ variables.  
Such a formulation is already meaningful for Electromagnetism; 
it amounts here to modelling the space of transverse $\mA_i$ for which the Gauss constraint has already been taken into account.  
The Wilson loop variables are here of the form 

\ni\beq
H_{\sA}(\gamma) = \mbox{exp}\left(i\oint_{\gamma}\d x^i \mA_i(x) \right)
\eeq
for $\gamma$ a loop path.
The somewhat more involved Yang--Mills version of Wilson loop variables are of the form

\ni\beq
H_{\sA}(\gamma) = \mbox{Tr}\left(P \, \mbox{exp}\left( i g \oint_{\gamma} \d x^i \mA_{iP}(x) \mg^P(x) \right)\right)
\eeq 
for $\FrG$ group generators $\mg^P$ and path-ordering symbol $P$.
$H_{\sA}(\gamma)$ are indeed holonomy variables in the sense of fibre bundles \cite{Nakahara}; 
see \cite{GPBook} for an extensive (if nonrigorous) development of these.  

The curves in question can be taken to be continuous and piecewise smooth, and, for now, to be living on $\mathbb{R}^3$.
In fact, use of equivalence classes of curves is required \cite{GPBook}.

\mbox{ } 

\ni If modelled in this way, the corresponding loop space is a topological group.
It is not however a Lie group, though it is contained within a Lie group: the so-called {\it extended loop group}.
See Gambini and Pullin's book \cite{GPBook} for further details of these loop spaces at a heuristic level.

\section{Reduced configuration spaces for GR}\label{GR-Config-2}

\subsection{Topology of $\FrG$ = Diff($\bupSigma$)}

Diff($\bupSigma$) can be -- matchingly with Riem($\bupSigma$) \cite{Fischer70} -- taken to be modelled by use of Fre$_{(1, 0)}$(${\cal C}^{\infty}$) [(1, 0) are vector fields].
Indeed, diffeomorphisms are commonly modelled in terms of Fr\'{e}chet manifolds, a fortiori as Fr\'{e}chet Lie groups \cite{Hamilton82}.  

\mbox{ } 

\ni Next consider the group action Diff($\bupSigma$) $\times$ Riem($\bupSigma$) $\rightarrow$ Riem($\bupSigma$).
The group orbits of this are then 

\ni $\mbox{Orb}(\bh) := \{\phi^*\mh \, | \, \phi \in \mbox{Diff}($\bupSigma$)\}$.
Then metrics -- points in Riem($\bupSigma$) -- lying on the same orbit amounts to these being isometric.
Thus the Diff($\bupSigma$)-orbits partition Riem($\bupSigma$) into isometric equivalence classes \cite{Fischer70}.
The corresponding stabilizers 

\ni $\mbox{Stab}(\bupSigma, \bh) = \{\phi\,|\,\phi \in \mbox{Diff}(\bupSigma) \mbox{ such that } \phi^*\bh = \bh\}$ 
constitute the isotropy group        Isot($\bupSigma, \bh$). 
Moreover, Isot($\bupSigma, \bh$) coincides with \cite{Fischer70} Isom($\bupSigma, \bh$); I mark this by using I($\bupSigma, \bh$) to denote this coincident entity. 
The Lie algebra corresponding to this is isomorphic to that of the Killing vector fields of $\langle \bupSigma$, $\mh_{\mu\nu} \rangle$. 
An interesting result is that I($\bupSigma$, $\bh$) is compact if $\bupSigma$ is \cite{MS39}.
Finally, since $\mI(\bupSigma, \, \bh)$ comes in multiple sizes, there are multiple dimensions of the corresponding orbits, pointing to the space of orbits not being a manifold.

\subsection{Topology of Superspace($\bupSigma$)}\label{Superspace-Top}

Fischer showed that Superspace($\bupSigma$) = Riem($\bupSigma$)/Diff($\bupSigma$) \cite{Fischer70} can be taken to possess the corresponding quotient topology.
Superspace($\bupSigma$) additionally admits a metric in the metric space sense of the form \cite{Fischer70} 

\ni\beq
\mbox{Dist}([\bh_1], [\bh_2]) = \inf_{\phi \in \mbox{\scriptsize Diff}(\Sigma)}\left(\mbox{Dist}(\phi^*\bh_1, \phi^*\bh_2)\right) \mbox{ } .
\label{Fischer-Dist}
\eeq
In this manner, Superspace($\bupSigma$) is a metrizable topological space and thus obeys all the separation axioms and thus in particular Hausdorffness; 
it is also second countable \cite{Fischer70}.
Thus Superspace is `2/3rds of a manifold' in Appendix \ref{Quotients}'s sense.\footnote{On the other hand, 
the space of spacetimes is an example of a non-Hausdorff `space of spaces' \cite{Fischer70}.\label{Not}}
 
\mbox{ } 
 
\ni However, unlike Riem($\bupSigma$), Superspace($\bupSigma$) fails to possess the infinite-dimensional analogue of the locally-Euclidean property. 
Wheeler \cite{Battelle} credits Smale with first pointing this out.
Fischer \cite{Fischer70} then worked out the details of the structure of Superspace($\bupSigma$) as a stratified manifold.  
In particular, the appearance of nontrivial strata occurs for $\bupSigma$ that admit metrics with non-trivial $\mI(\bupSigma, \bh)$.
In these cases Diff($\bupSigma$) clearly does not act freely upon these metrics. 
Rather, the Superspace($\bupSigma$) quotient space is here a stratified manifold of nested sets of strata ordered by dim(I($\bupSigma, \bh$)).
(Indeed, Fischer \cite{Fischer70} tabulated the allowed isometry groups on various different spatial topologies.)  
In this way, Superspace($\bupSigma$) is not a manifold in the sense of Fr\'{e}chet (corresponding to the underlying function spaces used in Fischer's mathematical modelling).  

\mbox{ } 

\ni A further useful concept is the {\it degree of symmetry} of $\bupSigma$, 

\ni \beq
\mbox{deg}(\bupSigma) := \sup_{\sbh \in \mbox{\scriptsize Riem}(\Sigma)} \big( \mbox{dim}(\mI(\bupSigma, \bh)) \big) \mbox{ } . 
\eeq
Fischer \cite{Fischer70} listed 3-manifolds with deg($\bupSigma$) $>$ 0, and further characterized deg($\bupSigma$) = 0 manifolds in collaboration with Moncrief, \cite{FM96}.
N.B. that for deg = 0 $\bupSigma$, Superspace($\bupSigma$) {\sl is} a manifold. 

\mbox{ }  

\ni Ebin \cite{Ebin} established that Diff($\bupSigma$) is not compact. 
However, Ebin and Palais furthermore showed that Diff($\bupSigma$) acting on Riem($\bupSigma$) is one of the cases for which a slice (Appendix \ref{Lie-Slice}) does none the less exist.

\mbox{ } 

\ni {\bf Ebin--Palais Slice Theorem} \cite{Ebin}.
For each $\bh \in \mbox{Riem}(\bupSigma)$ $\exists$ a contractible submanifold $\FrS$ containing $\bh$ such that

\mbox{ } 

\ni i)    For $\phi$ a diffeomorphism [in Diff($\bupSigma)$, $\phi \in \mI(\bupSigma, \bh)       \, \Rightarrow \, \phi^*\FrS = \FrS$; here the upstairs * denotes pull-back.

\ni ii)   $\phi \not{\!\in} \,\, \mI(\bupSigma, \bh) \, \Rightarrow \, \phi^*\FrS \bigcap \FrS = \emptyset$.  

\ni iii)  $\exists$ in Orb($\bh$) an open set $\FrO$ itself containing $\bh$, and 
                                  a local cross-section $\Gamma: \FrP \rightarrow \mbox{Diff}(\bupSigma)$ such that 
                                  $\phi(p, s) = \{\Gamma(p)\}^*s$ is a diffeomorphism of $\FrP \times \FrS$ onto an open neighbourhood $\FrU_{\sbh}$ of $\bh$.

Here $\FrP$ is an open neighbourhood of I($\bupSigma, \bh$)'s identity in the coset space: 
Diff($\bupSigma$)/I($\bupSigma, \bh$), and `diffeomorphism' and `submanifold' are in the sense of Fr\'{e}chet(${\cal C}^{\infty}$). 

\mbox{ } 

\ni The following is then a ready consequence \cite{Fischer70}. 

\mbox{ } 

\ni{\bf Superspace Decomposition Theorem}. 
The decomposition of Superspace($\bupSigma$) into orbits is a countable partially-ordered Fr\'{e}chet(${\cal C}^{\infty}$) manifold partition. 

\mbox{ } 

\ni Then via the preceding and Appendix \ref{Strat}'s definition of inverted stratification, the following also holds \cite{Fischer70}.  

\mbox{ } 

\ni{\bf Superspace Stratification, Stratum and Strata Theorems}. 
The manifold partition of superspace is an inverted stratification (Appendix \ref{Quotients}) indexed by symmetry type.  
Fischer then classifies the superspace topologies and the strata into two large tables.
The Stratum Theorem includes \cite{Fischer70} that a stratum of superspace is finite dimensional iff the group action on the manifold is transitive (corresponding to a homogeneous space).  

\mbox{ }  

\ni See e.g. \cite{Fischer70, Fischer86, Giu94, Giu95} for further topological studies                           of superspace, 
         and \cite{Giu09}                              for further difficulties with putting a Riemannian metric on superspace.

\subsection{Comparison between Theories. 1. Theorems.}

Let us next further compare the GR, Gauge Theory and Mechanics cases.
Firstly, Slice Theorems are known for each. 
See above for GR, Sec \ref{GOS} for Gauge Theory and e.g. \cite{OR02Schmah} for the case of Mechanics.  
Secondly, see the same Sections for the GR and Gauge Theory cases of Stratification Theorems; Mechanics also has a such, at least in the symplectic setting \cite{MR99}.
The above two results provide further directions in which to take Sec 2's model arena study.
Thirdly, the Decomposition Theorem that we have seen arise for GR also has a Gauge Theory's version of Decomposition Theorem \cite{RSV02}.  
This is for orbit spaces and is more mathematically standard (based on a `Hodge--de Rham decomposition').

\subsection{Comparison between Theories. 2. Handling dynamical trajectories exiting a stratum.}\label{starr}

Stratification becomes an issue as regards continuations of dynamical trajectories.

In the GR case, Leutwyler and Wheeler \cite{Battelle} appear to have been the first to ask about initial or boundary conditions on superspace.  
DeWitt, Fischer and Misner then suggested \cite{DeWitt70, Fischer70, Misref} 
that when the edge of one of the constituent manifolds -- i.e. where the next stratum starts -- is reached, 
the path in Superspace that represents the evolution of the 3-geometry could be {\it reflected}.
Simpler such reflection conditions were also previously considered for Mechanics; 
Misner's considerations were in Minisuperspace; 
DeWitt considered a further simple model arena \cite{DeWitt70}.  

\mbox{ }

\ni A subsequent alternative proposal of Fischer involved extending such motions via working instead with a nonsingular extended space.
This no longer encounters the stratified manifold's issues as regards differential equations for motion becoming questionable at the junctions between strata.

He explicitly built such an extended space \cite{Fischer86} by use of an unfolding which permits access to fibre bundle methods.
The unfolding involved is parametrized by I($\bupSigma$), as anticipated in the Mechanics case in Sec \ref{CPST}
This unfolding improves on previous such constructs by being generally covariant.
It provides the right amount of information at each geometry (the space's notion of point) to make the space of geometries into a manifold.
The unfolding attains this by making use of the {\it bundle of linear frames over} $\bupSigma$, $\mF(\bupSigma)$.   
Then no nontrivial isometries fix a frame.  
Thus the group action on the unfolded space Riem($\bupSigma$) $\times \mF(\bupSigma)$ is free.
In fact, by applying 1-point compactification to the open case, the open and closed cases are closer to each other than might be expected. 
In particular, Fischer \cite{Fischer86} established that Superspace$_{\sF}$(1-point compactified $\bupSigma$) is diffeomorphic to Superspace($\bupSigma$).

Fischer \cite{Fischer86} also pointed to Superspace possessing a `natural minimal resolution' of the resultant singularities.
This is based upon using the frame bundle quotient space Riem($\bupSigma$) $\times$ F($\bupSigma$)/ Diff($\bupSigma$).   
In this particular case, one can then regarding Riem($\bupSigma$) as a principal fibre bundle P(Superspace$_{\sF}$($\bupSigma$), Diff$_{\sF}$($\bupSigma$)).  
I.e.

\ni $\mbox{Diff}_{\sF}(\bupSigma) \stackrel{i}{\rightarrow} \mbox{Riem}(\bupSigma) \stackrel{\pi}{\rightarrow} \mbox{Superspace}_{\sF}(\bupSigma)$ 
for $i$ an inclusion map (Appendix C.1) and $\pi$ the Fibre Bundle Theory's projection map.

However, the above unfolding runs against relationalism, 
due to the $\mF(\bupSigma)$ involved being a mathematical construct that does not correspond to more detailed modelling of physical entities.

\mbox{ } 

\ni Within the alternative `accept' strategy to strata, see firstly item 1) of Sec \ref{Fur-App} as regards sheaf methods for extending geodesics between strata, 
for use in both GR and Mechanics.  
Secondly, I point out here that stratifolds (Appendix \ref{Stratifolds}) happen to further model a number of configuration spaces of interest.
This is firstly via Appendix \ref{H2LC}'s statement about Mechanics configuration spaces, and secondly via Appendix \ref{Stratifolds}'s statement about infinite-dimensional stratifolds 
moving toward being able to model GR configuration spaces.
On the other hand, the space of spacetimes modulo spacetime diffeomorphisms not being Hausdorff leaves this space outside the scope of stratifolds, as are some loop spaces.
Thus the reason I mention the stratifold construct is its applicability to a range of physically interesting examples, 
rather than as some full resolution of all stratified manifolds that arise in Physics.  

\mbox{ } 

\ni It is also worth pointing out that, as Fischer and Moncrief pointed out \cite{FM96}, the deg($\bupSigma$) = 0 case of Superspace($\bupSigma$) avoids having strata in the first place, 
thus not necessitating any boundary conditions or extension procedure thereat.
On the one hand, this deg($\bupSigma$) = 0  carries connotations of genericity, upon which general relativists place much weight. 
On the other hand, there is considerable interest in studying the simpler superspaces which are based on spaces with Killing vectors, such as $\mathbb{S}^3$ and $\mathbb{T}^3$, 
for which reduced approaches do encounter stratification.

\mbox{ }

\ni Another research direction arises if the actual Universe is acknowledged to at most have approximate Killing vectors; this can be considered as a realization of Strategy D).
This would however come at the price of significant amounts of standard techniques becoming inapplicable. 
E.g. 1) perturbation theory that is {\sl centred about} an exact solution with exact Killing vectors might cease to apply.
2) Ab initio averaging issues enter the modelling. 
1) would be covered by modelling the universe on some specific deg 0 spatial topology: in this case we know there are no Killing vectors for strata to arise from.
This would greatly complicate calculations as compared to those we are accustomed to on e.g. $\mathbb{S}^3$.
2) however would be manifested through us not knowing which deg 0 spatial topology to take; 
one would now have to average over all plausible such, and quite possibly allow for these to change over evolution.
By this stage one would be modelling with `big superspace' and it would be a `higher level excision' to exclude the superspaces with Killing vectors. 
One might still hope that sufficiently accurate analysis of the dynamical path would reveal it to avoid deg $\neq$ 0 topologies, or at least the solutions with Killing vectors therein.
However, issues remain as regards whether each of the nongeneric structures (deg($\bupSigma$) $\neq$ 0, $\mh_{ab}$ possessing Killing vectors) 
can come to possess a dynamical attractor role, by which generic {\sl paths} could be forced to have endpoints in, or go arbitrarily close to, non-generic points. 

\mbox{ } 

\ni Finally see e.g. \cite{DSVV09} for boundary condition considerations for the Gribov regions of Gauge Theory.

\subsection{Locality in the Thin Sandwich amounts to excision}\label{Thin-San}

\ni The Thin Sandwich Theorem of Bartnik and Fodor is subject to two locality conditions, one of which involves locality in configuration space 
and amounts to staying away from solutions with Killing vectors.
Thus it amounts to an excision.

\mbox{ }  

\ni On the other hand, conformal mathematics theorems associated with the below formulation are global in character \cite{IVP}.

\subsection{CS($\bupSigma$)}\label{CS-Geom}

Conf($\bupSigma$) and Diff($\bupSigma$) combine according to   Conf($\bupSigma$) $\rtimes$ Diff($\bupSigma$) \cite{FM77}.   
Note that Conf($\bupSigma$) $\bigcap$ Diff($\bupSigma$) $\neq \emptyset$ due to the existence of conformal isometries. 
However, quotienting something out twice is clearly the same as quotienting it out once, so this does not unduly affect the implementation.
Also note that Conf($\bupSigma$) is contractible, 
            so 
			
\ni            Conf($\bupSigma$) $\rtimes$ Diff($\bupSigma$) has the same topology as Diff($\bupSigma$), 
               Conf($\bupSigma$) $\rtimes$ Diff$_{\sF}$($\bupSigma$)                 as Diff$_{\sF}$($\bupSigma$), 
                 CS($\bupSigma$) as Superspace($\bupSigma$)
and      CS$_{\sF}$($\bupSigma$) as Superspace$_{\sF}$($\bupSigma$) (see e.g. \cite{Giu09}).

Fischer and Marsden \cite{FM77} extended Ebin's work by considering the action of the ${\cal C}^{\infty}$ version of Conf($\bupSigma$) on Riem($\bupSigma$), as motivated by York's work.  
They obtained a Conf($\bupSigma$) $\rtimes$ Diff($\bupSigma$) analogue of the Ebin--Palais Slice Theorem.  
They also demonstrated that CS($\bupSigma$) is an infinite-$d$ weak symplectic manifold near those points ($\bh$, $\bp$) with no simultaneous conformal Killing vectors. 
That implies sensible topology other than as regards being stratified.  
\cite{York73} includes a linearized version of the stratification.  
That stratification occurs carries over from Superspace($\bupSigma$) to CS($\bupSigma$), along with many results that follow from contractibility.  
In fact, Fischer and Marsden \cite{FM77} already had a CS($\bupSigma$) analogue of the stratification theorem.   
Fischer and Moncrief's superspace results \cite{FM96} carry over to CS($\bupSigma$) as well.  
Thus for the deg($\bupSigma$) = 0 case, one gets each of orbifolds, manifolds and contractible manifolds twice over.

\mbox{ } 

\ni CS($\bupSigma$) must be positive-definite since it is contained within CRiem($\bupSigma$).
CRiem($\bupSigma$) is better-behaved than Riem($\bupSigma$) along lines already established by DeWitt \cite{DeWitt67}.
One might hope that CS($\bupSigma$) is better-behaved than Superspace($\bupSigma$), in parallel to relational space containing a better-behaved shape space.

\mbox{ } 

\ni Finally note that Dil can be quotiented out of Superspace($\bupSigma$), giving a VPSuperspace($\bupSigma$) configuration space (volume-preserving Superspace).

\subsection{Notions of distance for geometrodynamics}\label{Gdyn-Dist}

Referring back to Sec \ref{Dist}, $||\mbox{ }||_{\sbM}$ is not a notion of distance for $\bM$ indefinite, e.g. for GR or its minisuperspace. 
The same restriction occurs again for path metrics.
Due to this, the Kendall, Barbour and DeWitt comparers do not carry over to GR as notions of distance. 
[Whereas the DeWitt comparer originates from geometrodynamics, it did not arise there as a distance, 
but rather as a {\sl metric functional} from which an indefinite geometry follows by double differentiation.]
Four ways out of this situation are as follows.  

\mbox{ }

\ni 1) Consider CRiem($\bupSigma$) and CS($\bupSigma$), which are positive-definite so that the Barbour and DeWitt comparers do carry over as notions of distance \cite{FileR}. 

\ni 2) Use an inf implementation instead (c.f. \ref{Fischer-Dist} and the double-inf Gromov--Hausdorff notion of distance \cite{Gromov}.

\ni 3) Use inhomogeneity quantifiers; density contrast is a simple such; see e.g. \cite{Zala12} for more complicated ones.

\ni 4) Use spectral notions of distance.  
The basic idea here is to consider the spectrum of some natural differential operator on the manifold.
Problems with this include non-uniqueness of such natural operators and the `isospectral problem' that `drums of different shapes' can none the less sound exactly the same, 
by which another axiom of distance fails.

\section{Reduced perturbative Midisuperspace}\label{SIC-Red-Geom}

This can be taken to arise from a particular example of the sandwich, for which the sandwich manoeuvre by itself fails to factor in the Diff($\bupSigma$) content.  
In these models, Diff($\mathbb{S}^3$) start to have effect at first order.

\subsection{Vacuum case}

This turns out to be more straightforward to handle \cite{AHH2, AHH3}.  
Now solving the thin sandwich equations gives 
(for $v_{\sn}$ having components $s_{\sn} := a_{\sn} + b_{\sn}$: the {\it scalar mode sum} ubiquitous quantity, $\md^{\so}_{\sn}$ and $\md^{\se}_{\sn}$)

\ni\beq
\mbox{$\frac{2}{\mbox{\scriptsize exp}(3\Omega)}$} \d s\mbox{}^2 = \{-1 + A_{\sn}\}\d\Omega^2 + \mbox{$\frac{2}{3}$}\d\Omega\d A_{\sn} + ||\d v_{\sn}||^2 \mbox{ } .  
\eeq
This is of dimension 4 + 1: $a_{\sn}$ drops out of the line element, so it is only short by 1 in removing the Diff($\mathbb{S}^3$) degrees of freedom.
Moreover, geometrically this is just flat $\mathbb{M}^5$.
Indeed,

\ni $$
T_{\sn} := \mbox{$\frac{2}{3}$}\sqrt{A_{\sn} - 1} \, \mbox{cosh}\big(\Omega + \mbox{$\frac{1}{3}$}\mbox{ln}(A_{\sn} - 1)\big) \mbox{ } ,
$$

\ni $$ 
X_{\sn} := \mbox{$\frac{2}{3}$}\sqrt{A_{\sn} - 1} \, \mbox{sinh}\big(\Omega + \mbox{$\frac{1}{3}$}\mbox{ln}(A_{\sn} - 1)\big)
$$
cast the line element takes the familiar form 

\ni \beq
\mbox{$\frac{2}{\mbox{\scriptsize exp}(3\Omega)}$} \d s\mbox{}^2 = - \d T_{\sn}^2 + \d X_{\sn}^2 + ||\d v_{\sn}||^2 \mbox{ } .
\eeq 

\mbox{ } \mbox{ } One can proceed from here by the V part of $\scH$ separating out to give an equation

\ni\beq
5\dot{\Phi}_{\sn}^{2}   -   16 \dot{\Phi}_{\sn} \dot{\Omega}   +   8\dot{\Omega}^{2}   +   \mbox{exp}(-2\Omega) = 0  \mbox{ }  
\eeq
(for dot denoting label-time derivative and $\Phi_{\sn} := \mbox{exp}(3\Omega)\{1 + A_{\sn}\}/3$) 
to be solved for the thus only temporarily convenient mixed-SVT variable $A_{\sn}$.  
This leads to a fully Diff($\mathbb{S}^3$)-reduced line element of the form 

\ni \beq
\mbox{$\frac{2}{\mbox{\scriptsize exp}(3\Omega)}$} \d s\mbox{}^2 = \{-1 + f_{\sn}(\Omega)\}\d \Omega^2 + ||\d v_{\sn}||^2 \mbox{ } ,
\eeq
for $f_{\sn}(\Omega) := A_{\sn}(\Omega) + \frac{2}{3}\frac{\d A_{\sn}(\Omega)}{\d \Omega}$.   
This is conformally flat.
Finally, define a new scale variable $\zeta_{\sn} := \int\sqrt{f_{\sn}(\Omega) - 1}\,\d\Omega$ to absorb the first term's prefactor.    
This leaves, up to a conformal factor,

\ni \beq
\d s^2 = - \d\zeta_{\sn}^2 + ||\d v_{\sn}||^2 \mbox{ } , 
\eeq
which is a spatially infinite slab of $\mathbb{M}^4$.

$\pa/\pa v^{\sfV}_{\sn}$'s components $\pa/\pa s_{\sn}, \pa/\pa d^{\so}_{\sn}, \pa/\pa d^{\se}_{\sn}$ are then among the 10 conformal Killing vectors; 
the others are $\pa/\pa\zeta$, 3 $v_{\sn}^{\sfV}\pa/\pa v_{\sn}^{\sfV^{\prime}} - v_{\sn}^{\sfV^{\prime}}\pa/\pa v_{\sn}^{\sfV}$ 
                           and 3 $v_{\sn}^{\sfV}\pa/\pa\zeta + \zeta\pa/\pa v_{\sn}^{\sfV}$.

Finally the corresponding shape space is also clearly flat, in this case $\mathbb{R}^3$: $\d s^2 = ||\d v_{\sn}||^2$.

\subsection{Minimally coupled scalar field case}

In the minimally-coupled scalar field case \cite{SIC1}, the outcome of the thin sandwich elimination is the undisturbed $\d s_0^2$ of the densitized version of (\ref{M-MINI}) alongside 
\ni $$
\d s_{\sb\sm}^{\sn}\mbox{}^2 = \mbox{$\frac{\mbox{\scriptsize exp}(3\Omega)}{2}$}
\left\{  
||\d v_{\sn}||^2 + \d f_{\sn}^2 +  
\left\{ 
\left\{
3 \d a_{\sn}  + \sqrt{3\{\mn^2 - 4\}} \d s_{\sn}
\right\}                                     
f_{\sn} +  6    a_{\sn} \d f_{\sn}
\right\}  \,
\d\phi 
\right.
$$

\ni\beq
\left.
+ \mbox{$\frac{2}{3}$}\d A_{\sn} \d\Omega - A_{\sn}\{ - \d\Omega^2 + \d\phi^2 \} 
\right\} \mbox{ } .  
\label{SIC-dsbm}
\eeq
But since this is of dimension 6 + 2, it is not yet Superspace.
In removing the Diff($\mathbb{S}^3$) degrees of freedom, the thin-sandwich manoeuvre has fallen short by 2. 
We do not know for now how to progress from here with the reduction.

However, the geometry of the currently attained `halfway house' has been further explored.
Its configuration space block structure can be tidied up by removing as many off-diagonal terms as possible can be done separately in each of the first two blocks. 
Diagonalize the one by using (\ref{Straightener}) again, whilst applying 

\ni \beq
\phi_{\sn} = \phi - \mbox{$\frac{3}{2}$} b_{\sn} f_{\sn} \mbox{ } 
\label{Exchanger}
\eeq
to the other to set the coefficient of $\d a_{\sn} \d \phi_{\sn}$ to zero.

This example serves to illustrate that the         configuration space metric split (\ref{grav-oplus-mcm}) and consequently                                   
                                                   the Hamiltonian constraint metric--matter split \cite{AHH2} are {\sl not in general preserved} by reduction procedures.  
Thus one has to face the complication that even minimally-coupled matter influences the form of the gravitational sector's reduced configuration space geometry.

The Ricci scalar for the partly reduced geometry is 

\ni\beq
R = 7 \, {\mbox{exp(-3$\Omega$)}/f_{\sn}^2} \mbox{ } , 
\eeq
by which the matter perturbation going to zero -- a physically inocuous situation --  gives a curvature singularity.
I previously commented that \cite{SIC1} this is not an unexpected phenomenon, 
paralleling for instance the situation with the collinear configurations in the $N$-body reduced configuration space. 
However, I now furthermore comment that the latter is based on stratification, whereas the latter is not known to be.

Also ${\pa}/{\pa \phi_{\sn}}$, ${\pa}/{\pa d_{\sn}^{\se}}$ and ${\pa}/{\pa d_{\sn}^{\so}}$ are Killing vectors; 
whereas ${\pa}/{\pa f_{\sn}}$ has lost this status upon reduction, the tensor mode directions have gained this property.  
As ever, ${\pa}/{\pa \Omega}$ is a conformal Killing vector.  
However, none of the above respect this model's potential (see \cite{SIC1} for its form).  

\mbox{ }

\ni Finally, scaled perturbative minisuperspace does not have a bona fide configuration space metric based notion of distance.
On the other hand, the space of pure inhomogeneities is positive-definite.
The vacuum case admits a 3-$d$ Euclidean metric with the $\underline{v}_{\sn}$ as the corresponding coordinates.

\section{Loop and knot spaces for GR}

\subsection{Loops and loop spaces}\label{Loop}

The loops of Sec \ref{Loop-G} are now taken to involve i) paths embedded in GR's topological notion of space: $\bupSigma$, and ii) the specific gauge group $SU(2)$.

\mbox{ } 

\ni The heuristic outline of the loop spaces being loop groups carries over to this case. 
For more rigorous treatments, see e.g. \cite{AL93, BF07, F10}.  
The theory of {\it cylindrical measures} is usually evoked.
The form taken by the stratification of the gauge orbit space in the GR case is covered in \cite{F00}.

\mbox{ } 

\ni See e.g. \cite{BF07} for the LQC equivalent of diagonal anisotropy.

\subsection{Knots and knotspace}\label{Knot}

A {\it knot} $K$ is an embedded closed curve in a closed orientable 3-manifold $\bupSigma$ (most usually $\mathbb{S}^3$).
We restrict attention to smoothly or piecewise linearly embedded curves to avoid `wild knots' \cite{Armstrong}.  
Two knots $K_1$, $K_2$ in a given $\bupSigma$ are equivalent if there exists an orientation-preserving automorphism such that $\mbox{Im}(K_1) = K_2$

{            \begin{figure}[ht]
\centering
\includegraphics[width=0.55\textwidth]{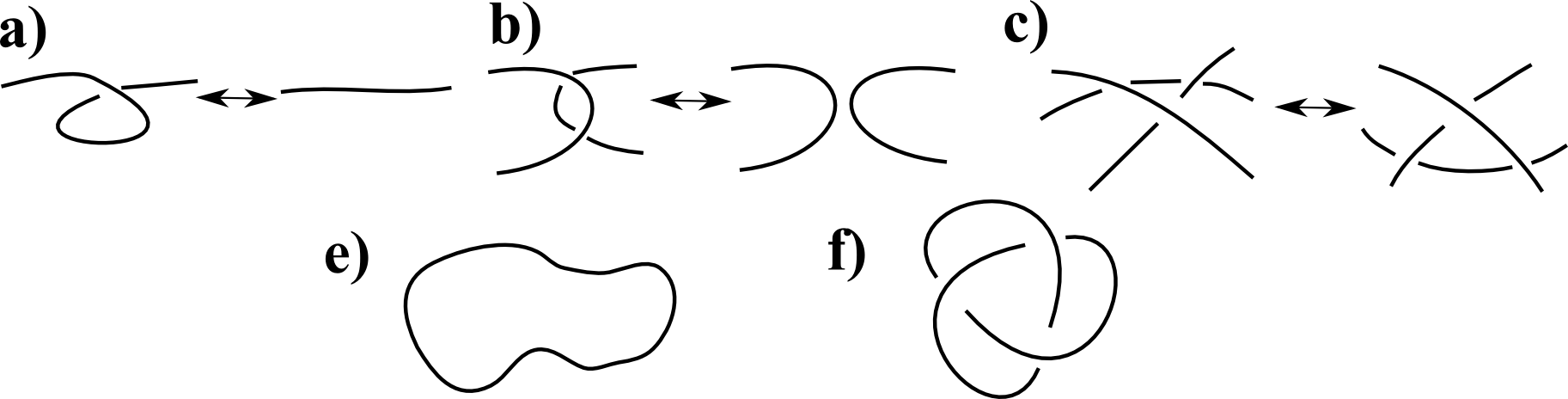}
\caption[Text der im Bilderverzeichnis auftaucht]{        \footnotesize{Representation of knots as planar graphs with an over-and-under crossing designation.  
The {\it Reidemeister moves} that preserve knots are a) twist/untwist, b) pull back/push under, and c) slide string up/down underneath a crossing.   
d) The unknot -- alias trivial knot -- has no crossings, or can be continuously deformed -- by the so-called `ambient isotopy' notion -- into having none.
e) The trefoil knot is the simplest nontrivial knot.  } }
\label{Knot-Fig} \end{figure}          }

\ni Whereas knot equivalence can be investigated using the Reidemeister moves (Fig \ref{Knot-Fig}), 
we do not know of an upper bound on how many such moves are needed to bring knots into obvious equivalence, so these moves are of limited practical use.
Rather, we seek characterization in terms of {\it knot invariants} (a subset of topological properties).
The obvious routes to such are homotopy and homology; 
these give respectively the {\it knot group} (fundamental group of the knot complement) \cite{Armstrong} and the {\it Alexander polynomial} \cite{Livingston} respectively. 
However, neither of these serve to discern between even some of the simplest knots.
The advent of the Jones polynomial \cite{KauffmanBook, Livingston, Graphs, GPBook} revived the subject; a number of further knot polynomials were subsequently discovered at short order.  
However, these still do not suffice to classify knots.  
See e.g. \cite{Witten89, RS88, GPBook, KauffmanBook}  for some applications of knots in Physics.
Also note the rather obvious topological manifold level background dependence in this formulation of knots.

\mbox{ }

\ni The mathematical form of the corresponding `Knotspace' remains an open problem.  
One approach is to view knot space as Emb($\mathbb{S}^1, \, \mathbb{S}^3$): embeddings of the circle in the 3-sphere. 
Vassiliev's work \cite{Vass, CDM} along these lines has close parallels to Arnol'd's study of the topological properties of $N$ points in the plane,  
in the sense of each $\FrQ$ being a subspace of a more tractable mapping space \cite{Budney}.
This turns the topological problem into one concerning singular maps, which is then {\sl aided} by these forming a {\sl stratified space}.  
See also \cite{Brylinski} for a mathematical account of spaces of knots.

\subsection{Another naming: not `Loop Quantum Gravity' but `Nododynamics'!}

There is the following analogy between preshapes and loops.
RPM                    preshapes are arrived at by quotienting out the dilations Dil  but not the more physically significant and mathematically harder rotations Rot($d$).  
Loop quantum gravity's loops     are arrived at by quotienting out $SU(2)(\bupSigma)$ but not the more physically significant and mathematically harder Diff($\bupSigma$).
Thus, whilst neither are the most redundant configurations of use in their theory, both are still partly redundant. 
Nor are they even `halfway houses' in each's passage to non-redundant `physical' kinematical variables, since both are prior to the {\sl main part} of that passage, 
both physically and in terms of the remaining parts of each's passage being far more mathematically complex than the parts already undertaken.
This has long been reflected in the former theories long having been named not after preshapes but after the shapes themselves: 
`shape geometry', `Shape Statistics', `dynamics of pure shape', one sense of `shape dynamics',  \cite{Kendall84, Kendall, B03, FileR}. 
This suggests that it would be clearer to name the latter theory not after loops but after knots. 
A suggestion then is {\it Nododynamics}, from the Latin {\it nodus} for `knot'.
This additionally makes sense at both the classical and quantum levels, just like `Geometrodynamics' does.
I end by noting that Geometrodynamics itself is indeed another naming based on identifying the non-redundant `physical' kinematical variables.

\section{Extensions} 

\subsection{Some further applications of this Article's examples}\label{Fur-App} 

Application 1) The idea of dynamics as a (para)geodesic principle on configuration space runs into global issues -- one of the many Global Problems of Time \cite{ABook} -- 
upon realizing that in general configuration space is not a manifold but rather a stratified manifold.
One possible way out of these is stratified manifold geometry's own notion of geodesic \cite{Pflaum} feeding into a notion of geodesic principle thereupon.
This approach is based on the use of Sheaf Methods \cite{Pflaum}, 
thus exemplifying that there are benefits from generalizing from Fibre Bundle Methods to Sheaf Methods as regards the Global Problems of Time.

\ni Application 2) If $\FrQ$ is stratified, then so are $\FrT(\FrQ)$ \cite{Pflaum2, Pflaum} and the symplectic version of $\FrT^*(\FrQ)$ (see e.g. \cite{Pflaum2, IY05} 
in the case of Mechanics). 
[Indeed, Whitney himself had already considered symplectic stratified spaces...]
Thus passage from a configuration space based approach to a configuration-and-change of configuration space perspective, or to the more habitual phase space perspective, 
does not affect arguments concerning stratified manifolds arising.

\ni Application 3) Each version of triangleland presented in this Article has a different kinematical quantization, involves different representations in its quantum theory 
and has mathematically distinct wavefunctions, due to QM's global sensitivities \cite{I84}. 
Thus Fig \ref{Tutti-Tri-2c}'s trichotomy has quantum consequences.  
Some classical consequences can also be expected from differences between the boundary value problems for each case. 
This provides yet further reasons to study this range of models 
(\cite{FileR} already argued for them to be model arenas for affine geometrodynamics and for foundational issues with Ashtekar variables approaches).
 
\ni Application 4) One can expect further interplay between sheaves and Problem of Time aspects \cite{ABook}, 
due to sheaves being well-suited to handle constraint algebraic structures, observables algebraic structures, records and histories \cite{K92I93APoT123}.
E.g. in being tools for tracking locally defined entities by attachment to open sets within a topological space, sheaves are well-placed for handling Records Theory.  

\ni Application 5) As regards Probability Theory and Statistics based upon configuration spaces in the role of sample or probability spaces, 
some cases that happen to be of relevance to RPMs \cite{Records} were worked out at the level of shape spaces by Kendall \cite{Kendall84, Kendall89, Kendall}. 
Further shape spaces' cases -- corresponding to a wider range of geometries -- have now also bee outlined in \cite{AMech}.  
On the other hand, the current paper makes it clear that the known area of Probability Theory and Statistics on $\mathbb{R}^n$ and $\mathbb{M}^n$ \cite{MinkProb} 
in some senses suffice to cover also Probability Theory and Statistics on some examples of Minisuperspace, Anisotropyspace, 
modewise perturbative Midisuperspace and its pure-shape counterpart.
A remaining caveat preventing just uplifting some techniques is that physically these configuration spaces come paired with specific potential functions, 
whereas traditional roles for Probability and Statistics on $\mathbb{R}^n$ and $\mathbb{M}^n$ solely involve the metric geometry.
Finally, modewise perturbative Midisuperspace is a {\sl slab of} $\mathbb{M}^4$. 

\ni Application 6) The simple geometries laid out in 5) also facilitate further quantum models, 
though the slab condition together with QM's global sensitivities would be expected to produce a kinematical quantization other than the standard (whole) Minkowski spacetime one.

\subsection{Further range of examples of configuration spaces of interest} 

\ni Example 1) Spaces of beins from approaching GR in first-order form (needed for subsequent incorporation of fermions).  

\ni Example 2) The status of configuration spaces in theories including fermions (whether flat-space, curved-space or coupled to GR) are of further interest, 
due to fermions' blurring of the configuration--momentum distinction.  

\ni Example 3) One could then furthermore consider the status and geometry of configuration spaces in Supergravity.

\ni Example 4) One could also consider configuration spaces for string and brane configurations.

\ni Example 5) The existing discrete Quantum Gravity programs' configuration spaces remain under-studied.  

\ni Example 6) There is a much larger assortment of generalized configuration spaces than 5)'s, as per Sec \ref{Q-Motiv}'s point IX). 

\mbox{ }

\ni{\bf Acknowledegments}.  I thank those close to me for support. 
I thank Chris Isham, Jeremy Butterfield, Sean Gryb, Gabriel Herczeg, Tim Koslowski, Matthias Kreck, Flavio Mercati, Ozgur Acik and Przemyslaw Malkiewicz for discussions and references, 
and the fqXi for a Travel Grant for 2014.

\vspace{10in}

\begin{appendices}

\section{Manifolds}\label{Geometry}

\subsection{Topological manifolds}\label{Top-Manifold}

A topological space $\langle \FrX, \tau\rangle$ is {\it locally Euclidean} if every point $x \in \FrX$ has a neighbourhood $\FrN_x$ homeomorphic to $\mathbb{R}^p$: Euclidean space.
If $\langle \FrX, \tau\rangle$ obeys 

\mbox{ }

\ni Topological Manifold 1) local Euclideanness, 

\ni Topological Manifold 2) Hausdorffness, and 

\ni Topological Manifold 3) second-countability, 

\mbox{ } 

\ni it is a {\it real topological manifold}, which I denote by $\bFrM$; see in particular \cite{Lee1} for more about these. 
I term the above trio of topological space properties `manifoldness'; moreover, this trio implies paracompactness as well \cite{Lee1}.   
Second countability ensures sequences suffice to probe most topological properties, 
whereas Hausdorffness ensures that neighbourhoods retain many of the intuitive properties of their metric space counterparts.
In these ways, much of Analysis can be carried over to manifolds (e.g. notions of continuity at the topological manifold level of structure). 

\mbox{ } 

\ni Next introduce the notion of {\it chart} \index{chart} alias {\it local coordinate system}  
in $\bFrM$ is a 1 to 1 map $\phi: \FrU \rightarrow \phi(\FrU) \subset \mathbb{R}^n$ for $\FrU$ an open subset of $\bFrM$.  
Each chart does not in general cover the whole manifold. 
The main idea is then to consider a suitable collection of charts.
These then serve as the homeomorphisms that guarantee the locally Euclidean property.
One can loosely think of these as `deformations of a rubber sheet', with continuous stretching but no guarantee of smoothness.
Appendix \ref{Diff-Manifold} then concerns adding in a further level of structure to model the smoothness.  
What we want to do is compare those charts that overlap, leading to the 2-chart Fig \ref{Top-Man}.b), 
with $\phi_1: \FrU_1 \rightarrow \mathbb{R}^n$, 
     $\phi_2: \FrU_2 \rightarrow \mathbb{R}^n$ which do indeed overlap: $\FrU_1 \bigcup \FrU_2 \neq 0$. 
There is then a composite map $\phi_2 \circ \phi_1^{-1}$ sends $\FrU_1 \bigcup \FrU_2$ to itself.
This is a locally defined map of $\mathbb{R}^n \rightarrow \mathbb{R}^n$; it is a local coordinate transformation, and is termed a {\it transition function} $t_{12}$.  
An {\it atlas} \index{atlas} for a topological manifold is then a collection of charts that between them cover the whole manifold.
%
{\begin{figure}[ht]
\centering
\includegraphics[width=0.6\textwidth]{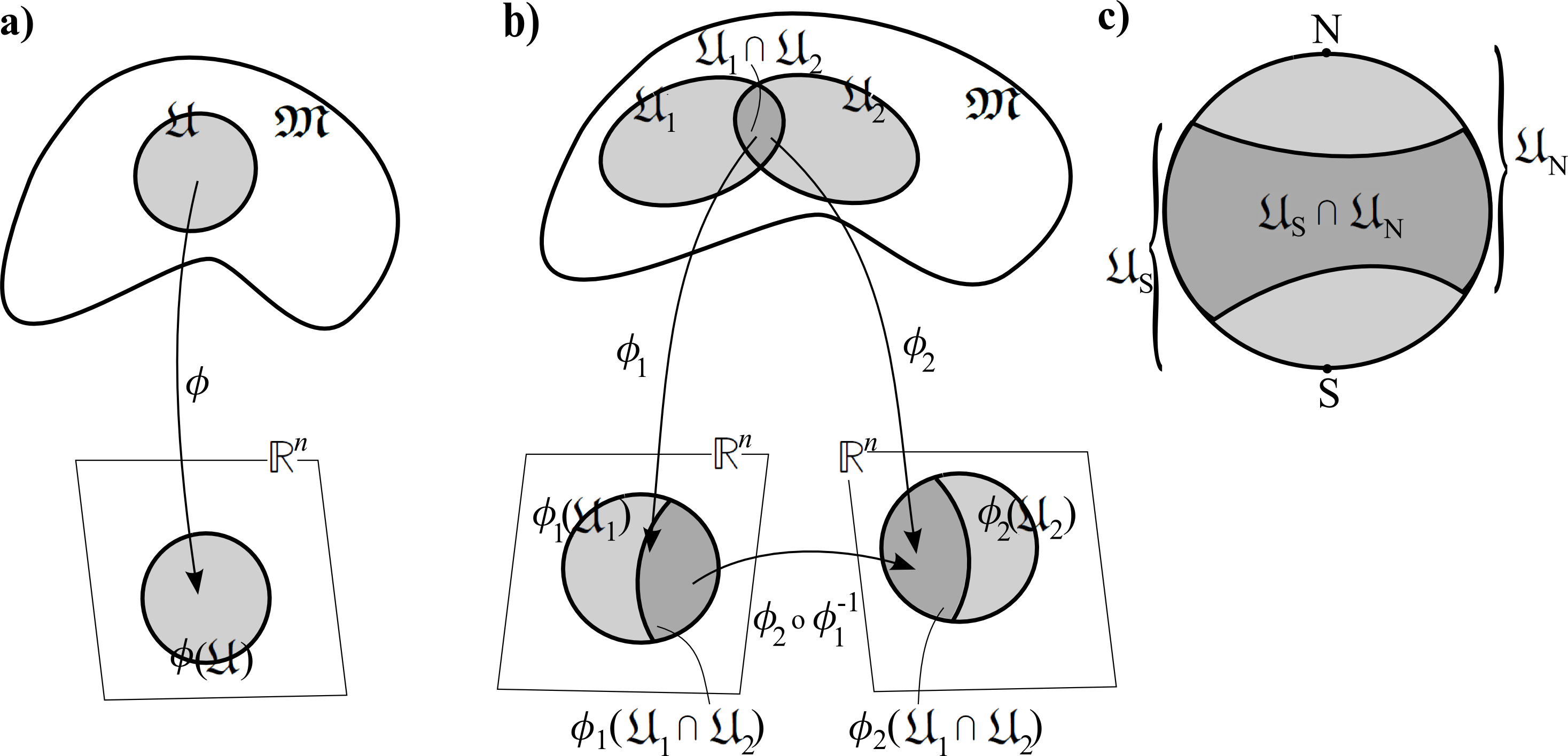}
\caption[Text der im Bilderverzeichnis auftaucht]{\footnotesize{a) A chart. 
b) Overlapping charts and transition function.  
c) A 2-chart cover for $\mathbb{S}^2$: from N to the lower curve and from S to the upper curve.  
}} 
\label{Top-Man}\end{figure}} 

\subsection{Differentiable manifolds}\label{Diff-Manifold}

On some topological manifolds, charts furthermore permit tapping into the standard $\mathbb{R}^p \longrightarrow \mathbb{R}^q$ Calculus.
In this way, differentiable structure can be established upon these manifolds.   
The main idea here that these manifolds possess a notion of global differentiable structure, rather than just a local differentiable structure in each coordinate patch $\FrU_i$.
This is held together by the `{\it meshing condition}' \index{meshing condition} on the coordinate patch overlaps (Fig Top-Man.b).
The transition function $t_{12} = \phi_2 \circ \phi_1^{-1}$ can now be interpreted in terms of a Jacobian matrix of derivatives of one local coordinate system with respect to another: 
${\mL^{\sfA}}_{\sfB} = \pa (x^{\sfA})/\pa(\bar{x}^{\sfB})$.  

\mbox{ }

\ni The topological-level notion of atlas can furthermore be equipped with differentiable structure.
This permits Calculus to be performed throughout the manifold, as is required for studying {\it differential equations} that represent physical law. 
Moreover, our main interest here is really in equivalence classes of atlases.
Differentiable structure is then studied using a convenient small atlas [such as in Fig \ref{Top-Man}.d)'s 2-chart approach to the 2-sphere].
In contrast to the previous Sec's atlas being ${\cal C}^0$ (the continuous functions) the present Sec's is usually taken to be ${\cal C}^{\infty}$ (the smooth functions).  
In fact, weakening ${\cal C}^{\infty}$ to ${\cal C}^k$ $k \geq 1$ makes little difference, 
since each such differentiable structure is uniquely smoothable \cite{Whitney36}

\section{General and fibre bundles}\label{Bundles}
%
{            \begin{figure}[ht]
\centering
\includegraphics[width=0.8\textwidth]{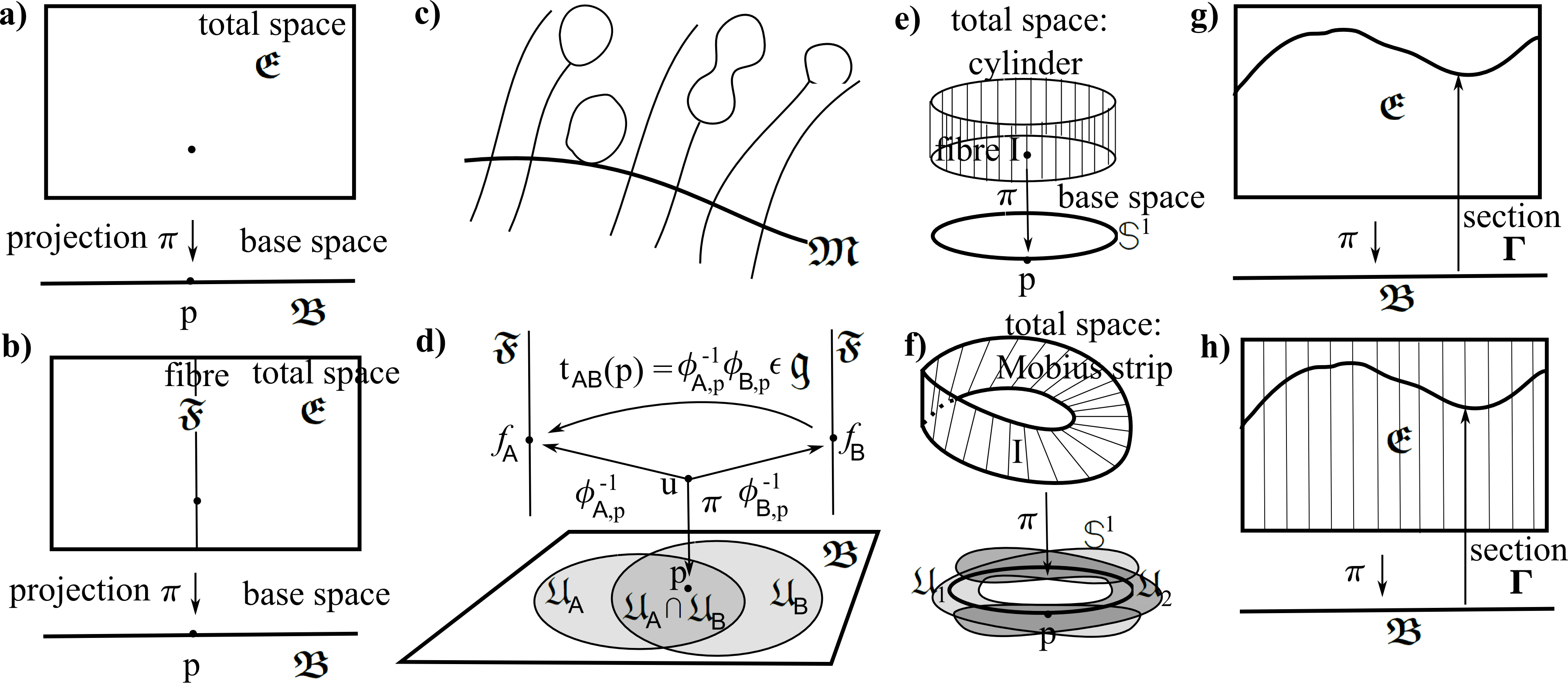}
\caption[Text der im Bilderverzeichnis auftaucht]{        \footnotesize{a) General bundle at the topological level.
b) For a fibre bundle, the total space $\FrE$ is made up of identical disjoint copies of a fibre manifold $\FrF$ at each point in the base space $\FrB$.
c) The general bundle structure, on the other hand, can be thought of as being more general due to permitting attachment of distinct manifolds to different parts of the base space. 
d) The fibre bundle additionally involves transition functions (c.f. the manifold) and a structure group $\lFrg$. 
e) and f) are simple examples of trivial and nontrivial bundle respectively. 
g) and h) illustrates the further key notion of cross sections for general and fibre bundles. } }
\label{Bundle} \end{figure}          }

\ni Consider first topological spaces which project down continuously onto lower-$d$ topological spaces, $\pi: \bFrE \longrightarrow \bFrB$. 
Such can be viewed in reverse\foo{\ni Bundles were originally considered from a perspective of total space primality by Seifert; 
Whitney \cite{Whitney35}, however, switched attention to base space primality, meaning that $\FrB$ is an a priori known manifold $\bFrM$.} 
as higher-$d$ bundle {\it total spaces} $\bFrE$, each built over a lower-$d$ {\it base space} $\bFrB$; $\pi$ is a {\it projection map}.  
This is the {\it general bundle} notion; see Fig \ref{Bundle}.a), \cite{IshamBook} for an outline and \cite{Husemoller} for an advanced account.

Suppose that one further introduces a local product structure, in which the total space is made up of identical copies of a {\it fibre space} (alias just {\it fibre}) $\bFrF$, 
itself for now regarded as a topological space. 
Then one has a topological-level \ni {\it fibre bundle}; 
see Fig \ref{Bundle}.b) \cite{IshamBook, Nakahara, AMP, NSBook} for introductory treatment and \cite{Husemoller, Wells} for more advaced accounts.
Moreover, from a global perspective, fibre bundles are in typically `twisted versions' of product spaces, 
whereas, conversely, global product spaces are the trivial cases of fibre bundles. 
Figs \ref{Bundle}.c-d) are simple examples of these respectively.  
The inverse image $\pi^{-1}(\mp)$ is the fibre $\bFrF_{\sp}$ at $\mp$ (Fig \ref{Bundle}.a).  
That all fibres are the same is mathematically encoded by $\bFrF_{\sp}$ homeomorphic to $\bFrF$, with extra isomorphic equivalence if and when required.

In fact, fibre bundles are also taken to have a {\it structure group} $\lFrg$ acting upon the fibres $\bFrF$, 
%
%
by which they are denoted $\langle \bFrE, \pi, \bFrB, \bFrF, \lFrg \rangle$.

\mbox{ } 

\ni Example 1) In the important case of a {\it principal fibre bundle} $\bFrP(\bFrM, \lFrg)$ alias $\lFrg$-{\it bundle}, $\lFrg$ and $\bFrF$ coincide, 
so that $\lFrg$ now just acts on itself.
On the other hand, for an {\it associated fibre bundle}, $\lFrg$ acts on a distinct type of fibre $\bFrF$, giving a somewhat more general and complicated structure.

\mbox{ } 

\ni Taking an open cover $\{\FrU_{\sfA}\}$ of $\bFrB$, each $\FrU_{\sfA}$ is equipped with a homeomorphism $\phi_{\sfA}: \FrU_{\sfA} \times \bFrF \rightarrow \pi^{-1}(\FrU_{\sfA})$.  
This is such that $\pi \phi_{\sfA}$ sends $(\mp, \mf)$ -- for $\mf$ a point on $\bFrF_{\sp}$ -- down to $\mp$.
$\phi_{\sfA}$ is then termed a {\it local trivialization} since its inverse maps $\pi^{-1}(\FrU_{\sfA})$ onto $\FrU_{\sfA} \times \bFrF$ which is a trivial product structure.
{\sl Local} triviality refers to {\sl globally} nontrivial fibre bundles encoding information in excess of that in the also globally trivial product space.
Figs \ref{Bundle}.c-d) are, in more detail of a M\"{o}bius strip viewed as a nontrivial fibre bundle as compared to the cylinder viewed as a trivial bundle.
Both of these have circles for base spaces and line intervals for fibres.
In this case, the extra global information is the non-orientability.  
Moreover, this ongoing definition of fibre bundle can furthermore be shown to be independent of the choice of covering, so I do not enumerate this paragraph as part of the definition.

As a final piece of structure, consider $\FrU_{\sfA}$ and $\FrU_{\sfB}$ -- an arbitrarily chosen pair of open sets except that nontrivial overlap between them is guaranteed: 
$\FrU_{\sfA} \bigcup \FrU_{\sfB} \neq \emptyset$.
Somewhat simplify the notation according to $\phi_{\sfA}(\mp, \mf) = \phi_{\sfA, \sp}(\mf)$, $\phi_{\sfA, \sp}$ is the homeomorphism sending $\bFrF_{\sp}$ to $\bFrF$.  
Then the {\it transition functions} $t_{\sfA\sfB}(\mp) := \phi_{\sfA, \sp}^{-1}\phi_{\sfB, \sp}: \bFrF \rightarrow \bFrF$ 
corresponding to the overlap region as per Fig \ref{Bundle}.b) are elements of $\lFrg$.
$\phi_{\sfA}$ and $\phi_{\sfB}$ are moreover related by a continuous map $t_{\sfA\sfB}: \FrU_{\sfA} \bigcup \FrU_{\sfB} \rightarrow \lFrg$
according to $\phi_{\sfA}(\mp, f) = \phi_{\sfA}(\mp, t_{\sfA\sfB}(\mp^f)$: Fig \ref{Bundle}.b).
Note the parallels between this and the meshing condition for topological manifolds.

\mbox{ } 

\ni Topological {\it fibre bundle morphisms} are then continuous maps between fibre bundles $\langle \bFrE_1, \pi_1, \bFrB_1, \bFrF_1, \lFrg_1 \rangle$ and 

\ni                                                                                         $\langle \bFrE_2, \pi_2, \bFrB_2, \bFrF_2, \lFrg_2 \rangle$ 
that map each fibre $\bFrF_1$ onto a fibre $\bFrF_2$.  

\ni A {\it section} \index{section} alias {\it cross-section} of a topological fibre bundle is a continuous map in the opposite direction to $\pi$,
$\Gamma: \bFrB \rightarrow \bFrE$ such that $\pi(\Gamma(x)) = x \mbox{ } \mbox{  } \forall \,  x \in \bFrB$. 
This is to cut each fibre precisely once.
N.B. that not all fibre bundles possess a global such; whether they do is often insightfully expressible in cohomological terms \cite{Husemoller}
and gives rise to the theory of characteristic classes.

The above definitions of fibre bundle -- and of the corresponding morphisms and sections -- can furthermore be elevated to the case of differentiable manifolds, 
now with smooth maps in place of continuous maps and diffeomorphisms in place of homeomorphisms.

\mbox{ }

\ni Example 2) Tangent space, cotangent space and the general space of tensors can be thought of as {\it tangent}, {\it cotangent} and {\it tensor fibre bundles} respectively. 

\ni Example 3) Gauge theory can be formulated in terms of fibre bundles (using both principal and more general associated fibre bundles); 
see e.g. \cite{IshamBook, NSBook, Nakahara, Bleecker, AMP} for details.   
This requires  considering {\it connections} \index{connections} on fibre bundles.
One can then indeed interpret Gauge Theory's potential $\mA_{\mu}$ as a connection, alongside corresponding notions of parallel transport and of covariant derivative $\mD_{\mu}$.

\mbox{ } 

\ni The space $\bFrl_{\sp}(\bFrM)$ of {\it loops} \index{loop} at a point $\mp \in \bFrP(\bFrM, \lFrg)$: curves $\gamma: [0, 1] \rightarrow \bFrM$ starting and ending at $\mp$; 
these define transformations $\phi_{\gamma}: \pi^{-1}(\mp) \rightarrow \pi^{-1}(\mp)$ on $\bFrF = \lFrg$.  
Then for $\muu \in \bFrP(\bFrM, \lFrg)$ such that $\muu$ projects down to $\mp$ [= $\pi(\muu)$] 
the {\it holonomy group} at $\muu$ is $\mbox{Hol}_{\su} := \{g \in \lFrg \, | \, \phi_{\gamma}(\muu) = \muu \, g, \gamma \in \bFrl_{\sp}(\bFrM)\}$; it is a subgroup of $\lFrg$.
Finally indeed then the field strength $\mF_{\mu\nu}$ corresponding to $\mA_{\mu}$ plays the corresponding role of curvature.  

\mbox{ } 

\ni Finally returning to the notion of general bundles, 
these can be viewed as a generalization in which there need no longer be a notion of identical fibre at each point of the base space. 
This is useful since assuming such identical fibres throughout turns out to be a significantly restrictive assumption in some kinds of modelling required by Theoretical Physics 
(see e.g. Appendix \ref{Quotients}).

\section{Outline of supporting Functional Analysis}\label{Fun}

\subsection{From Hilbert to Banach and Fr\'{e}chet spaces}\label{Hilb-Ban-Fre}

\ni The purpose of these Appendices is to provide a basic outline of mathematics relevant to stratified configuration spaces, with some mentions of related notions 
which are more widely familiar in Theoretical Physics. 
Let us start with a ladder of increasingly general topological vector spaces which are infinite-$d$ function spaces.
    A {\it Hilbert space} Hilb is a complete inner product space, 
    a {\it Banach space}       is a complete normed space,
and a {\it Fr\'{e}chet space}  is a complete metrizable locally convex topological vector space \cite{Hamilton82}.

\mbox{ } 
	
\ni Whilst Hilbert Spaces are the most familiar in Theoretical Physics due to their use in QM, Functional Analysis has also been extensively developed for Banach spaces \cite{AMP}.  
Major results here are the Hahn--Banach Theorem, the Uniform Boundedness Principle and the Open Mapping Theorem; see \cite{DS88} for details and proofs.    
The second and third of these follow from from Baire's Category Theorem; also the Inverse Function Theorem \cite{Lee2} extends to Banach spaces, following from the Open Mapping Theorem.

\mbox{ } 

\ni However, treatment of GR configuration spaces involves the even more general  Fr\'{e}chet spaces. 
Let us first explain their definition. 
A topological vector space is {\it metrizable} if its topology can be induced by a metric -- in the Analysis metric space sense -- 
and that is furthermore translation-invariant: $d(x, y) = d(x + w, y + w)$.    
This qualification is required since for topological vector spaces, one uses a collection of neighbourhoods of the origin (vector space $0$).
Then from this, translation (by the vector space +) establishes the collection of neighbourhoods at each other point.
Next, a {\it base} in a topological vector space $\FrV$ is a linearly-independent subset $\FrA$ such that $\FrV$ is the closure of the linear subspace with Hamel basis $\FrA$.
A {\it Hamel basis} itself is a maximal linearly-independent subset of $\FrV$].
Finally, a topological vector space $\FrV$ is {\it locally convex} if it admits a base that consists of convex sets.

Next, many important results in Functional Analysis -- in particular the Hahn--Banach Theorem, the Uniform Boundedness Principle and the Open Mapping Theorem -- 
further carry over from Banach spaces to Fr\'{e}chet spaces \cite{Hamilton82}.
On the other hand, be warned that there is no longer in general a Inverse Function Theorem here, 
though the Nash--Moser Theorem \cite{Hamilton82} is a replacement of this for a subclass of Fr\'{e}chet spaces.  
One further consequence of this is that the usual local existence theorem for ODEs does not hold either.
See e.g. \cite{Hamilton82} as regards Calculus on Fr\'{e}chet spaces.

\subsection{Hilbert, Banach and Fr\'{e}chet Manifolds}\label{Inf-Manifolds}

Topological manifolds' local Euclideanness and ensuing $\mathbb{R}^p$-portion charts 
extend well to infinite-$d$ cases, for which the charts involve portions of Hilbert, Banach and Fr\'{e}chet spaces.  
See e.g. \cite{Lang95} for Hilbert manifolds, \cite{AMP} for Banach manifolds and \cite{Hamilton82} for Fr\'{e}chet manifolds.
Banach manifolds are the limiting case as regards having a very broad range of analogies with finite manifolds.
Fre\'{e}chet manifolds remain reasonably tractable \cite{AMP} despite losing in general the Inverse Function Theorem.
Fre\'{e}chet Lie groups can also be contemplated \cite{Hamilton82}.

\mbox{ } 

\ni Finite manifolds' incorporation of differentiable structure also has an analogue in each of the above cases.
So e.g. one can consider differentiable functions and tangent vectors for each, 
and then apply multilinearity to set up whichever rank (p, q) and symmetry type $S$ of tensor versions. 
In particular, applying this construction to a Fr\'{e}chet manifold with tangent space Fre(${\cal C}^{\infty}$) produces another Fr\'{e}chet space Fre$_{(p,q)}$(${\cal C}^{\infty}$).
These are used in Sec \ref{GR-Config} and \ref{GR-Config-2}.

\section{Quotient spaces and stratified manifolds}\label{Quotients}

\subsection{Quotienting out groups}

Quotienting has a number of subtleties.
For instance, it is well-known that attempting to quotient one group by another does not in general produce yet another.  
Indeed, quotienting is more generally an operation under which only some mathematical structures and properties are inherited; see the next SubAppendix for further examples.

\mbox{ } 

\ni For $\FrG$ a group with an element $g$ and a subgroup $\FrH$, $g\FrH := \{gh \, | \, h \in \FrH\}$ is a {\it (left) coset}, 
and the corresponding {\it (left) coset space} is the set of all of these for that particular $\FrG$ and $\FrH$.  
Suppose one has a {\it group action} $\alpha$ on a set $\FrX$: a map $\alpha: \FrG \times \FrX \rightarrow \FrX$ such that 

\mbox{ }

\ni i) $\{g_1 \circ g_2\} x = g_1 \circ \{g_2 x\}$ (compatibility) and 

\ni ii) $e x = x$                                  (identity)           $\forall \, x \in \FrX$.

\mbox{ } 

\ni Then {\it orbits} are defined as                       Orb($x$) := $\{ g x \, | \, g \in \FrG\}$: the set of images of $x$,
and  {\it stabilizers} alias {\it isotropy groups} by 

\ni Stab($x$) := $\{ g   \, | \, g x = x\}$:    the set of $g \in \FrG$ that fix $x$. 

\mbox{ } 

\ni Finally, the {\it quotient of the action of a group} $\FrG$ on a space $\FrS$, $\FrS/\FrG$ is the set of all orbits, 
which is usually termed {\it orbit space}.

\subsection{Quotient topologies}\label{Quot-Top}

Next consider quotienting a topological space by an equivalence relation, $\langle \FrX, \tau\rangle/\,\widetilde{\mbox{ }}$, 
so as to produce the corresponding {\it quotient topology} \cite{Lee1, Munkres00}.

\mbox{ }

N.B. that this does not in general preserve a number of topological properties, in particular none of the three manifoldness properties.
A simple counter-example to preservation of Hausdorffness is as follows.
Let $\FrX = \{(x, y) \in \mathbb{R}^2 \, | \, y = 0 \mbox{ or } 1\}$ with the obvious topology, and $(x, y) \sim (z, w)$ iff either $(x, y) = (z, w)$ or $x = z \neq 0$: 
the line with two origins which cannot be separated.
As regards non-preservation of dimension, quotienting is capable of decreasing or increasing topological dimension.
Whereas the decreasing case is obvious, space-filling curves \cite{Armstrong} provide examples of it increasing.
Quotienting can furthermore produce dimension that varies from point to point in its quotient, of which Sec \ref{Q-Geom} already presented simple examples.
Moreover, in the physical examples below, Hausdorffness and second-countability {\sl are} often retained, so quotienting here leads to entities which are `2/3 of a manifold'.

\mbox{ }

\ni On the other hand, quotienting does preserve connectedness, path connectedness and compactness (see \cite{Lee1} for these three),  
albeit not simple connectedness (e.g. passage to nontrivial universal covering group) or contractibility (e.g. $\mathbb{R}^2/Dil = \mathbb{S}^1$).
Moreover, if $\FrS/\FrG$ arises by a group $\FrG$ acting on a space $\FrS$ freely and properly, then $\FrS/\FrG$ is Hausdorff \cite{Lee2}.
One application of this result is in protecting 1- and 2-$d$ RPM shape spaces.

\subsection{Orbifolds}

{\it Orbifolds} are locally quotients $\FrMgen/\FrG$ following from a properly discontinuous action of a finite Lie group $\FrG$ on a manifold $\FrMgen$ 
This construction can moreover be applied to equipped manifolds such as (semi-)Riemannian manifolds.
Orbifolds are more general than manifolds, since quotients do not in general preserve manifoldness, by which some orbifolds carry singularities.
See e.g. \cite{Orbi} for more on orbifolds.  

\mbox{ } 

\ni $\FrMgen$ itself admits an open cover $\FrU_{\sfC}$.
Then each constituent $\FrU_{\sfC}$ possesses an {\it orbifold chart}: a continuous surjective map 

\ni $\phi_{\sfC}: \FrV_{\sfC} \rightarrow \FrU_{\sfC}$ for $\FrV_{\sfC}$ open $\underline{\subset} \, \mathbb{R}^n$ 
-- for n = dim($\FrMgen$) -- , with $\FrV_{\sfC}$ and $\phi_{\sfC}$ invariant under the action of $\FrG$.  
One can then define a notion of gluing between such charts and finally a notion of orbifold atlas, in close parallel to that for manifolds.  

\mbox{ } 

\ni The everyday notion of cone can be thought of as a simple example of orbifold,
Another is Fig \ref{Tutti-Tri-2c}.b), in the context of a 3-body problem configuration space.
More generally, orbifolds are common in $N$-body problem configuration spaces, indeed including the generalized sense of cone that applies to relational spaces.
The 2-$d$ $N$-body problem's simplest shape spaces $\mathbb{CP}^{n - 1}$ are best thought of as complex manifolds.
There is then indeed a notion of complex orbifold as well as of real orbifold, in parallel to how there are real and complex manifolds \cite{Nakahara}. 
Well-known elsewhere in Theoretical Physics, many of the orbifolds which occur in String Theory are also complex; 
in particular, these occur in the study of Calabi--Yau manifolds \cite{Orbi}.  
Furthermore, some simpler models of this last example are closely related to the penultimate example, though both being discrete quotients of $\mathbb{CP}^k$ spaces \cite{FileR}.

\subsection{Quotienting by Lie group action, and slices.}\label{Lie-Slice}

For the action of a Lie group $\FrG$ on a space $\FrX$, the generalized {\it slice} $S_{x}$ at $x \in \FrX$ is a manifold transverse to the orbit $Orb(x)$ (see e.g. \cite{IM82}).
This generalizes the fibre bundle notion of local section to the case involving compact transformation groups in place of principal bundles.
[The corresponding generalization of the fibre bundle notion of local trivialization is termed a {\it tube}.]

\mbox{ } 

\ni The slice can be taken to exist in the above compact case.  
However, in other cases one can sometimes prove Slice Theorems to this effect (Appendices \ref{GOS} and \ref{Superspace-Top}).
Of subsequent relevance below, the Implicit Function Theorem \cite{Lee2} -- a close relative of the Inverse Function Theorem -- enters these proofs. 
A slice $S_x$ gives a local chart for $\FrX/\FrG$: the space of orbits near $x$, so the slice notion, when available, is an important tool in the study of the corresponding orbit spaces.  

\mbox{ } 

\ni Slices carry information concerning the amount of isotropy of points near $x$ \cite{IM82}.
Let us illustrate `amount of isotropy' using Sec \ref{Q-Geom}'s examples. 
Whereas 2-$d$ mechanics configurations have just the one isotropy group $SO(2)$, 3-$d$ ones have 3 possible isotropy groups: id, $SO(2)$ and $SO(3)$.  
These have corresponding orbits of the form $SO(3)$, $\mathbb{S}^2$ and 0 respectively.
This correspondence follows from the isotropy group also being known as the stabilizer group, and there being well-known ties between orbits and stabilizers.
{\sl Multiple dimensions of isotropy groups Isot point to multiple dimensions of orbits.
Thus to orbit spaces are not manifolds -- entities of unique dimension -- but rather collections of manifolds that span various dimensions.}
This motivates consideration of further generalizations of manifolds as follows.

\subsection{Stratified manifolds} \label{Strat}

Manifolds do not cover enough cases of quotients $\FrMgen/\FrG$ for the purpose of studying physical reduced or relational configuration spaces $\FrQ/\FrG$.
We saw above that these more generally produce unions of manifolds of in general different dimensions.
Moreover, some of these of further physical relevance -- such as reduced configuration spaces in Mechanics and GR, spaces of orbits in Gauge Theory -- 
`fit together' according to some fairly benevolent rules. 
The constituent manifolds here are known as {\it strata}, and each collection `fitted together' in this manner is known as a {\it stratified manifold}. 
Simple examples include the following.

\mbox{ } 

\ni Example 1) \ni A {\it topological manifold with boundary} is locally homeomorphic to open sets in the half-space $\{(x_1, ..., x_p) \in \mathbb{R}^p \, | \, x_p \geq 0\}$.
Charts ending on the half-space's boundaries are describing part of the manifold that is adjacent to its boundary. 
Manifolds with boundary can also be equipped with smooth structure \cite{Lee2}.
This example can furthermore be interpreted as a simple type of stratified manifold.
Here the manifold and its boundary are the two constituent strata, the former possessing the full dimension and the latter codimension-1.
Fig \ref{Strat-Sheaf} illustrates a particular case of this.
%
{            \begin{figure}[ht]
\centering
\includegraphics[width=0.4\textwidth]{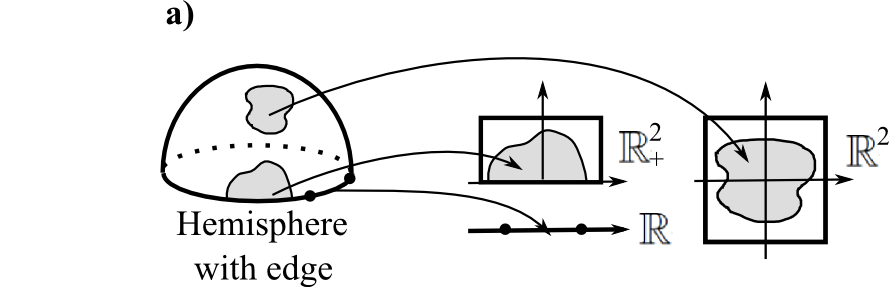}
\caption[Text der im Bilderverzeichnis auftaucht]{        \footnotesize{a) This configuration space -- a manifold with boundary -- has three types of chart.
On the other hand, D and C configurations have the same isotropy group, so the D's do not constitute distinct strata.
b) Conceptual depiction of a general bundle or (the beginnings of) a sheaf.} }
\label{Strat-Sheaf} \end{figure}          }

\ni Example 2) {\it Manifolds with corners}. 
These have, in addition to the previous example's strata, the codimension-2 strata that are the corners themselves.

\mbox{ } 

\ni Historically, the original formulation of stratified manifolds was of differentiable stratified manifolds by Whitney \cite{Whitney46} (and reviewed in \cite{Whitney65}). 
Subsequently, Thom formed a theory of stratified topological manifolds as an arena for dealing with singularities \cite{Thom55}. 
Thom \cite{Thom69} additionally showed that every stratified space in the sense of Whitney was also one of his own stratified spaces and with the same strata.

\mbox{ } 

\ni Here is a brief outline of some basic concepts in the theory of stratified manifolds.
Let $\FrV$ be a topological space that is not presupposed to be a topological manifold.
Suppose this can be split according to $\FrV = \FrV_{\sp} \bigcup \FrV_{\sq}$ \cite{Whitney65}. 
Here $\FrV_{\sp} := \{p \in \FrV, p \mbox{ simple }, \mbox{dim}_p(\FrV) = \mbox{dim}(\FrV)$ with `simple' meaning `regular' and `ordinary', and $\FrV_{\sq} := \FrV - \FrV_{\sp}$.  
In fact, one considers a recursion of such splittings, so e.g. $\FrV_{\sq}$ can subsequently be split into $\{\FrV_{\sq}\}_{\sp}$ and $\{\FrV_{\sq}\}_{\sq}$.
Then setting $\FrMgen_1 = \FrV_{\sp}$, $\FrMgen_2 = \{\FrV_{\sq}\}_{\sp}$, $\FrMgen_3 = \{\{\FrV_{\sq}\}_{\sq}\}_{\sp}$ etc gives 
             $\FrV = \FrMgen_1 \bigcup \FrMgen_2 \bigcup ... , \mbox{dim}(\FrV) = \mbox{dim}(\FrMgen_1) > \mbox{dim}(\FrMgen_2) > ...$, where each $\FrMgen_i$ is a manifold.
The point of this procedure is that it provides the partition of $\FrV$ by dimension. 
Then indeed $\FrV$ is only a topological manifold is this is a trivial partition: involving a single dimension only.
On the other hand, a {\it strict} partition of a topological space is a (locally finite) partition into strict manifolds. 
[In outline, a manifold $\FrMgen$ in the $m$-dimensional open set ${\cal O}$ is ${\cal O}${\it -strict} if the ${\cal O}$-{\it closure} 
$\overline{\FrMgen}$ := ${\cal O} - \mbox{clos}\,\FrMgen$ and the ${\cal O}$-frontier $\overline{\FrMgen} - \FrMgen$ are topological spaces in  ${\cal O}$.]

\mbox{ } 

\ni Next, a set of manifolds in ${\cal O}$ has the {\it frontier property} if, for any two of them, say $\FrMgen$, $\FrMgen^{\prime}$ with $\FrMgen \neq \FrMgen^{\prime}$

\ni\beq
\mbox{ if } \FrMgen^{\prime} \bigcap \overline{\FrMgen} \neq \emptyset \mbox{ } , \mbox{ } 
\mbox{ then } \FrMgen^{\prime} \subset \overline{\FrMgen} \mbox{ } \mbox{ and } \mbox{ } \mbox{ } \mbox{dim}(\FrMgen^{\prime}) < \mbox{dim}({\FrMgen}) \mbox{ } .
\label{Frontier}
\eeq
[A partition into manifolds itself has the frontier property if the corresponding set of manifolds does.]

\mbox{ } 

\ni One definition of a {\it stratification} of $\FrV$  
is finally then \cite{Whitney65} a strict partition of $\FrV$ which has the frontier property. 
The corresponding set of manifolds are then known as the {\it strata} of the partition.

\mbox{ } 

\ni The variant that in particular Fischer \cite{Fischer70} also makes use of the {\it inverse frontier property}, i.e. (\ref{Frontier}) with primed and unprimed switched over,  
which then feeds into the corresponding notion of inverted stratification.  
Another sometimes useful \cite{RSV02} property is the {\it regular stratification property}, which involves

\ni \beq
\FrX_i \cup \overline{\FrX_j} \neq \emptyset \mbox{ } \Rightarrow \mbox{ } \FrX_i \, \underline{\subset} \, \overline{\FrX_j} \mbox{ } \forall \, i, j \in I \mbox{ } .  
\eeq 
Whitney \cite{Whitney65} also established that a locally finite partition of $\FrV$ with the frontier property is a stratification.
Moreover, for each stratum $\FrMgen$, then $\overline{\FrMgen} - \FrMgen$ is the union of the other closed strata in $\overline{\FrMgen}$.
Indeed, any strict partition of $\FrV$ admits a refinement which is a stratification into connected strata.  
Take this to be a brief indication that the theory of refinements of partitions (a type of `graining') plays a role in the theory of stratified manifolds.

\mbox{ } 

\ni Due to nontrivial stratified manifolds having strata with a range of different dimensions, 
clearly the locally Euclidean property of manifolds has broken down, and with it the standard notions of charts and how to patch charts together.
These notions still exist for stratified manifolds, albeit in a {\sl more complicated form} (see Fig \ref{Strat-Sheaf}).
Also, in general losing Hausdorffness and second-countability leaves stratified manifolds `more to the left' than topological manifolds in the diagram of the levels of structure.  
Moreover, this Article considers in any detail only Hausdorff second-countable stratified manifolds, i.e. entities which are`2/3rds of a manifold'.  

\mbox{ } 

\ni Since Whitney, stratified manifolds have additionally been considered in the case furthermore equipped with differentiable structure.
Furthermore, individual strata being manifolds, some are metrizable. 
E.g. Pflaum \cite{Pflaum} then considers Riemannian metric structures on stratified spaces (Kendall \cite{Kendall} also makes use of this level of structure).
This permits Pflaum to give furthermore a definition of geodesic distance.  
Pflaum also considers the morphisms corresponding to stratified manifolds.

\mbox{ } 

\ni Next note that stratified manifolds and bundle theory do not fit well together due to stratified manifolds' local structure varying from point to point. 
Three distinct strategies to deal with this are outlined in Sec 1.3.  
Relational considerations point to the strategy of accepting the stratified manifold.  
This points to seeking a generalization of Fibre Bundle Theory, for which Sheaf Theory (Appendix \ref{PreSheaves}) is a strong candidate.  

\mbox{ } 

\ni Finally note that stratified orbifolds also make sense, and indeed occur in configuration space study: the 3-$d$ case of Fig \ref{Tutti-Tri-2c}.f).

\subsection{Hausdorff second-countable locally compact (H2LC) spaces}\label{H2LC}

$\FrX$ is {\it locally compact} if each point in $\FrX$ is contained in a compact neighbourhood lying within $\FrX$.  

\mbox{ }

\ni In particular, these include Hausdorff second-countable compact spaces, and the outcome of the coning construction.
Thus many of Sec \ref{Q-Geom}'s configuration spaces from Mechanics fall within this remit.

\mbox{ } 

\ni Furthermore, H2LC spaces are rather well-behaved from an Analysis point of view.
As well as being at least `2/3rds of a manifold', HLC suffices to have an analogue of Baire's Category Theorem (c.f. Sec \ref{Hilb-Ban-Fre}), 
and various further Analysis results of note hold for HLC or H2LC \cite{Lee1}.

\subsection{Stratifolds}\label{Stratifolds}

A {\it differential space} is a pairing ($\FrX$, ${\cal C}$) of a topological space $\FrX$ and a function space ${\cal C}$ endowed with algebraic structure; the functions act on $\FrX$.  
The ${\cal C}$ generalizes the standard use of smooth functions in elementary algebraic topology.

\mbox{ } 

{\it Sikorski spaces} \cite{SniBook} are a prominent and quite general example of such a pairing from early on in the literature.
Here $\FrX$ is any topological space and ${\cal C}$ a certain type of subalgebra of the continuous functions $\FrX \rightarrow \mathbb{R}$).

\mbox{ }

\ni However, in this Article I concentrate on the more recent differential spaces of Kreck \cite{Kreck}, which are termed {\it stratifolds}.
These are rather well-behaved through the $\FrX$ half of the pair being HL2C in this case. 
Moreover, the ${\cal C}$ half of the stratifold's pair receives a sheaf interpretation (outlined in Appendix \ref{Sheaves}).    

\mbox{ }

\ni As regards modelling with infinite-$d$ stratifolds, work in this direction has started \cite{EwaldKT15}, centering around sheaf methods and study of cohomology, 
in a Hilbert and Fr\'{e}chet space setting that does extend to manifolds in these senses.

\section{(Pre)sheaves}\label{PreSheaves}

\subsection{Categories}\label{Categories}

\ni A very brief outline of those aspects used in setting up the theory of (pre)sheaves is as follows; see e.g. \cite{Lawvere, MacLane, Sheaves1} for further details.

\mbox{ }

\ni{\bf Categories} $\underset{^\sim}{\mC} = (\mO, M)$ consist of objects $\mO$ and {\it morphisms} $M$ (the maps between the objects, 
$M: \mO \longrightarrow \mO$, obeying the axioms of domain and codomain assignment, identity relations, associativity relations and book-keeping relations.

\ni {\bf Functors} are then maps $\mbox{\Large $F$}: \mC_1 \longrightarrow \mC_2$ 
that obey various further axioms concerning domain, codomain, identity and action on composite morphisms.  

\mbox{ } 

\ni For example, $\underset{^\sim}{\mbox{Sets}}$ is the category of sets; per se this is foundationally trivial, though indirectly it plays a repeated role in the below developments.
$\underset{^\sim}{\mbox{Vec}}$ is the category of vector spaces, and $\underset{^\sim}{\mbox{Top}}$ is the category of topological spaces. 
Finally {\it functor categories} are categories of maps between categories.  

\mbox{ } 

\ni For $\FrY \subset \FrX$, the corresponding {\it inclusion map} is the injection $\iota: \FrY \rightarrow \FrX$ with $\iota(y) = y \mbox{ } \forall \, y \in \FrY$.

\subsection{Presheaves}\label{Presheaves}

{\it Presheaves} are then functors $\mbox{\Large $E$}: \underset{^\sim}{\mbox{Top}} \rightarrow \underset{^\sim}{\mbox{Sets}}$ 
(or sometimes some other category such as $\underset{^\sim}{\mbox{Vec}}$) such that the following holds.

{            \begin{figure}[ht]
\centering
\includegraphics[width=0.45\textwidth]{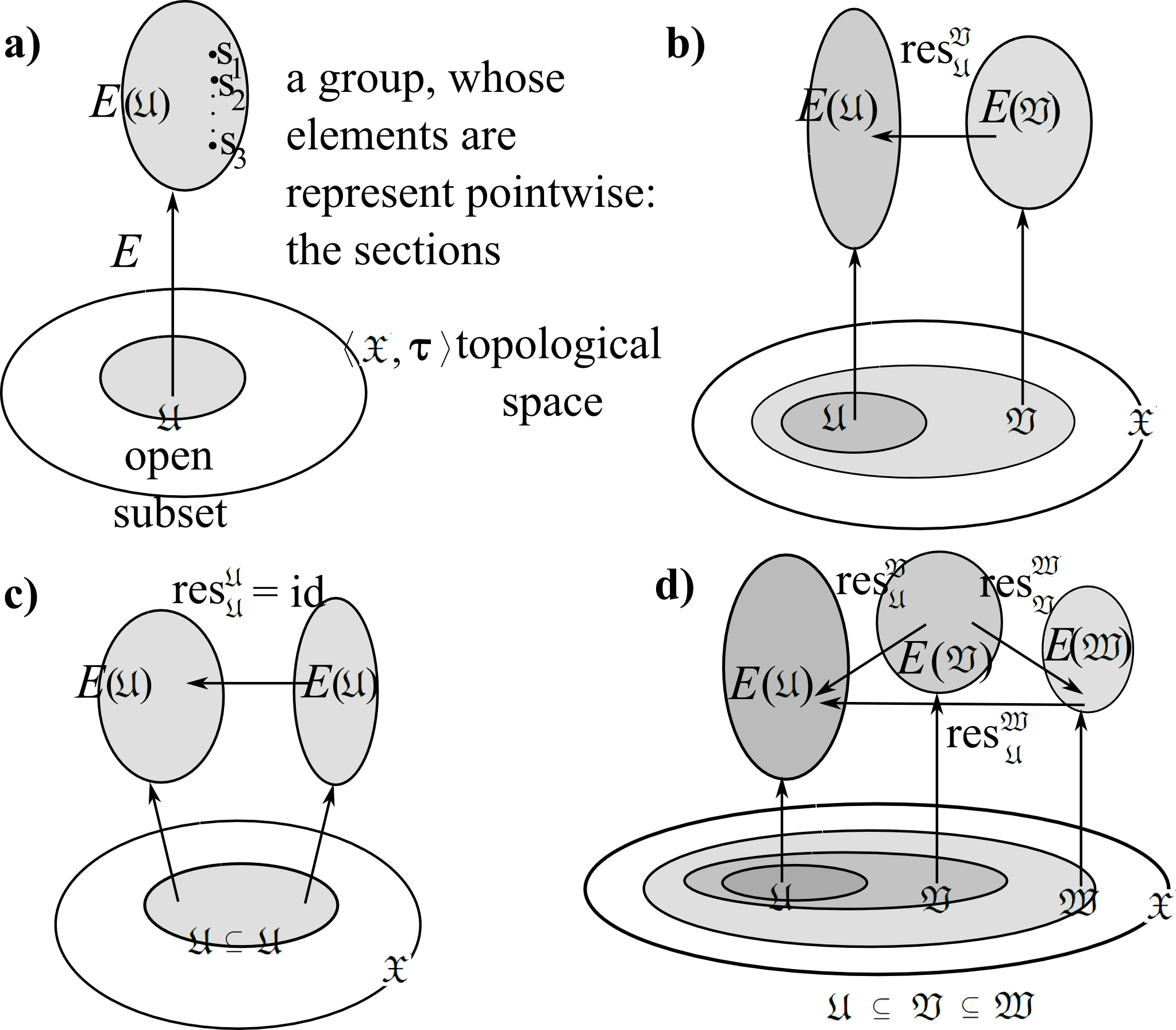}
\caption[Text der im Bilderverzeichnis auftaucht]{        \footnotesize{a) Maps from each open subset $\FrU \in \langle \FrX, \tau \rangle$ to groups of sections over $\FrU$.  
b) These are equipped with restriction maps $\mbox{res}^{\sFrV}_{\sFrU}$ for each $\FrU$ included within each $\FrV$.   
c) The restriction of an open subset to itself is just the identity.
d) Restriction is independent of whether one goes via an intermediate subset: the drawn maps form a commuting triangle.} }
\label{Presheaf-Ax}\end{figure}            }

\ni Presheaf-1) Each inclusion of open sets $\FrV \, \underline{\subset} \, \FrU$ corresponds to a 
{\it restriction morphism} $\mbox{res}_{\sFrV, \sFrU}: \mbox{\Large $E$}(\FrU) \rightarrow  \mbox{\Large $E$}((\FrV)$ in $\underset{^\sim}{\mbox{Sets}}$.\footnote{I subsequently use 
the standard notation for restriction $s|_{\sFrV}$ to denote $\mbox{res}_{\sFrV, \sFrU}(s)$.}

\ni Presheaf-2) $\mbox{res}_{\sFrU, \sFrU}$ is the identity morphism.

\ni Presheaf-3) $\mbox{res}_{\sFrW, \sFrV} \circ \mbox{res}_{\sFrV, \sFrU} = \mbox{res}_{\sFrW, \sFrU}$ (transitivity).  

\mbox{ }

\ni For $\FrU$ an open subset of $\FrX$ (upon which the topological space $\langle \FrX, \tau \rangle$ is based), 
$\mbox{\Large $E$}(\FrU)$ is the {\it section} of $\mbox{\Large $E$}$ over $\FrU$.  
It is a {\it global section} if it is over the whole of $\FrX$ itself.  
Use of the fibre bundle notation $\Gamma$ for sections carries over to presheaves; 
moreover, we now write $\Gamma(\mbox{\Large $E$}, \FrU)$, which is a useful notation since the case in which $\FrU$ rather than $\mbox{\Large $E$}$ is fixed is common.
This notion of section indeed generalizes that of fibre bundles as regards being the gateway to a more general range of global methods.

\subsection{Sheaves}\label{Sheaves}

For a presheaf to additionally be a {\it sheaf} \cite{Sheaves1, Sheaves2, Iversen86, Banagl, Pflaum}, two further conditions are required.

{            \begin{figure}[ht]
\centering
\includegraphics[width=0.7\textwidth]{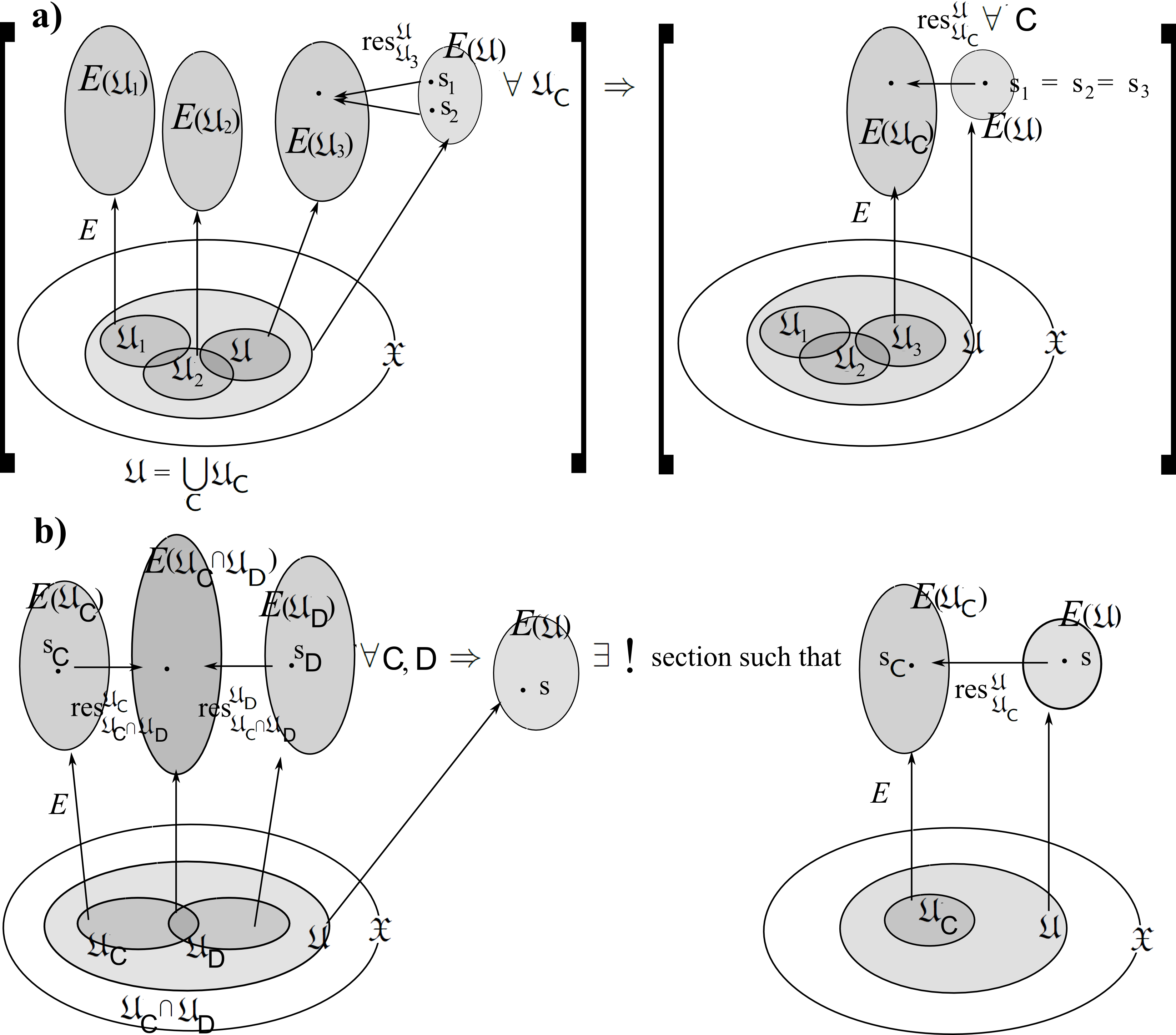}
\caption[Text der im Bilderverzeichnis auftaucht]{        \footnotesize{a) A section is determined by its local restriction in the sense depicted.
b) A section over all of $\FrU$ can be glued together from sections on $\FrU_{\sfC}$ such that $\FrU = \bigcup_{\sfC} \FrU_{\sfC}$ under the depicted circumstances.} }
\label{Sheaf-Ax}\end{figure}            }

\ni Sheaf-1) ({\it local condition}): let \{$\FrU_{\sfC}$\} be an open cover of an open set $\FrU$.
If $r, s \in \mbox{\Large $E$}(\FrU)$ obey $r|_{\sFrU_{\sfC}} = s|_{\sFrU_{\sfC}}$ for each $\FrU_{\sfC}$, then $r = s$.

\ni Sheaf-2) ({\it gluing condition}): let $s_{\sfC} \in      \mbox{\Large $E$}(\FrU_{\sfC})$ be sections that agree on their pairwise overlaps 
$s_{\sfC}|_{\sFrU_{\sfC} \cap \sFrU_{\sfD}} = s_{\sfD}|_{\sFrU_{\sfC} \cap \sFrU_{\sfD}}$.

\ni Then there exists a section $s \in \mbox{\Large $E$}(\FrU_{\sfC})$ with $s|_{\sFrU_{\sfC}} = s_{\sfC}$ for each $i$ in the cover.\footnote{Another manner in which sheaves generalize 
fibre bundles is in possessing a notion of connection.}  

\mbox{ }  

\ni Example 1) Each of the sets of: smooth, real-analytic and complex-analytic functions can be viewed as sheaves. 
This includes in the setting of these being defined ove suitable manifolds \cite{Wells}.
The reader might wish to verify this statement and to show that the bounded functions on $\mathbf{C}$ do not form a sheaf.  

\ni Example 2) Each of the sets of smooth, real-analytic and complex-analytic sections of a vector bundle form a sheaf \cite{Wells}.
This illustrates how bundles themselves can carry sheaf structure.

\mbox{ } 

\ni Sheaves additionally have a notion of {\it section} \index{section!-sheaf|\textbf} $s \in \mbox{\Large $E$}(\FrU_{\sfC})$ 
with $\ms|_{\sFrU_{\tfC}} = \ms_{\sfC}$ for each $\fC$ in the cover.

\mbox{ }

\ni Sheaves are thus the basis for more general patching constructs.
One can now attach heterogeneous objects to different base space points rather than attaching homogeneous fibres in the formation of a fibre bundle. 
A very simple application of this is to the heterogeneous types of chart involved in the study of a given nontrival stratified manifold as per Fig \ref{Strat-Sheaf}. 
See e.g. below and \cite{Kreck, Banagl, Pflaum} for a wider range of applications to stratified configuration spaces and phase spaces.

\mbox{ } 

\ni By possessing the gluing constuction, sheaves provide a means of formulating obstructions 
that generalizes the topological treatment using fibre bundles of a number of obstructions that are already familiar in Theoretical Physics.
In each case, the notion of section has an associated notion of cohomology concerning obstructions to the presence of global sections. 
In the case of sheavesw, this has the logical name of {\it sheaf cohomology}, and indeed turns out to be widely useful from a computational perspective \cite{Iversen86}.
This ensures the sheaf encodes the topological level of structure of generalized spaces as well as their geometrical structure.

\mbox{ } 

\ni On paracompact Hausdorff spaces, sheaf cohomology and the somewhat more familiar \v{C}ech cohomology coincide \cite{Brylinski}.  
However, more generally, sheaf cohomology  {\sl extends} \v{C}ech cohomology; indeed, this is how the former was arrived at by Serre \cite{Serre} and by Grothendieck \cite{Groth}.
[Historically, sheaves originated in the French School of mathematics through the works of Leray, Henri Cartan, Serre and Grothendieck.]

\mbox{ } 

\ni Moreover, for all that sheaves were not originally developed with singular spaces in mind, Whitney and Thom's work on the latter proved to be a further place to apply sheaf methods.
The more recent development of stratifolds by Kreck is a further variation on this theme.   
The other half of the stratifold pair is an algebraic structure of continuous functions ${\cal C}$ 
which can be interpreted as an algebraic structure of global sections in the sheaf-theoretic sense.  

\mbox{ }

\ni [As regards whether there is relation between slices and sheaves, there is, though it involves the theory of \'etale spaces, which lies outside of the scope of the current Article.]

\mbox{ } 

\ni As a concluding punchline, {\sl configuration spaces are not in general manifolds, nor are fibre bundle methods always applicable to them either. 
They are more generally stratified manifolds, for which sheaf methods are more natural and more generally applicable.}

\end{appendices}



\begin{thebibliography}{99}

\footnotesize


\bibitem{Lanczos}             C. Lanczos, {\it The Variational Principles of Mechanics} (University of Toronto Press, Toronto 1949). 

\bibitem{Marchal}             See e.g. C. Marchal, {\it Celestial Mechanics} (Elsevier, Tokyo 1990).

\bibitem{LR97}                R.G. Littlejohn and M. Reinsch, ``Gauge Fields in the Separation of Rotations and Internal Motions in the $N$-Body Problem", 
                              Rev. Mod. Phys. {\bf 69} 213 (1997).  

\bibitem{Goldstein}           H. Goldstein, {\it Classical Mechanics} (Addison-Wesley, Reading, Massachusetts 1980).  

\bibitem{Dirac}               P.A.M. Dirac, {\it Lectures on Quantum Mechanics} (Yeshiva University, New York 1964). 

\bibitem{HTBook}              M. Henneaux and C. Teitelboim, {\it Quantization of Gauge Systems} (Princeton University Press, Princeton 1992).   

\bibitem{FileR}               E. Anderson, ``The Problem of Time and Quantum Cosmology in the Relational Particle Mechanics Arena", arXiv:1111.1472.  

\bibitem{Kendall}             D.G. Kendall, D. Barden, T.K. Carne and H. Le, {\it Shape and Shape Theory} (Wiley, Chichester 1999).  

\bibitem{FORD}                E. Anderson, ``Foundations of Relational Particle Dynamics", Class. Quant. Grav. {\bf 25} 025003 (2008), arXiv:0706.3934.

\bibitem{Cones}               E. Anderson, ``Relational Mechanics of Shape and Scale", arXiv:1001.1112.
 
\bibitem{Moeckel}             See e.g. R. Moeckel's Lecture Course, available at http://www.math.umn.edu/~rick/notes/CMNotes.pdf ;
%
                              F. Diacu, {\it Singularities of the $N$-body Problem} (Les Publications CRM, Montr\'{e}al 1992). 
							  
\bibitem{BGS}                 R.A. Battye, G.W. Gibbons and P.M. Sutcliffe, ``Central Configurations in Three Dimensions", Proc. R. Soc. Lond. A {\bf 459} 911 (2003), hep-th/0201101. 
							  
\bibitem{Merzbacher}          E. Merzbacher, {\it Quantum Mechanics} (Wiley, New York 1998). 

\bibitem{BB82}                J.B. Barbour and B. Bertotti, ``Mach's Principle and the Structure of Dynamical Theories", Proc. Roy. Soc. Lond. {\bf A382} 295 (1982). 

\bibitem{B03}                 J.B. Barbour, ``Scale-Invariant Gravity: Particle Dynamics", Class. Quant. Grav. {\bf 20} 1543 (2003), gr-qc/0211021.
		
\bibitem{QuadI}               E. Anderson, ``Relational Quadrilateralland. I. The Classical Theory", Int. J. Mod. Phys. {\bf D23} 1450014 (2014), arXiv:1202.4186.

\bibitem{Mont93}              R. Montgomery, ``Gauge Theory of the Falling Cat", Fields Institute Communications {\bf 1} 193 (1993).

\bibitem{AG}                  A. Abrams and R. Ghrist. "Finding Topology in a Factory: Configuration Spaces." American Mathematical Monthly 140 (2002); 
%
%
                              S.M. LaValle, {\it Planning Algorithms} (Cambridge University Press, Cambridge 2006); 
%
					          R. Ghrist, ``Configuration Spaces, Braids and Robotics", 
							  in {\it Braids}, in: Lect. Notes Ser. Inst. Math. Sci. Natl. Univ. Singap. {\bf 19} (2010).

\bibitem{AMP}                 Y. Choquet-Bruhat, C. DeWitt-Morette and M. Dillard-Bleick, {\it Analysis, Manifolds and Physics} Vol. 1 (Elsevier, Amsterdam 1982).  
 							  
\bibitem{Nakahara}            M. Nakahara, {\it Geometry, Topology and Physics} (Institute of Physics Publishing, London 1990).   

\bibitem{Battelle}            J.A. Wheeler, in {\it Battelle Rencontres: 1967 Lectures in Mathematics and Physics} ed. C. DeWitt and J.A. Wheeler (Benjamin, New York 1968). 

\bibitem{DeWitt67}            B.S. DeWitt, ``Quantum Theory of Gravity. I. The Canonical Theory.", Phys. Rev. {\bf 160} 1113 (1967).

\bibitem{Fischer70}           A.E. Fischer, ``The Theory of Superspace", in {\it Relativity} (Proceedings of the Relativity Conference in the Midwest, held at Cincinnati, 
                              Ohio June 2-6, 1969), ed. M. Carmeli, S.I. Fickler and L. Witten (Plenum, New York 1970). 

\bibitem{Magic}               C.W. Misner, ``Minisuperspace", in {\it Magic Without Magic: John Archibald Wheeler} ed. J. Klauder (Freeman, San Francisco 1972).

\bibitem{York74}              J.W. York Jr., ``Covariant Decompositions of Symmetric Tensors in the Theory of Gravitation", Ann. Inst. Henri Poincar\'{e} {\bf 21} 319 (1974).  

\bibitem{Fischer86}           A.E. Fischer, ``Resolving the Singularities in the Space of Riemannian Geometries", J. Math. Phys {\bf 27} 718 (1986).  

\bibitem{Giu95}               D. Giulini, ``On the Configuration Space Topology in General Relativity", Helv. Phys. Acta {\bf 68} 86 (1995), gr-qc/9301020.   

\bibitem{Giu95b}              D. Giulini, ``What is the Geometry of Superspace?", Phys. Rev. {\bf D51} 5630 (1995), gr-qc/9311017. 

\bibitem{FM96}                A.E. Fischer and V. Moncrief, ``A Method of Reduction of Einstein's Equations of Evolution and a Natural Symplectic Structure on the Space of Gravitational 
                              Degrees of Freedom", Gen. Rel. Grav. {\bf 28}, 207 (1996).

\bibitem{Giu09}               D. Giulini, ``The Superspace of Geometrodynamics", Gen. Rel. Grav. {\bf 41} 785 (2009) 785, arXiv:0902.3923.  

\bibitem{RWRAM13}             J.B. Barbour, B.Z. Foster and N. \'{o} Murchadha, ``Relativity Without Relativity", Class. Quant. Grav. {\bf 19} 3217 (2002), gr-qc/0012089; 
%
                              E. Anderson and F. Mercati, ``Classical Machian Resolution of the Spacetime Construction Problem", arXiv:1311.6541. 
							  
\bibitem{ADM}                 R. Arnowitt, S. Deser and C. Misner, ``The Dynamics of General Relativity", in {\it Gravitation: An Introduction 
                              to Current Research} ed. L. Witten (Wiley, New York 1962), arXiv:gr-qc/0405109.  

\bibitem{K92I93APoT123}       K.V. Kucha\v{r}, ``Time and Interpretations of Quantum Gravity", 
                              in {\it Proceedings of the 4th Canadian Conference on General Relativity and Relativistic Astrophysics} 
                              ed. G. Kunstatter, D. Vincent and J. Williams (World Scientific, Singapore, 1992),
%
                              reprinted as Int. J. Mod. Phys. Proc. Suppl. {\bf D20} 3 (2011);  
%
                              C.J. Isham, ``Canonical Quantum Gravity and the Problem of Time", in {\it Integrable Systems, Quantum Groups and Quantum Field Theories} 
                              ed. L.A. Ibort and M.A. Rodr\'{\i}guez (Kluwer, Dordrecht 1993), gr-qc/9210011; 
%
                              E. Anderson, ``The Problem of Time in Quantum Gravity", in {\it Classical and Quantum Gravity: Theory, Analysis and Applications} 
                              ed. V.R. Frignanni (Nova, New York 2012), arXiv:1009.2157;  
%
                              ``Problem of Time in Quantum Gravity", Annalen der Physik, {\bf 524} 757 (2012),  arXiv:1206.2403;    
%
                              ``Problem of Time and Background Independence: the Individual Facets", arXiv:1409.4117.  

\bibitem{ABook}               E. Anderson, {\it Problem of Time between Quantum Mechanics and General Relativity}, forthcoming Book.

\bibitem{I91I03}              C.J. Isham, ``Canonical Groups And The Quantization Of Geometry And Topology", in {\it Conceptual Problems of Quantum Gravity} ed. 
                              A. Ashtekar and J. Stachel (Birkh\"{a}user, Boston, 1991); 
%
                              ``A New Approach to Quantising Space-Time: I. Quantising on a General Category", Adv. Theor. Math. Phys. 7 (2003) 331-367 arXiv:gr-qc/0303060. 

\bibitem{BONew}              J.B. Barbour and N. \'{o} Murchadha, ``Conformal Superspace: the Configuration Space of General Relativity", arXiv:1009.3559. 
							  						  
\bibitem{AMech}              E. Anderson,``The Future Shape of Theoretical Physics and Beyond", arXiv:1505.00488. 
							  							  
\bibitem{ASoS}               E. Anderson, ``Spaces of Spaces", arXiv.1412.0239.

\bibitem{York72}             J.W. York Jr., ``Role of Conformal Three-Geometry in the Dynamics of Gravitation", Phys. Rev. Lett. {\bf 28} 1082 (1972).

\bibitem{ABFKO}              E. Anderson, J.B. Barbour, B.Z. Foster, B. Kelleher and N. \'{o} Murchadha, ``The Physical Gravitational Degrees of Freedom", 
                             Class. Quant. Grav {\bf 22} 1795 (2005), gr-qc/0407104.   

\bibitem{IVP}                J.W. York Jr., ``Conformally Invariant Orthogonal Decomposition of Symmetric Tensors on Riemannian Manifolds and the 
                             Initial-Value Problem of General Relativity", J. Math. Phys. {\bf 14} 456 (1973);
%
                             Y. Choquet-Bruhat and J.W. York 1980, in {\it General Relativity and Gravitation} ed. A Held Vol 1 (Plenum Press, New York, 1980);  
%
                             J. Isenberg, ``Constant Mean Curvature Solutions of the Einstein Constraint Equations on Closed Manifolds", Class. Quantum Grav. {\bf 12} 2249 (1995); 
%
							 T.W. Baumgarte and S.L. Shapiro, ``Numerical Relativity and Compact Binaries", Phys. Rept. {\bf 376} 41 (2003), gr-qc/0211028;
%
                             R. Bartnik and J. Isenberg, ``The Constraint Equations", in {\it The Einstein Equations and the Large Scale Behavior of Gravitational Fields} 
                             ed. P. Chrusciel and H. Friedrich (Birkh\"{a}user, Basel 2004); 
%
                             E. Gourgoulhon, {\it 3+1 Formalism in General Relativity: Bases of Numerical Relativity} (Lecture Notes in Physics, Vol. 846, Springer, Berlin 2012); 
                             an earlier version of this is available at gr-qc/0703035.
							 
\bibitem{BI75HH83}           W.F. Blyth and C.J. Isham, ``Quantization of a Friedmann Universe Filled with a Scalar Field"  Phys. Rev. {\bf D11} 768 (1975); 
%
                             J.B. Hartle and S.W. Hawking, ``Wave Function of the Universe", Phys. Rev. {\bf D28} 2960 (1983).   

\bibitem{Mis68}               C.W. Misner, ``The Isotropy of the Universe", Astrophys. J. {\bf 151} 431 (1968).  

\bibitem{SIC1}               E. Anderson, ``Problem of Time in Slightly Inhomogeneous Cosmology", arXiv:1403.7583. 
							 
\bibitem{HallHaw}            J.J. Halliwell and S.W. Hawking, ``Origin of Structure in the Universe", Phys. Rev. {\bf D31}, 1777 (1985).

\bibitem{Ashtekar}           A. Ashtekar, {\it Lectures on Nonperturbative Canonical Gravity} (World Scientific, Singapore 1991).
		
\bibitem{Armstrong}          M.A. Armstrong, {\it Basic Topology} (Springer-Verlag, New York 1983).

\bibitem{Livingston}          C. Livingston, {\it Knot Theory} (Math. Assoc. of America, Washington 1993).  

\bibitem{Graphs}			     B. Bollob\'{a}s, {\it Modern Graph Theory}, (Springer-Verlag, New York 1998).		
		
\bibitem{KauffmanBook}       L.H. Kauffman, {\it Knots and Physics} (World Scientific, Singapore 1994).
							 
\bibitem{TSC2}               R. Bartnik and G. Fodor, ``On the Restricted Validity of the Thin-Sandwich Conjecture", Phys. Rev. {\bf D48} 3596 (1993). 
 
\bibitem{Records}            E. Anderson, ``Records Theory", Int. J. Mod. Phys. {\bf D18} 635 (2009), arXiv:0709.1892; 
%
							``Kendall's Shape Statistics as a Classical Realization of Barbour-type Timeless Records Theory approach to Quantum Gravity", 
                             Stud. Hist. Phil. Mod. Phys. {\bf 51} 1 (2015), arXiv:1307.1923. 
	
\bibitem{Small}              C.G.S. Small, {\it The Statistical Theory of Shape} (Springer, New York, 1996).  
		
\bibitem{NSIPW83B94IIEOTPage12} S.W. Hawking and D.N. Page, ``Operator Ordering and the Flatness of the Universe", Nucl. Phys. {\bf B264} 185 (1986); 
%
                             D.N. Page and W.K. Wootters,  ``Evolution Without Evolution: Dynamics Described by Stationary Observables", Phys. Rev. {\bf D27}, 2885 (1983); 
%
                             J.B. Barbour, ``The Timelessness of Quantum Gravity. II: The Appearance of Dynamics in Static Configurations", Class. Quant. Grav. {\bf 11} 2875 (1994); 
%
                             {\it The End of Time} (Oxford University Press, Oxford 1999);
%
                             D.N. Page, Int. J. Mod. Phys. D5 583 (1996), quant-ph/9507024; 
%
                             arXiv:1102.5339.  	
							 							 								
\bibitem{B94I}               J.B. Barbour, ``The Timelessness of Quantum Gravity. I. The Evidence from the Classical Theory",               Class. Quant. Grav. {\bf 11} 2853 (1994).

\bibitem{Rovelli}            C. Rovelli, {\it Quantum Gravity} (Cambridge University Press, Cambridge 2004).  
							 
\bibitem{APoB}               E. Anderson, ``Beables/Observables in Classical and Quantum Gravity", SIGMA {\bf 10} 092 (2014), arXiv:1312.6073. 
																							
\bibitem{GMHHartleILH03H09}  M. Gell-Mann and J.B. Hartle, ``Decoherence as a Fundamental Phenomenon in Quantum Dynamics", Phys. Rev. {\bf D47} 3345 (1993);        
%
                             J.B. Hartle, ``Spacetime Quantum Mechanics and the Quantum Mechanics of Spacetime", in {\it Gravitation and Quantizations: 
                             Proceedings of the 1992 Les Houches Summer School} ed. B. Julia and J. Zinn-Justin (North Holland, Amsterdam 1995),  gr-qc/9304006. 
%
                             C.J. Isham and N. Linden, ``Continuous Histories and the History Group in Generalized Quantum Theory", J. Math. Phys. {\bf 36} 5392 (1995), gr-qc/9503063;  
%
                             J.J. Halliwell, ``The Interpretation of Quantum Cosmology and the Problem of Time", in {\it The Future of Theoretical Physics and Cosmology} 
                             (Stephen Hawking 60th Birthday Festschrift volume) ed. G.W. Gibbons, E.P.S. 
                             Shellard and S.J. Rankin (Cambridge University Press, Cambridge 2003), arXiv:gr-qc/0208018; 
%
                            ``Probabilities in Quantum Cosmological Models: A Decoherent Histories Analysis Using a Complex Potential", Phys. Rev. {\bf D80} 124032 (2009), arXiv:0909.2597.

\bibitem{Woodhouse}          N.M.J. Woodhouse, {\it Geometric Quantization} (Clarendon Press, Oxford 1997).  
							
\bibitem{I84}                C.J. Isham, ``Topological and Global Aspects of Quantum Theory", 
                             in {\it Relativity, Groups and Topology {II}}, ed. B. DeWitt and R. Stora (North-Holland, Amsterdam 1984). 

\bibitem{DeWitt57}           B.S. DeWitt, ``Dynamical Theory in Curved Spaces. [A Review of the Classical and Quantum Action Principles.]", Rev. Mod. Phys. {\bf 29} 377 (1957).  
							 

\bibitem{ArchRatMontgomery2}  R. Montgomery, ``Infinitely Many Syzygies", Arch. Rat. Mech. Anal. {\bf 164} 311 (2002); 
%
                              ``Fitting Hyperbolic Pants to a 3-Body Problem", Ergod. Th. Dynam. Sys. {\bf 25} 921 (2005), math/0405014.

\bibitem{+Tri}                E. Anderson, ``Shape Space Methods for Quantum Cosmological Triangleland", Gen. Rel. Grav. {\bf 43} 1529 (2011), arXiv:0909.2439.  

\bibitem{AF}                  E. Anderson and A. Franzen, ``Quantum Cosmological Metroland Model", Class. Quant. Grav. {\bf 27} 045009 (2010), arXiv:0909.2436. 
 
\bibitem{Kendall89}           D.G. Kendall, ``A Survey of the Statistical Theory of Shape", Statistical Science {\bf 4} 87 (1989).

\bibitem{DragtIwai87}         A.J. Dragt, ``Classification of Three-Particle States According to $SU_3$, J. Math. Phys. {\bf 6} 533 (1965); 
%
                              T. Iwai, ``A Geometric Setting for Internal Motions of the Quantum Three-Body System", J. Math. Phys. {\bf 28} 1315 (1987).  

\bibitem{08I}                 E. Anderson, ``Triangleland. I. Classical Dynamics with Exchange of Relative Angular Momentum", Class. Quant. Grav. {\bf 26} 135020 (2009), arXiv:0809.1168.  

\bibitem{DeWitt70}            B.S. DeWitt, ``Spacetime as a Sheaf of Geodesics in Superspace", in {\it Relativity} (Proceedings of the Relativity Conference in the Midwest, 
                              held at Cincinnati, Ohio June 2-6, 1969), ed. M. Carmeli, S.I. Fickler and L. Witten (Plenum, New York 1970). 

\bibitem{Gromov}              M. Gromov, ``Metric Structures for Riemannian and Non-Riemannian Spaces", (Birkh\"{a}user, Boston 1999).

\bibitem{RS}                  M. Reed and B. Simon {\it Methods of Modern Mathematical Physics. II. Fourier Analysis, Self-Adjointness} (Academic Press, New York 1975).  

\bibitem{KieferBook}          C. Kiefer, {\it Quantum Gravity} (Clarendon, Oxford 2004).  

\bibitem{Teitelboim}          C. Teitelboim, in {\it General Relativity and Gravitation} Vol 1 ed. A. Held (Plenum Press, New York 1980).

\bibitem{MacCallum}           H. Stephani, D. Kramer, M.A.H. MacCallum, C.A. Hoenselaers, and E. Herlt
                              {\it Exact Solutions of Einstein's Field Equations} 2nd Edition (Cambridge University Press, Cambridge 2003).

\bibitem{BKL}                 V.A. Belinskii, I.M. Khalatnikov and E.M. Lifshitz, ``Oscillatory Approach to a Singular Point in the Relativistic Cosmology", 
                              Adv. Phys. {\bf 19} 525 (1970).

\bibitem{AHH2}                E. Anderson, ``Origin of Structure in the Universe: Quantum Cosmology Reconsidered", arXiv:1501.02443.	

\bibitem{KR86}                W. Kondracki and J. Rogulski, ``On the Stratification of the Orbit Space for the Action of Automorphisms on Connections", Diss. Math. {\bf 250} (1986).

\bibitem{RSV02}               G. Rudolph, M. Schmidt and I.P. Volobuev, ``On the Gauge Orbit Space Stratification: a Review", J. Phys. A. Math. Gen. {\bf 35} R1 (2002).  

\bibitem{Schmidt03}           M. Schmidt, ``How to Study the Physical Relevance of Gauge Orbit Space Singularities?" Rep. Math. Phys. {\bf 51} 325 (2003). 

\bibitem{GPBook}              R. Gambini and J. Pullin {\it Loops, Knots, Gauge Theories and Quantum Gravity} (Cambridge University Press, Cambridge 1996).

\bibitem{Hamilton82}          R.S. Hamilton, ``The Inverse Function Theorem of Nash and Moser", Bull. Amer. Math. Soc. {\bf 7} 65 (1982). 

\bibitem{MS39}                S.B. Meyers and N.E. Steenrod. ``The Group of Isometries of a Riemannian Manifold", Ann. Math. {\bf 40} 400 (1939). 

\bibitem{Ebin}                D.G. Ebin, ``The Manifold of Riemannian Metrics", Proc. Symp. Pure Math. AMS {\bf 15} 11 (1970).  

\bibitem{Giu94}               D. Giulini, ``Properties of 3-Manifolds for Relativists", Int. J. Theor. Phys. {\bf 33} 913 (1994), gr-qc/9308008. 

\bibitem{OR02Schmah}          P. Ortega and T.S Ratiu, ``Optimal Momentum Map", in {\it Geometry, Mechanics and Dynamics: Volume in Honour of the 60th Birthday of Jerrold Marsden} 
                              ed P. Newton, P. Holmes and A. Weinstein (Springer--Verlag, New York 2002);
%
    						  T. Schmah, ``A Cotangent Bundle Slice Theorem", Diff. Geom. Appl.{\bf 25} 101 (2007), math/0409148.

\bibitem{MR99}                J.E. Marsden and T.S. Ratiu, {\it Introduction to Mechanics and Symmetry} (Springer, New York 1999).

\bibitem{Misref}              C.W. Misner, ``Classical and Quantum Dynamics of a Closed Universe",, in {\it Relativity (Proceedings of the Relativity Conference in the Midwest, held at Cincinnati, Ohio June 2-6, 1969)} 
                              ed. M. Carmeli, S.I. Fickler and L. Witten (Plenum, New York 1970).

\bibitem{DSVV09}              D. Dudal, S.P. Sorella, N. Vandersickel and H. Verschelde,``Gribov no-pole condition, Zwanziger horizon function, Kugo-Ojima confinement criterion, 
                              boundary conditions, BRST breaking and all that", Phys. Rev. {\bf D79} 121701 (2009), arXiv:0904.0641.
   														  
\bibitem{FM77}                A.E. Fischer and J.E. Marsden, ``The Manifold of Conformally Equivalent Metrics", Can. J. Math. {\bf 1} 193 (1977).

\bibitem{York73}              J.W. York Jr., ``Conformally Invariant Orthogonal Decomposition of Symmetric Tensors on Riemannian Manifolds and the 
                              Initial-Value Problem of General Relativity", J. Math. Phys. {\bf 14} 456 (1973).

\bibitem{AHH3}                E. Anderson, ``Kinematical Quantization", forthcoming.                            

\bibitem{Zala12}              R. Zalaletdinov, ``Averaging out the Einstein equations and Macroscopic Spacetime Geometry", Gen. Rel. Grav. {\bf 24} 1015 (1992);  
%
                              ``Towards a Macroscopic Theory of Gravity", 673 {\bf 25} (1993). 
							  
\bibitem{AL93}                A. Ashtekar, and J. Lewandowski, J., ``Representation Theory of Analytic Holonomy $C^*$ algebras", 
                              in {\it Knots and Quantum Gravity, Proceedings of Workshop held at UC Riverside on May 14-16, 1993, 
							  Oxford Lecture Series in Mathematics and its Applications} {\bf 1} ed. J.C. Baez (Clarendon, Oxford and OUP, New York, 1994), arXiv:gr-qc/9311010.   

\bibitem{BF07}                J. Brunnemann and C. Fleischhack, ``On the Configuration Spaces of Homogeneous Loop Quantum Cosmology and Loop Quantum Gravity", arXiv:0709.1621.

\bibitem{F10}                 C. Fleischhack, ``Loop Quantization and Symmetry: Configuration Spaces", arXiv:1010.0449.
							
\bibitem{F00}                 C. Fleischhack, ``Stratification of the Generalized Gauge Orbit Space", Commun. Math. Phys. {\bf 214} 607 (2000), math-ph/0001006.  

\bibitem{Witten89}            E. Witten, ``Quantum Field Theory and the Jones Polynomial", Comm. Math. Phys. {\bf 121} 351 (1989).  

\bibitem{RS88}                C. Rovelli and L. Smolin, ``Knot Theory and Quantum Gravity", Phys. Rev. Lett. {\bf 61} 1155 (1988).  

\bibitem{Vass}                V.A. Vassiliev, ``Cohomology of Knot Spaces", in {\it Theory of Singularities (Advances in Soviet Math., vol. 1)} (American Math. Society, 1990).
 
\bibitem{CDM}                 S. Chmutsov, S. Duzhin and A. Mostovoy, {\it Introduction to Vassiliev Knot Invariants} (Cambridge University Press, Cambridge 2012). 
 
\bibitem{Budney}              A.R. Budney, ``Little Cubes and Long Knots", Topology {\bf 46} 1 (2007). 

\bibitem{Kendall84}           D.G. Kendall, ``Shape Manifolds, Procrustean Metrics and Complex Projective Spaces", Bull. Lond. Math. Soc. {\bf 16} 81 (1984). 

\bibitem{Pflaum2}             M.J. Pflaum, ``Smooth Structures on Stratified Spaces" Progress in Mathematics {\bf 198} 231 (2001).

\bibitem{Pflaum}              M.J. Pflaum, {\it Analytic and Geometric Study of Stratified Spaces}, Lecture Notes in Mathematics {\bf 1768} (Springer, Berlin 2001).  

\bibitem{IY05}                T. Iwai and H. Yamaoka, ``Stratified Reduction of Classical Many-Body Systems with Symmetry", J. Phys. A. Math. Gen {\bf 38} 2415 (2005).  

\bibitem{MinkProb}            J. Franchi and Y. Le Jan, ``Relativistic Diffusions and Schwarzschild Geometry", Commun. Pure Appl. Math. {\bf 60} 187 (2007).
  

\bibitem{Lee1}                 J.M. Lee, {\it Introduction to Smooth Manifolds} (Springer, New York 2003).   

\bibitem{Whitney36}           H. Whitney, ``Differentiable Manifolds", Ann. Math. {\bf 37} 645 (1936).

\bibitem{Whitney35}           H. Whitney, ``Sphere Spaces", Proc. Nat. Acad. Sci. USA {\bf 21} 464 (1935), 464. 

\bibitem{IshamBook}           {\it Modern Differential Geometry for Physicists} (World Scientific, Singapore 1999).

\bibitem{Husemoller}          D. Husemoller, {\it Fibre Bundles} (Springer, New York 1994).							  
 
\bibitem{NSBook}              C. Nash and S. Sen, {\it Topology and Geometry for Physicists} (Dover, New York 2011). 
 
\bibitem{Wells}               R.O. Wells, ``Differential Analysis on Complex Manifolds" (Springer, New York 2008).  
 							  
\bibitem{Bleecker}            D. Bleecker, {\it Gauge Theory and Variational Principles} (Dover, New York 2001). 

\bibitem{DS88}                N. Dunford and J.T. Schwarz {\it Linear Operators. Part I. General Theory}, (Wiley, Hoboken N.J. 1988).			

\bibitem{Lee2}                J.M. Lee, {\it Introduction to Smooth Manifolds} 2nd Ed. (Springer, New York 2013).

\bibitem{Lang95}              S. Lang, {\it Differential and Riemannian Manifolds} (Springer, New York 1995). 

\bibitem{Munkres00}           J.R. Munkres, {\it Topology} (Prentice--Hall, Upper Saddle River, New Jersey 2000).

\bibitem{Orbi}                J. Davey, A. Hanany, R.-K. Seong, JHEP {\bf 06} 010 (2010),  arXiv:1002.3609; 
%
                              M. Green, J. Schwarz and E. Witten {\it Superstring Theory. Volume 2. Loop Amplitudes, Anomalies and Phenomenology} 
                             (Cambridge University Press, Cambridge 1987).

\bibitem{IM82}                J. Isenberg and J.E. Marsden, {\it A Slice Theorem for the Space of Solutions of Einstein's Equations}, Phys. Rep. {\bf 89} 179 (1982).  

\bibitem{Whitney46}           H. Whitney, ``Complexes of Manifolds", Proc. Nat. Acad. Sci. USA {\bf 33} 10 (1946).  
							 
\bibitem{Whitney65}           H. Whitney, ``Tangents to an Analytic Variety", Ann. Math. {\bf 81} 496 (1965). 

\bibitem{Thom55}              R. Thom, ``Les Singularit\'{e}s des Applications Diff\'{e}rentiables" (Singularities in Differentiable Maps), Ann. Inst. Fourier (Grenoble) {\bf 6} 43 (1955).

\bibitem{Thom69}              R. Thom, ``Ensembles et Morphismes Stratifi\'{e}s" (Stratified Spaces and Morphisms), Bull. Amer. Math. Soc. (N.S.) {\bf 75} 240 (1969).
 
\bibitem{SniBook}             J. \'{S}niatycki, {\it Differential Geometry of Singular Spaces and Reduction of Symmetry} (Cambridge University Press, Cambridge 2013).

\bibitem{Kreck}               M. Kreck, {\it Differential Algebraic Topology: From Stratifolds to Exotic Spheres} (American Mathematical Society, Providence 2010).							 

\bibitem{EwaldKT15}           C-0. Ewald, {\it Hochschild Homology and De Rham Cohomology of Stratifolds}, (PhD Thesis, University of Bonn 2002), http://archiv.ub.uni-heidelberg.de/volltextserver/2918/ ;
%
                              H. Tene, {\it Some geometric equivariant cohomology theories} (PhD Thesis, University of Bonn 2010), arXiv:1210.7923; 
%
                              M. Kreck and H. Tene, ``Hilbert Stratifolds and a Quillen Type Geometric Description of Cohomology for Hilbert Manifolds", arXiv:1506.07075.   			

\bibitem{Lawvere}             F.W. Lawvere and S.H. Schanuel, {\it Conceptual Mathematics. A First Introduction to Categories} (Cambridge University Press, Cambridge 2009); 
%
                              F.W. Lawvere and R. Rosebrugh {\it Sets for Mathematics} (Cambridge University Press, Cambridge 2003).

\bibitem{MacLane}             S. Mac Lane, {\it Categories for the Working Mathematician} (Springer-Verlag, New York 1998). 
 
\bibitem{Sheaves1}            J.L. Bell, {\it Toposes and Local Set Theories} (Dover, New York 2008).  

\bibitem{Sheaves2}            G.E. Bredon, {\it Sheaf Theory} (McGraw--Hill, New York 1997).

\bibitem{Brylinski}           J-L. Brylinski, {\it Loop Spaces, Characteristic Classes and Geometric Quantization} (Springer--Verlag, New York 2007).

\bibitem{Iversen86}           B. Iversen, {\it Cohomology of Sheaves} (Springer--Verlag, Berlin 1986).

\bibitem{Banagl}              M. Banagl, {\it Topological Invariants of Stratified Spaces} (Springer--Verlag, Berlin 2007).

\bibitem{Serre}               J-P. Serre, ``Faisceaux Algebraiques Coherents" (Coherent Algebraic Sheaves), Ann. Math. {\bf 61} 197 (1955).

\bibitem{Groth}               A. Grothendieck, ``Sur Quelques Points d'Algebre Homologique" I. Tohoku Math. J. {\bf 9} 119 (1957).
 
\end{thebibliography}
\end{document}